\documentclass[12pt,cite,epsfig]{article}
\usepackage{times,epsfig}
\usepackage{cancel}                                                              

\newcommand{\nc}{\newcommand}
\nc{\bb}{\bibitem}
\nc{\be}{\begin{equation}}
\nc{\ee}{\end{equation}}
\nc{\pa}{\partial}
\nc{\parsym} {\stackrel{\leftrightarrow}{\pa}}
\nc{\ra}{\rightarrow}
\nc{\la}{\leftarrow}
\nc{\etp}{{\eta^\prime}}
\nc{\omg}{\omega}
\nc{\ggam}{\gamma \gamma}
\nc{\gam}{\gamma }
\nc{\bea}{\begin{eqnarray}}
\nc{\eea}{\end{eqnarray}}
\nc{\beas}{\begin{eqnarray*}}
\nc{\eeas}{\end{eqnarray*}}
\nc{\non}{\nonumber}
\nc{\second}{{\prime\prime}}

\def\hhhb{\rule[-3.mm]{0.mm}{9.mm}}

\def\hhhc{\rule[-3.mm]{0.mm}{3.mm}}
\def\hhhd{\rule[-3.mm]{0.mm}{2.mm}}

\def\hhhu{\rule[-3.mm]{0.mm}{12.mm}}
\def\hhhv{\rule[-3.mm]{0.mm}{9.mm}}
\def\hhhw{\rule[-1.mm]{0.mm}{5.mm}}

\begin{document}
\begin{titlepage}
\vbox{~~~ \\
                                   \null \hfill LPNHE/2012-01\\
                                   \null \hfill HU-EP/12-40 \\
                                   \null \hfill DESY 12-190 \\
				       
\title{An Update of the HLS Estimate of the Muon $g-2$\\
   }
\author{
M.~Benayoun$^a$, P.~David$^a$, L.~DelBuono$^a$, F.~Jegerlehner$^{b,c}$ \\
\small{$^a$ LPNHE des Universit\'es Paris VI et Paris VII, IN2P3/CNRS, F-75252 Paris, France }\\
\small{ $^b$ Humboldt--Universit\"at zu Berlin, Institut f\"ur Physik, Newtonstrasse 15, D--12489 Berlin,
Germany }\\
\small{ $^c$ Deutsches  Elektronen--Synchrotron (DESY), Platanenallee 6, D--15738 Zeuthen, Germany}
}
\date{\today}
\maketitle
\begin{abstract}
A global fit of parameters allows us to pin down the Hidden Local
Symmetry (HLS) effective Lagrangian, which we apply for the prediction
of the leading hadronic vacuum polarization contribution to the muon
$g-2$. The latter is dominated by the annihilation channel $e^+e^- \to
\pi^+\pi^-$, for which data are available by scan (CMD-2 \& SND) and
ISR (KLOE-2008, KLOE-2010 \& BaBar) experiments. It is well known that
the different data sets are not in satisfactory agreement. In fact it
is possible to fix the model parameters without using the $\pi^+\pi^-$
data, by using instead the dipion spectra measured in the
$\tau$-decays together with experimental spectra for the
$\pi^0\gamma$, $\eta \gamma$, $\pi^+\pi^-\pi^0$, $K^+K^-$,
$K^0\overline{K^0}$ final states, 
supplemented by specific meson decay
properties. Among these, the accepted decay width for $\rho^0 \to
e^+e^-$ and the partial widths and phase information for the $\omega/\phi
\to \pi^+\pi^-$ transitions, are considered. It is then shown that,
relying on this global data set, the HLS model, appropriately broken,
allows to predict accurately the pion form factor
below 1.05 GeV. It is shown that the data
samples provided by CMD-2, SND and KLOE-2010 behave consistently with
each other and with the other considered data. Consistency problems
with the KLOE-2008 and BaBar data samples are substantiated. "All
data" global fits are investigated by applying reweighting the
conflicting data sets. Constraining to our best fit, the broken HLS
model yields $a_\mu^{\rm th} = (11\,659\,169.55 + \left[^{+1.26}_{-0.59} \right]_\phi  
+\left[^{+0.00}_{-2.00} \right]_\tau \pm 5.21_{th})~10^{-10}$
associated with a very good global fit probability. Correspondingly, we
find that $\Delta a_\mu=a_\mu^{\rm exp}-a_\mu^{\rm th}$ exhibits a
significance ranging between 4.7 and 4.9 $\sigma$.
\end{abstract}
}
\end{titlepage}

\section{Introduction}
\label{introduction}
\indent \indent

 The theoretical value for muon anomalous magnetic moment  $a_\mu$ is an
 important window in the quest for  new phenomena in particle physics.
 The predicted value is  the sum of several contributions and the most
 prominent ones are already derived from the Standard Model with  very high accuracies. 
 The QED contribution is thus estimated with
an accuracy of a few $10^{-12}$ \cite{Kinoshita1,Kinoshita2,Passera06} and the precision
of the electroweak contribution  is now of order $10^{-11}$ \cite{Fred09}. 
The light--by--light contribution  to $a_\mu$  is currently 
known  with an accepted accuracy of $2.6 \times 10^{-10}$ \cite{LBL}.  

Presently, the uncertainty of the Standard Model  prediction for $a_\mu$ is driven by
the uncertainty on the leading order (LO) hadronic vacuum 
polarization (HVP) up to  $ \simeq 2$ GeV \cite{DavierHoecker3,Fred11}. 
This region is covered by  the non--perturbative regime of QCD and
the leading order HVP $(LO-HVP)$ is evaluated by means of~: 
\be
\left \{
\begin{array}{lll}
a_\mu^{LO-HVP} &= \sum_{i} a_\mu(H_i)\\[0.5cm]
\displaystyle
a_\mu(H_i) &= 
\displaystyle
\frac{1}{4 \pi^3} \int_{s_{{H_i}}}^{s_{cut}} ds K(s) \sigma_{H_i}(s)~~,
\end{array}
\right .
\label{Eq1}
\ee
which relates the hadronic intermediate state contributions $\{H_i,~i=1 \cdots n\}$ to the annihilation
cross sections $\sigma(e^+ e^- \ra H_i) \equiv \sigma_{H_i}(s)$. $K(s)$ is a 
known kernel \cite{Fred09} enhancing the weight of  
the threshold region $s_{H_i}$ and $s_{cut}$ is some
energy squared where perturbative QCD  starts to be  applicable.
In the region where perturbative QCD holds\footnote{The charmonium and bottomium
regions carry uncertainties also in the range of  a few $10^{-11}$.}, its contribution to $a_\mu$ 
carries an uncertainty of the order of a few $10^{-11}$.

Up to very recently, the single method used to get the $a_\mu(H_i)$'s was
to plug  the experimental cross sections into Eq. (\ref{Eq1}). 
Among the most recent studies based on this method, let us quote 
\cite{DavierHoecker,DavierHoecker3,Fred11,Teubner}. When several
data sets cover the same cross section $\sigma_{H_i}(s)$, Eq. (\ref{Eq1})
is used with some appropriate  weighting   of the various spectra,
allowing to improve the corresponding $a_\mu(H_i)$.

On the other hand, it is now widely accepted that the Vector Meson
Dominance (VMD) concept applies to low energy physics \cite{Ecker1,Prades}. 
VMD based Effective Lagrangians have been proposed like the Resonance
Chiral Perturbation Theory or the Hidden Local Symmetry (HLS) Model;
it has been proven~\cite{Ecker2} that these are essentially equivalent.
Intrinsically, this means that there exist physics correlations between
the various $e^+ e^- \ra H_j$ annihilation channels. Therefore,
it becomes conceptually founded to expect improving each $a_\mu(H_i)$ by
means of the  data covering the other channels $e^+ e^- \ra H_j$ ($j\ne i$).

This is basically the idea proposed in \cite{ExtMod3} relying on the
HLS model \cite{HLSOrigin,HLSRef}. Using a symmetry breaking mechanism
based on the simple BKY idea \cite{BKY} and a vector meson mixing scheme,
the model has been developed 
stepwise \cite{Heath,taupaper,ExtMod1,ExtMod2} and its
most recent form  \cite{ExtMod3} has been shown to provide a successful
$simultaneous$ description of the $e^+ e^-$ annihilation into the
$\pi^+\pi^-$, $\pi^0\gamma$, $\eta\gamma$, $\pi^+\pi^-\pi^0$, $K^+K^-$, 
$K^0 \overline{K}^0$ final states as well as the  $\tau^\pm \ra \nu_\tau \pi^\pm \pi^0$ 
decay spectrum. Some more decays of the form\footnote{We denote by $V$ or $P$ resp. any 
meson belonging to the (basic) vector or pseudoscalar lowest mass nonets.
}  $V \ra P\gamma$ or $P \ra\gamma\gamma$ are considered. 
 
As higher mass meson nonets are absent from the
standard HLS model, its energy scope is {\it a priori} limited upwards 
by the $\phi$ meson  mass region ($ \simeq 1.05$ GeV). However, as this 
region contributes more than 80 \% to the total HVP, improvements
which can follow from the broken HLS model are
certainly valuable\footnote{The broken
HLS model does not include the $4\pi$, $5\pi$, $6\pi$, $\eta \pi\pi$
 and $\omg \pi$ annihilation channels. Therefore, the  (small) contribution of these
 missing channels  \cite{ExtMod3} to $a_\mu$ should be still evaluated by
 direct integration of the experimental cross sections; up to  the $\phi$ mass, 
 this amounts to  \cite{ExtMod3} $(1.55\pm 0.57_{tot})~10^{-10}$. }. 

 The global simultaneous fit of the data corresponding to the channels
quoted above allows to reconstruct the various cross sections $\sigma_{H_i}(s)$
taking automatically into account the physics correlations inside the
set   ${\cal H} \equiv \{H_i\}$ of possible final states and decay
processes. The fit
parameter values and the parameter error covariance matrix 
summarize optimally the full knowledge of ${\cal H}$. 
This has two important consequences~:

\begin{itemize}
\item One should get the $\{a_\mu(H_i), i=1,\cdots n\}$ with improved
uncertainties by integrating the $model$ cross sections instead of the
$measured$ ones. Indeed, the functional correlations
among the various cross sections turn  out to provide (much) larger statistics 
in $each$  channel and thus yield improved uncertainties for $each$
$a_\mu(H_i)$.

\item When several data samples cover the same process $ H_i$, one has a handle
to motivatedly examine the behavior of each within the global fit
context. Stated otherwise, the issue of the consistency of each data set 
with $all$ the others can be addressed with the (global) fit probability as a tool
to detect data samples carrying   problematic properties.   
\end{itemize}

Up to now, the broken HLS model (BHLS) \cite{ExtMod3} -- basically an empty shell -- has been fed
with all existing data sets\footnote{The full list of data sets can be found
in \cite{ExtMod1} or \cite{ExtMod3} together with a critical analysis of their individual behavior.}
for what concerns the annihilation channels $\pi^0\gamma$, $\eta\gamma$, $\pi^+\pi^-\pi^0$, 
$K^+K^-$, $K^0 \overline{K}^0$, with the spectra from ALEPH \cite{Aleph}, CLEO
\cite{Cleo} and BELLE \cite{Belle}
 for the $\tau$ dipion decay\footnote{The energy region used in the fits
has been limited to the $[2 m_\pi,~1 $ GeV] interval where the three data sets are in accord with 
each other. This should lessen the effect of some systematic effects.}
and with the $VP\gamma/P\gamma\gamma$ partial width information extracted from the Review of
Particle Properties (RPP) \cite{RPP2010}. This already represents more than 40 data sets collected
by different groups with different detectors; one may thus consider that 
the systematics affecting these data sets wash out to a 
large extent within a global fit framework.

For what concerns the crucial process $e^+ e^- \ra \pi^+\pi^-$,
the analysis in  \cite{ExtMod3} only deals with the data sets collected in the scan experiments 
performed at Novosibirsk and referred to globally hereafter
as NSK \cite{Barkov,CMD2-1995corr,CMD2-1998-1,CMD2-1998-2,SND-1998}. 
The main reason was, at this step, to avoid  discussing the reported tension 
\cite{DavierHoecker2,DavierHoecker} between the various existing $\pi^+\pi^-$
data sets~: the scan data sets just quoted, and the data sets collected using the
Initial State Radiation (ISR) method 
by KLOE \cite{KLOE08,KLOE10} and 
BaBar \cite{BaBar,BaBar2}, not to mention the pion form factor data collected
in the spacelike region \cite{NA7,fermilab2}.

It has thus been shown that the global fit  excluding the ISR data sets, allows
to yield a splendid  fit quality; this proves that the whole collection
of data sets considered in   \cite{ExtMod3} is self--consistent and may provide
a safe reference, {\it i.e.} a benchmark, to examine the behavior of other data samples. 

Using the fit results,  the uncertainty on the contribution to $a_\mu$ of each 
of the annihilation channels 
considered was improved by -- at least -- a factor of 2, compared to the standard
estimation method based on  the numerical integration
of the measured cross sections.
For the case of the $\pi^+\pi^-$ channel, the final uncertainty was even found slightly better
than those obtained with the standard method by merging scan and ISR data, {\it i.e.}
a  statistics about 4 times larger in the $\pi^+\pi^-$ annihilation channel. 

The main purpose of the present study is an update of the work in  \cite{ExtMod3} 
aiming at confronting all scan (NSK) and ISR  (from BaBar and KLOE) --
and even spacelike \cite{NA7,fermilab2} -- $\pi^+\pi^-$ data  
and reexamine the reported issues \cite{DavierHoecker2,DavierHoecker}. 
The framework in which our analysis is performed is the same as
the one motivated and developed in~\cite{ExtMod3}.

The broken HLS model described in \cite{ExtMod3} happens to provide
a tool allowing to compare the behavior of any of these $\pi^+\pi^-$ data sets
when confronted with the  $\pi^0\gamma$, $\eta\gamma$, $\pi^+\pi^-\pi^0$, $K^+K^-$, 
$K^0 \overline{K}^0$ annihilation data $and$ with the $\tau$ dipion spectra.  
Indeed, the latter data alone, supplemented with some limited information extracted from the
Review of Particle Properties\footnote{We occasionally refer to the RPP
as Particle Data Group (PDG).} (RPP) 
\cite{RPP2010}, allow to predict the pion
form factor with a surprisingly good precision. The additional  RPP information is supposed
to carry the Isospin Breaking (IB) information requested in order to derive reliably the
$\pi^+\pi^-$ information from the knowledge of the $\pi^\pm\pi^0$ spectrum.

We also take profit of the present work to update the numerical values
for some contributions to the muon anomalous moment $a_\mu$, all gathered in Table 10 
of \cite{ExtMod3}.
Thus, we update the QED entry by using the recent spectacular progress by Aoyama, 
Hayakawa, Kinoshita and Nio~\cite{Kinoshita1,Kinoshita2}. They have been able to 
perform a complete numerical calculation of the 5-loop QED corrections to 
$a_e$ and $a_\mu$. On the other hand, the electroweak contribution, which depends on
the Higgs mass at 2--loops is now better known if we accept that 
ATLAS \cite{ATLAS} and CMS \cite{CMS} have
observed at the LHC the Higgs boson at a mass of about 125 GeV in a narrow window.
Using this information slightly changes the central value as well as  
the uncertainty of the EW entry. We also have reevaluated the higher 
order HVP contribution (HO) within the standard
approach based on all $\pi^+\pi^-$ channels ({\it i.e.} all scan and ISR data).

The paper is organized as follows. Section \ref{reminder} reminds the motivations
of the BHLS model and a few basic topics concerning the $\pi\pi$ channel description
(from  \cite{ExtMod3}); we also reexamines how the isospin breaking corrections apply. 
In Section \ref{issue}, the detailed framework -- named "$\tau$+PDG" -- 
used to study the differential behavior of the scan and ISR data
is presented. Thanks to the (wider than usual) energy range covered
by the BaBar spectrum  \cite{BaBar,BaBar2}, a detailed study
of the $\pi^+\pi^-$ spectrum in
the $\phi$ region can be performed for the first time. This leads 
 to update the $\phi \ra \pi^+\pi^-$ treatment within our computer code; this is emphasized
 in Subsection \ref{Vdecays}. In Section \ref{tauPred}, one confronts the 
 "$\tau$+PDG" predictions with the available scan (NSK) and ISR data samples; it is shown
 that the NSK data and both KLOE data samples (referred to hereafter as
 KLOE08 \cite{KLOE08} and KLOE10 \cite{KLOE10}) have similar properties while
 BaBar behaves differently, especially in the $\rho-\omg$ interference region.
Section \ref{globalfits}, especially Subsection \ref{features}, reports on the global fits 
performed using the various
$\pi^+\pi^-$ data samples each in isolation or combined. Subsection \ref{physicsInfo}
collects some topics on various aspects of the physics covered by the HLS model.  
More precisely,  Subsection \ref{PhiRegion}
is devoted to studying the $\phi$ region of the pion form factor and Subsection 
\ref{FitNumRes} gives numerical fit information which may allow to compare with corresponding
results available from other studies performed using different methods.    
In Section \ref{gM2_studies}, 
we focus  on the consequences for the muon anomalous moment $a_\mu$ of the
various  scan and ISR $\pi^+\pi^-$ spectra and compare results with the BNL \cite{BNL,BNL2}
measurement. The $\pi^+\pi^-$  intermediate state contribution
to  $a_\mu$ from the invariant mass region $\left [ 0.630,0.958\right]$ GeV
is especially considered as it 
serves to examine the outcome of various fits with respect
to the experimental expectations. Finally, Section \ref{Conclusion}
is devoted to conclusions.

\section{A Brief Reminder of Concern for the $\pi\pi$ Channel}
\label{reminder}

\subsection{The General Context of the HLS Model}
\label{context}
\indent \indent 
At very low energies chiral perturbation theory (ChPT)\cite{GL1,GL2} is the
"from first principles" approach to low-energy hadron physics. Unfortunately,
ChPT ceases to converge at energies as low as about 400 MeV, and thus the
most important region of the spin 1 resonances fails to be in the scope of
ChPT. 

A phenomenologically well established description of the vector mesons
is the VMD model, which may be neatly put into a quantum field theory (QFT) framework.
This, however, has to be implemented in accord with the chiral structure of 
the low energy spectrum. It is now widely accepted that a low energy effective QFT of massive
spin 1 bosons must be a Yang-Mills theory supplemented with a
Higgs-Kibble mechanism. The general framework is the
Resonance Chiral Perturbation Theory (RChPT)~\cite{Ecker1}, an extension of
ChPT to vector mesons usually expressed in the (not very familiar)
antisymmetric tensor field formalism.
Like in ChPT, the basic fields are the unitary matrix
fields $\xi_{L,R}=\exp \left[\pm i\,P/f_\pi\right]$, where $P=P_8+P_0$
is the $SU(3)$ matrix of pseudoscalar fields, with $P_0$ and $P_8$ being
respectively the basic singlet and octet pseudoscalar field matrices.

The Hidden Local Symmetry (HLS) ansatz  \cite{HLSOrigin,HLSRef} is an extension of the 
ChPT non-linear sigma model to a non-linear chiral Lagrangian based on the symmetry pattern
$G_{\rm global}/H_{\rm local}$, where $G=SU(3)_L \otimes SU(3)_R$ is
the chiral group of QCD and $H=SU(3)_V$ is the vector subgroup. The hidden
local $SU(3)_V$ requires the vector meson fields, represented
by the $SU(3)$ matrix field $V_\mu$, to be gauge fields.  The
corresponding covariant derivative reads $D_\mu=\partial_\mu-i\,g\,V_\mu$ 
and can be naturally extended  \cite{HLSRef}  in order to include the couplings to the
electroweak gauge fields $A_\mu$, $Z_\mu$ and $W^\pm_\mu$. 

It has been proven in~\cite{Ecker2} that RChPT and HLS are 
equivalent provided consistency with the QCD asymptotic behavior is incorporated. 
Such an extension  of ChPT to include VMD structures is
fundamental. Although it is not yet established   which version is the true
 low-energy effective QCD, it is the widely accepted 
framework which includes all particles as effective fields up to the $\phi$ and
only confrontation with data can tell  to which extent 
such an effective  theory works.
This has been the subject of the study \cite{ExtMod3} which we update
by extending it.
Obviously, this approach is more complicated than a
Gounaris-Sakurai ansatz and requires elaborate calculations because the
basic symmetry group is not SU(2)
but SU(3)$\times$ SU(3) where the SU(3) vector subgroup must be gauged in order to
obtain the Yang-Mills structure for the spin 1 bosons.

All relevant states have to be incorporated in accord with the chiral 
structure of the low--energy hadron spectrum. In a low-energy expansion, 
one naturally expects the leading low--energy tail to be close to a 
renormalizable effective theory; however, this is not true for the pseudo
Nambu-Goldstone boson sector, which is governed by a non-linear $\sigma$ model,
rather than by a renormalizable linear $\sigma$ model. The reason is that the latter
 requires a scalar ($\sigma$) meson as a main ingredient. Phenomenologically, scalars
only play a kind of "next--to--leading" role. 

The situation is quite different for the spin 1
bosons, which naturally acquire a Yang-Mills effective structure in a low--energy
expansion {\it i..e.}, they naturally exhibit a leading local gauge symmetry structure
with masses as generated by a Higgs-Kibble mechanism. There is one important
proviso, however~: such a low-energy effective structure is pronounced
only to the extent that the effective expansion scale $\Lambda_{\rm eff}$ is high
enough, which is not clear at all for QCD unless we understand why
$\Lambda_{\rm eff} >> \Lambda_{OCD}\sim 400~\mbox{MeV}$. However,
there is a different approach, namely, a ``derivation'' of the
Extended Nambu--Jona-Lasinio (ENJL) model~\cite{ENJL85,ENJL86}, which has 
also been proved
to be largely equivalent to the Resonance Lagrangian Approach
(RLA)~\cite{Prades94,Prades99}.  Last but not least, large--$N_c$ 
QCD~\cite{tHooft74,Manohar01,leutw,leutwb} in fact
predicts the low energy hadron spectrum to be dominated by spin 1
resonances. These arguments are also the guidelines
for the construction of the HLS model \cite{HLSOrigin,HLSRef}. It
provides a specific way to incorporate the phenomenologically
known low energy hadron spectrum into an effective field theory.

Most frequently the RLA is applied to study
individual processes. In this paper as in  a few previous ones, we attempt to
fit the whole HLS Lagrangian by a global fit strategy. This is, in our opinion,
the only way to single out a phenomenologically acceptable low-energy effective
theory, which allows to make predictions which can be confronted with
experiments.

\subsection{The Broken HLS Lagrangian}
\label{direct_brk}
\indent \indent 
The (unbroken)  HLS Lagrangian is then given by
${\cal L}_{\rm HLS}={\cal L}_A+a {\cal L}_V$, where
\be
{\cal L}_{A/V}=-\frac{f_\pi^2}{4}\,{\rm Tr} \left[L\pm R\right]^2~,
~~~(\xi_{L,R}=\exp \left[\pm i\,P/f_\pi\right])
\label{Eq2}  
\ee
with $L=\displaystyle [D_\mu\xi_L]\xi_L^\dagger$ and 
$R=\displaystyle [D_\mu\xi_R]\xi_R^\dagger$;
 $a$ is a basic HLS parameter not fixed by the theory,  
which should be constrained by confrontation with the data. From standard VMD models,
one expects $a\simeq 2$.

It is well known that  the global chiral symmetry $G_{\rm global}$
is  not realized as an exact symmetry in nature, which
implies that the ideal HLS symmetry is evidently not a symmetry of
nature either. Therefore, it has obviously to be broken appropriately 
in order to
provide a realistic low energy effective theory mimicking low energy
effective QCD. 

Unlike in ChPT where one is
performing a systematic low energy expansion in low momenta
and the quark masses, here one introduces symmetry breaking as
phenomenological parameters to be fixed from appropriate data. Since a
systematic low energy expansions \`a la ChPT does not converge  
above about  $\simeq 400$ MeV, this is the only way 
to model phenomenology
up to, and including, the $\phi$ resonance region.

In our approach, the Lagrangian pieces in Eqs. (\ref{Eq2}) are broken
in a two step procedure. A first breaking  mechanism named BKY is used,
 originating from   \cite{BKY,HLSRef}. In order to avoid some undesirable 
properties \cite{BGP,BGPbis} of the original BKY mechanism, 
we have adopted the modified BKY scheme proposed in 
\cite{Heath}. In its original form, this modified BKY breaking scheme only covers 
the breaking of the SU(3) symmetry; following \cite{Hashimoto}, it has been extended 
in order to include isospin symmetry breaking effects. This turns out to modify  
Eqs. (\ref{Eq2}) by introducing  two constant diagonal matrices $X_{A/V}$~:
 \be
{\cal L}_{A/V}  
 \displaystyle \Longrightarrow
 {\cal L}_{A/V}^{\prime} = 
 -\frac{f_\pi^2}{4} {\rm Tr} \left\{[L \pm R]~ X_{A/V}  \right \}^2
\label{Eq3}  
\ee
and the (non--zero) entries in $X_{A/V}$ are fixed from fit to the data. The final
broken HLS Lagrangian can be written~:
 \be
 \displaystyle 
 {\cal L}_{HLS}^{\prime}={\cal L}_A^{\prime}+a{\cal L}_V^{\prime}
 +{\cal L}_{'tHooft}~~.
\label{Eq4}  
\ee
One has, here, included  ${\cal L}_{'tHooft}$ which provides determinant 
 terms \cite{tHooft}  breaking the nonet symmetry
 in the pseudoscalar sector and thus allowing  an improved account of the
 $\pi^0,\,\eta,\,\eta^\prime$  sector. ${\cal L}_{HLS}^{\prime}$ can be 
 found expanded in the various Appendices of \cite{ExtMod3}.

However, in order to account successfully for the largest possible set of data,
isospin symmetry breaking \`a la BKY
should be completed by a second step involving the kaon loop
mixing of the neutral vector mesons ($\rho^0_I$, $\omg_I$ and $\phi_I$) 
outlined just below. 
This implies a change of fields to be performed in the ${\cal L}_{HLS}^{\prime}$ Lagrangian.

\subsection{Mixing of Neutral Vector Mesons Through Kaon Loops}
\label{mix1}
\indent \indent It has been   shown \cite{ExtMod1,ExtMod2,ExtMod3} that ${\cal L}_{HLS}^{\prime}$
is insufficient in order to get a good simultaneous account of the $e^+e^- \ra \pi^+\pi^-$ annihilation 
data and of the dipion spectrum measured 
in the  $\tau^\pm \ra \nu_\tau \pi^\pm \pi^0$ decay. A consistent 
solution to this problem is provided by the vector field mixing mechanism first introduced in
\cite{taupaper}.

Basically, the  vector field mixing is motivated by the one--loop corrections to the vector
field squared mass matrix. These are generated by the following term of the broken HLS 
Lagrangian\footnote{
\label{simply}
For clarity, we have dropped out  the isospin breaking corrections 
generated by the BKY mechanism; 
the exact formula can be found in the Appendix A of \cite{ExtMod3}. 
The parameter $z_V$ corresponds to the breaking of
the SU(3) symmetry in the Lagrangian piece ${\cal L}_{V}$, while
 $z_A$ is associated with the SU(3) breaking of ${\cal L}_{A}$. $z_V$
 has no really intuitive value, while $z_A$  can be expressed
 in terms of the kaon and pion decay constants as $z_A=[f_K/f_\pi]^2$.
}
${\cal L}_{HLS}^{\prime}$ ~:
 \be
 \displaystyle
\frac{i a g}{4 z_A} \left \{
\left[\rho^0_I+ \omg_I -\sqrt{2} z_V ~\phi_I \right]K^- \parsym K^+
 +  \left[\rho^0_I-\omg_I
+\sqrt{2} z_V~\phi_I \right] K^0 \parsym \overline{K}^0 \right \}~,
\label{Eq5}
\ee
where $g$ is the universal vector coupling and the subscript $I$ indicates the ideal 
vector fields originally occurring in the Lagrangian.

Therefore, the vector meson squared mass matrix $M_0^2$, which is diagonal at tree level, undergoes
corrections at one--loop. The perturbation matrix $\delta M^2(s)$ \cite{taupaper,ExtMod1,ExtMod2}
depends on the square of the momentum  flowing through the vector meson lines. The diagonal entries
acquire self--mass corrections -- noticeably the $\rho^0$ entry absorbs the pion loop -- but 
non--diagonal entries are also generated which correspond to transitions among the ideal $\rho^0$, $\omg$ 
and $\phi$ 
meson fields which originally enter the HLS Lagrangian\footnote{These mixing functions were denoted resp. 
$\varepsilon_{\omega \phi}(s)$, $\varepsilon_{\rho \omega}(s)$  and $\varepsilon_{\rho \phi}(s)$
in Section 6 of \cite{ExtMod3}.}~: $ \Pi_{\omega \phi}(s)$, $\Pi_{\rho \omega}(s)$ 
and $\Pi_{\rho \phi}(s)$.  These  are linear combinations of the 
kaon loops\footnote{Other contributions than kaon loops, 
like $K^* K$ loops, take place \cite{taupaper,ExtMod1}
which are essentially real in the energy region up to the $\phi$ meson mass. These can be 
considered as numerically
absorbed by the subtraction polynomials of the kaon loops.}. 
Denoting by resp. $\Pi_c(s)$ and $\Pi_n(s)$ the charged and the neutral 
 kaon loops (including resp. the $\rho^0 K^+K^-$ and $\rho^0 K^0\overline{K^0}$ 
 coupling  constants squared), one defines two combinations of these~:
 
 \be
 \begin{array}{ll}
 \varepsilon_1(s)=\Pi_c(s)- \Pi_n(s)~~,&~~ \varepsilon_2(s)=\Pi_c(s)+ \Pi_n(s)
 \end{array}
 \label{Eq6}
 \ee
 
 In term of $\varepsilon_1(s)$ and $\varepsilon_2(s)$, the transition amplitudes write~:
 \be
 \begin{array}{lll}
  \Pi_{\omega \phi}(s)=-\sqrt{2} z_V \varepsilon_2(s)~~,&~~ 
  \Pi_{\rho \omega}(s)=\varepsilon_1(s)~~,&~~ 
  \Pi_{\rho \phi}(s)=-\sqrt{2} z_V \varepsilon_1(s)
 \end{array}
 \label{Eq7}
 \ee

Therefore,
at one--loop  order, the ideal vector field $V_I=[\rho^0_I,~\omg_I,~\phi_I]$  originally
occurring in    
${\cal L}_{HLS}^{\prime}$
are no longer mass eigenstates; the $physical$
vector fields are then (re)defined as the eigenvectors 
of $M^2=M^2_0+\delta M^2(s)$. This change
of fields should be propagated into the whole broken HLS Lagrangian  
${\cal L}_{HLS}^{\prime}$, extended in order to include 
the anomalous couplings \cite{FKTUY} as done in \cite{ExtMod3}. In terms
of the combinations $(V_{R1})$ of the original vector fields $V_I$ which diagonalize 
${\cal L}_{HLS}^{\prime}$  (see Section 5 in \cite{ExtMod3}), the physical
vector fields  -- denoted $V_{R}$ -- can be derived by inverting~:

\be
\left (
\begin{array}{lll}
\rho_{R1}\\[0.5cm]
\omg_{R1}\\[0.5cm]
\phi_{R1}
\end{array}
\right ) = 
\left (
\begin{array}{cll}
\displaystyle  1   		&-\alpha(s) 	&\beta(s) \\[0.5cm]
\displaystyle  \alpha(s)	& ~~1		& \gamma(s) \\[0.5cm]
\displaystyle -\beta(s) 	& -\gamma(s) 	& 1
\end{array}
\right ) 
\left (
\begin{array}{lll}
\rho_R\\[0.5cm]
\omg_R\\[0.5cm]
\phi_R
\end{array}
\right )
\label{Eq8}
\ee
where $\alpha(s)$, $\beta(s)$ and $\gamma(s)$ are the ($s$--dependent)
vector mixing angles and $s$ is the 4--momentum squared flowing
through the corresponding vector meson line. These functions are
proportional to the transition amplitudes reminded above.
In contrast to $\varepsilon_1(s)$ which identically vanishes in the Isospin
Symmetry limit, $\varepsilon_2(s)$ is always a (small) non--identically vanishing function. 
Therefore,
within our breaking scheme, the $\omg - \phi$ mixing is a natural
feature following from loop corrections and not from IB effects.
In contrast, the $\rho - \omg$ and $\rho - \phi$ mixings are pure 
effects of Isospin breaking in the pseudoscalar sector.

For brevity, the Lagrangian ${\cal L}_{HLS}^{\prime}$
expressed in terms of the physical fields is referred to as BHLS.

\subsection{The $V \pi \pi$ and $V-\gamma/W^\pm$ Couplings}
\label{mixMod}
\indent \indent As the present study focuses on  $e^+ e^- \ra \pi^+\pi^-$ data, it is worth
to briefly remind a few relevant pieces of the ${\cal L}_{HLS}^{\prime}$ Lagrangian.
In terms of  $physical$ vector fields, {\it i.e.} the eigenstates of  
$M^2=M_0^2 + \delta M^2(s)$,
the $V \pi \pi$ Lagrangian piece writes~:
\be
\hspace{-0.5cm}
\displaystyle \frac{i a g}{2}(1+ \Sigma_V)\left[
\left \{
  \rho^0 +\left[( 1-h_V)\Delta_V   -\alpha(s)\right]~\omg  +\beta(s)  ~\phi
\right \}\cdot \pi^- \parsym \pi^+ 
+\left \{ \rho^- \cdot \pi^+ \parsym \pi^0
- \rho^+ \cdot \pi^- \parsym \pi^0 \right \} 
\right],
\label{Eq9}
\ee
where $\Sigma_V$ and $ (1-h_V)\Delta_V$ are isospin breaking parameters 
generated by the BKY mechanism \cite{ExtMod3}, whereas
$\alpha(s)$ and $\beta(s)$ are (complex) "angles" already defined. 
Their expressions can be found in \cite{ExtMod3}. Eq. (\ref{Eq9}) 
shows how the IB decays $\omg/\phi \to \pi^+ \pi^-$ appear
in the BHLS Lagrangian. 

Another Lagrangian piece relevant for the present update is~:
\be
\displaystyle -e
\left [f_{\rho\gamma}(s) \rho^0 + f_{\omg \gamma}(s)\omg -f_{\phi \gamma}(s)\phi\right] \cdot A
-\frac{g_2V_{ud}}{2} f_{\rho W} \left [ W^+ \cdot \rho^- +  W^- \cdot \rho^+\right]~,
\label{Eq10}
\ee
where $g_2$ is the weak $SU(2)_L$ gauge coupling 
and $V_{ud}$ is the element of the $(u,d)$ entry in the CKM matrix. The $f_{V\gamma}(s)$ functions and 
$f_{\rho W}$ are the transition amplitudes of the physical vector mesons to the photon
and the $W$ boson, respectively. At leading order in the breaking parameters, they are given 
by \cite{ExtMod3}~:

\be 
\left \{
\begin{array}{l}
f_{\rho\gamma}(s)=\displaystyle a g  f_\pi^2 \left[ 
1+ \Sigma_V +h_V\frac{\Delta_V }{3} +\frac{\alpha(s)}{3} + \frac{\sqrt{2}z_V}{3} \beta(s) \right]~,\\[0.5cm]
\displaystyle 
f_{\omg \gamma}(s)=\frac{a g  f_\pi^2}{3}\left[ 1+ \Sigma_V +3 (1-h_V)\Delta_V -3\alpha(s) 
+ \sqrt{2}z_V~ \gamma(s)\right]~,\\[0.5cm]
\displaystyle 
f_{\phi \gamma}(s)=\frac{a g  f_\pi^2}{3}\left[-\sqrt{2}z_V+3\beta(s)+\gamma(s)\right]~,\\[0.5cm]
\displaystyle 
f_{\rho W} \equiv f_\rho^\tau = a g  f_\pi^2 \left[ 1+ \Sigma_V \right]~.
\end{array}
\right.
\label{Eq11}
\ee

Eqs. (\ref{Eq9}) and (\ref{Eq10}) exhibit an important property which should be noted. The 
functions $\alpha(s)$ and $\beta(s)$ providing the coupling of the physical $\omg$ and $\phi$ 
mesons to a pion pair also enter each of the $f_{V\gamma}(s)$ transition amplitudes, especially
into $f_{\rho\gamma}(s)$. Therefore, any change in the  conditions used in order
to account for the decays $\omg/\phi \ra \pi^-\pi^+$  correspondingly affects the whole
description of  the $e^+ e^- \ra \pi^+\pi^-$ cross section. Using
the  $\omg/\phi \to \pi^+\pi^-$ branching fractions in place of the $\pi^+\pi^-$ spectrum
in the corresponding regions has, of course, local consequences by affecting the corresponding
invariant mass regions; it has also quite global consequences~: indeed, it also affects
the description of the annihilation cross-sections to 
$\pi^0\gamma$, $\eta \gamma$, $\pi^+\pi^-\pi^0$, $K^+K^-$,
$K^0\overline{K^0}$ final states which all carry the $f_{V\gamma}(s)$ transition amplitudes.  

Another effect, already noted in \cite{ExtMod3}, is exhibited by 
Eqs. (\ref{Eq11})~: the ratio $f_{\rho\gamma}(s)/f_{\rho W}$ becomes
$s$--dependent, which is an important difference between $\tau$ decays
and $e^+e^-$ annihilations absent from all previous studies, except for
\cite{Fred11}. Figure 11  in \cite{ExtMod3} shows that the difference
between $f_{\rho\gamma}(s)$ and $f_{\rho W}$ is at the few percent level.

\subsection{The Pion Form Factor}
\indent \indent Here, we only remind the BHLS form of the pion form factor in
$\tau$ decay and in $e^+ e^-$  annihilation  and refer the interested reader 
to \cite{ExtMod3} for detailed information on the other channels.
  The pion form factor in the $\tau^\pm$ decay to $\pi^\pm \pi^0 \nu_\tau$ can be written~:
\be
\displaystyle 
F_\pi^\tau(s) = \left[
1-\frac{a}{2}(1+\Sigma_V) \right]- \frac{ag}{2}(1+\Sigma_V)F_\rho^\tau(s) 
\frac{1}{D_\rho(s)}~,
\label{Eq12}
\ee
where $a$ and $g$ are the basic HLS parameters \cite{HLSRef} already encountered; 
$\Sigma_V$ is one of the isospin breaking
parameters introduced by the (extended) BKY breaking scheme. The other quantities are~:
\be
\left \{
\begin{array}{lll}
\displaystyle 
F_\rho^\tau(s) =f_\rho^\tau - \Pi_{W}(s)~,\\[0.5cm]
\displaystyle 
D_\rho(s)=s-m^2_\rho -\Pi_{\rho \rho}^\prime(s)~,\\[0.5cm]
\displaystyle 
f_\rho^\tau =a g  f_\pi^2 (1+\Sigma_V) ~~~,~~ m^2_\rho=a g^2 f_\pi^2(1+\Sigma_V)~,
\end{array}
\right .
\label{Eq13}
\ee
where $\Pi_{W}(s)$ and $\Pi_{\rho \rho}^\prime(s)$  are, respectively, the
loop correction to the $\rho^\pm-W^\pm$ transition amplitude
 and the charged $\rho$ self--mass (see   \cite{ExtMod3}).                                                                              
                                                                                                           
   The pion form factor in $e^+ e^-$ annihilation is more complicated and writes~:
\be
\displaystyle
F_\pi^e(s) = \left [ 1-\frac{a}{2}(1+\Sigma_V+\frac{h_V \Delta_V}{3})\right] - 
F_{\rho \gamma}^e(s) \frac{g_{\rho \pi \pi}}{D_\rho(s)}
- F_{\omega \gamma}^e(s) \frac{g_{\omega \pi \pi}}{D_\omega(s)}
- F_{\phi \gamma}^e(s) \frac{g_{\phi \pi \pi}}{D_\phi(s)}~,
\label{Eq14}
\ee 
where $g_{\rho \pi \pi}$, $g_{\omega \pi \pi}$ and $g_{\phi \pi \pi}$ can be read off
Eq. (\ref{Eq9}) and the $F_{V \gamma}^e$ are given by~:
 \be
\displaystyle F_{V\gamma}^e(s)= f_{V\gamma}(s) - \Pi_{V\gamma}(s)~~,~~
(V=\rho^0_R,~\omg_R,~\phi_R)~~,
\label{Eq15}
\ee
with the $f_{V\gamma}(s)$ given by  Eqs. (\ref{Eq11}) above and the $\Pi_{V\gamma}(s)$ 
being loop corrections \cite{ExtMod3}. $D_\rho(s)=s-m^2_\rho -\Pi_{\rho \rho}(s)$
is the inverse $\rho^0$ propagator while $D_\omega(s)$ and $D_\phi(s)$  are the modified
fixed width Breit--Wigner functions defined in \cite{ExtMod3}; these have been chosen in order to 
cure the violation of $F_\pi(0)=1$ produced by the usual fixed width Breit--Wigner approximation 
formulae.  

\subsection{IB Distortions of the Dipion $\tau$ and $e^+e^-$ Spectra }
\label{IB_effects}
\indent \indent IB effects in the $\pi\pi$ channel are of various kinds.
In the breaking model developed in \cite{ExtMod3} and outlined just above,  
 the IB effects following from the neutral vector meson mixing,
together or with the photon (see also \cite{Fred11}), are dynamically generated from the HLS
Lagrangian. The most relevant effects have been reminded
in  Subsections \ref{mix1} and \ref{mixMod} 
for the $\pi\pi$ channel. Indeed, Eq. (\ref{Eq9}) exhibits the
generated coupling of the $\omg$ and $\phi$ mesons to a pion pair and
Eqs. (\ref{Eq10}) and (\ref{Eq11}) show how the $V-\gamma$ couplings are modified
by the extended BKY breaking and vector mixing mechanisms. Therefore, in
principle, all breaking effects\footnote{Except for a possible nonet symmetry
breaking in the vector meson sector.} of  concern for $e^+e^-$
annihilations are exhausted.

Some IB effects affecting the dipion $\tau$ spectrum are also generated
by the breaking mechanism, which modifies the $W-\rho^\pm$ transition amplitude 
and the $\rho^\pm \pi^\mp \pi^0$ coupling. In some sense, 
the breaking  mechanism decorrelates
the universal coupling $g$  as it occurs in the anomalous sector
from those in the non--anomalous sector, where $g$ appears 
in  combinations reflecting IB effects, like $g (1+\Sigma_V)$ for the simplest
form \cite{ExtMod3}.

On the other hand, and as a general statement, the effects generated by 
the pion mass difference 
$m_{\pi^\pm} -m_{\pi^0}$ do not call for any specific IB treatment, as 
the appropriate pion masses are utilized at the corresponding places
inside the model formulae derived from BHLS; this concerns, in particular, 
the pion 3--momentum which appears, for instance, in the 
phase space terms of the charged and neutral $\rho$ widths.  

\vspace{0.5cm}

However, there are IB breaking effects in $\tau$ decay, which have
not yet been taken into account. Indeed, known distortions of the
dipion $\tau$ spectrum relative to $e^+e^-$ are produced by the
radiative corrections due to photon emission. The long distance effects have
been calculated in
\cite{Cirigliano2,Bijnens,Cirigliano3,Mexico2,Mexico1,Mexico4} and the
short distance contributions in
 \cite{Marciano,Braaten:1990,Braaten:1991,Erler}. 

We have adopted the corresponding corrections, $G_{EM}(s)$ and $S_{EW}~(=1.0235\pm 0.0003)$, 
respectively, as specified in \cite{ExtMod3}. In Ref.~\cite{Mexico2} the contribution of
the sub-process $\tau \to
\omega\pi^-\nu_\tau(\omega \to \pi^0\gamma)$ has been evaluated to
substantially shift the correction $G_{EM}(s)$ (see Fig.~2 
in \cite{Mexico1}). This sub-process has been subtracted in
the Belle data \cite{Belle} which supposes that the 
corresponding correction has not to apply\footnote{
It is stated in \cite{DavierHoecker} that this subtraction
has also been performed in the ALEPH and CLEO data.}.
 Hence, we applied to the three dipion $\tau$
spectra the correction as given in \cite{Cirigliano2}, as in our
previous analysis \cite{ExtMod3}.

These IB corrections distort the dipion spectra 
from the $\tau$ decay. They are accounted for  
by submitting to the global fit the
experimental dipion $\tau$  distributions \cite{Aleph,Cleo,Belle} 
using the HLS expression for $d\Gamma_{\pi \pi}(s)/ds$ 
(see Eqs. (73) and (74) in 
\cite{ExtMod3} and the present Eq.~(\ref{Eq14})) 
corrected -- as usual -- in the following way~:

\be
 {\cal B}_{\pi \pi} \frac{1}{N}\frac{dN(s)}{ds} =
\displaystyle \frac{1}{\Gamma_{\tau}} \frac{d\Gamma_{\pi \pi}(s)}{ds} 
S_{EW} G_{EM}(s) 
\label{Eq16}
\ee
where $\Gamma_{\tau}$ is the full $\tau$ width and ${\cal B}_{\pi \pi}$
its branching fraction to $\pi^\pm \pi^0 \nu_\tau$, both 
extracted from the RPP \cite{RPP2010}. Indeed, as our fitting range is
bounded by 1.05 GeV, both pieces of information are beyond the scope 
of our model. These corrections represent, by far, the most important
corrections specific of the $\tau$ decay not accounted for within
the HLS framework.

\vspace{0.5cm}

Another source of isospin breaking which may distort the $\tau$ spectrum
compared to $e^+e^-$  is due to the $\rho$ mass
difference $\delta M_\rho=m_{\rho^\pm} - m_{\rho^0}$. We note that the
Cottingham formula, which provides a rather precise prediction of the
$m_{\pi^\pm} - m_{\pi^0}$ electromagnetic mass difference, predicts for 
the $\rho$ an electromagnetic mass difference~: 
$$\delta M_\rho \equiv m_{\rho^\pm} - m_{\rho^0} \simeq
\frac12 \frac{\Delta m^2_\pi}{M_{\rho^0}}\simeq 0.814~\mbox{MeV}~~.$$
In
principle, within the HLS model a $\rho^0-\rho^\pm$ mass shift is also
generated by the Higgs-Kibble mechanism (corresponding to the well
known shift $\delta M^2=M_Z^2-M_W^2=g'^2\,v^2/4$ in the
Electroweak Standard Model). 
In the HLS model $m^2_{\rho^\pm}= a\,f_\pi^2\,g^2$
while $m^2_{\rho^0}= a\,f_\pi^2\,(g^2+e^2)$ due to $\rho^0-\gamma$
mixing. This leads to a Higgs-Kibble shift of about 
$m_{\rho^0}-m_{\rho^\pm}\simeq
\frac{e^2}{2g} \sqrt{a}\, f_\pi\sim 1~\mbox{MeV}$ (see \cite{HLSRef}), 
which essentially
compensates the electromagnetic shift obtained from the Cottingham
formula. In addition, the masses  are subject to modifications by
further $\rho-\omega-\phi$ mixing effects, obtained from diagonalizing
the mass matrix after including self-energy effects. The mentioned
effects have been estimated in \cite{Bijnens}   and lead to~: 
$$ -0.4~\mbox{MeV} < m_{\rho^0}-m_{\rho^\pm} < +0.7~\mbox{MeV}\,.$$ 
When evaluating the anomalous magnetic moment from $\tau$ data, 
several choices have been made; for instance, the analysis in  
\cite{DavierHoecker} assumes $\delta M_\rho \simeq 1.0\pm 0.9~\mbox{MeV}$,
while Belle \cite{Belle} preferred 
$\delta M_\rho \simeq 0.0\pm 1.0~\mbox{MeV}$.

In our study, we followed Belle and have adopted
$\delta M_\rho \simeq 0\pm 1~\mbox{MeV}$, 
consistent with the estimate by Bijnens and Gosdzinsky  just reminded 
and with most experimental values reported in the 
RPP\footnote{The average value proposed by the PDG is 
$ m_{\rho^0}-m_{\rho^\pm}=(-0.7 \pm 0.8)$ MeV.} \cite{RPP2010}.
As noted elsewhere \cite{ExtMod2,ExtMod3}, based on
the available data, auxiliary HLS fits do not improve by letting 
$\delta M_\rho$ floating. This justified reducing the model freedom 
by fixing this additional parameter  to 0.

\vspace{0.5cm} 
Yet another source of isospin breaking which  may somewhat distort the 
$\tau$ dipion spectrum compared to its $e^+e^-$ partner is the
width difference $ \delta \Gamma_\rho=\Gamma_{\rho^0}-\Gamma_{\rho^\pm}$
between the charged and the neutral $\rho$; however, this can be
expected to be  
small as  the accepted average \cite{RPP2010},
 $\delta \Gamma_\rho=(0.3 \pm 1.3)$ MeV, is consistent with 0. 

The expected dominant contribution to $\delta \Gamma_{\rho}$, comes
from the radiative $\rho$ decays 
$\delta \Gamma_\rho^\gamma=\Gamma(\rho^0 \to \pi^+ \pi^- \gamma)
-\Gamma(\rho^\pm \to \pi^\pm \pi^0 \gamma)$. A commonly
used estimation \cite{Cirigliano3,Belle} for this unmeasured quantity is 
 $\delta \Gamma_\rho^\gamma=0.45 \pm 0.45$ MeV; other values have been
proposed, the largest one \cite{DavierHoecker}
being  $\delta \Gamma_\rho^\gamma  \simeq 1.82 \pm 0.18$ MeV. However,
summing up all contributions always leads to $\delta \Gamma_\rho$ 
in accord with the RPP average.

Usually, the  evaluation of the ($\delta M_\rho$, $\delta \Gamma_\rho$) effects
is performed using the Gounaris--Sakurai (GS) parametrization \cite{Gounaris}
of the pion form factor.  However, the
GS formula does not parametrize the radiative corrections 
expected to affect the measurement of the pion form factor. Therefore, the
correction for the radiative width may not be well taken into account
by just shifting the width in the GS formula. In
Refs.~\cite{Mexico4,DavierHoecker} an effective shift of
$\delta \Gamma_\rho^\gamma \simeq 1.82~\mbox{MeV}$ has been estimated by
subtracting $\rho^+\to \pi^+\pi^0 \gamma$ in the $\tau$ channel and
adding $\rho^0\to \pi^+\pi^- \gamma$ in the $e^+e^-$ channel.
The question is how this affects $|F_\pi(s)|^2$. Usually one adopts
the GS formula to parametrize the undressed data, which is not
precisely what is measured. If one assumes the GS formula to represent
the dressed data as well, one may just modify the width for undressing
the $\tau$ spectrum and redressing the radiative effects in the
$e^+e^-$ channel, as an IB correction. 

An increase of the width in the
GS formula has two effects. One is to broaden the $\rho$ shape, which
results in an increase of the cross section. The second, working in
the opposite direction, is to lower the peak cross section. In the
standard form of the GS formula (see e.g. CMD-2 \cite{cmd2:1999} or Belle \cite{Belle}) 
the second effect wins and one gets a substantial reduction of the muon $g-2$
integral by $\delta a_\mu^{\rm had,LO}[\pi\pi,\tau]=(-5.91\pm 0.59)
\times 10^{-10}$~\cite{DavierHoecker}, a large reduction of the
$\tau$-based evaluation. Looking at the Breit-Wigner peak cross
section given by $$ \sigma_{\rm peak}=\frac{12\pi \Gamma_{\rho \to ee}
\Gamma_{\rho \to \pi\pi}}{M_\rho^2
\Gamma_{\rho \to {\rm all}}^2}$$
it is not a priori clear, which of the different widths are
affected. If one keeps fixed the branching fractions for $\rho \to ee$ and $\rho
\to \pi\pi$, the peak cross section would not change at
all. Therefore, the correction for radiative events via the GS
parametrization is not unambiguous. In the standard GS parametrization
$\Gamma^{\rm
GS}_{ee}=\frac{\alpha^2\beta^3_{\rho}M_\rho^2}{36\Gamma_\rho}\,(1+d
\Gamma_\rho/M_\rho)$ ($d M_\rho \Gamma_\rho=-\Pi^{\rm
ren}_{\rho\rho}(0)$) is a derived quantity and
depends on $\Gamma_\rho$ in an unexpected way. We therefore consider
this standard procedure of correcting for radiative decays as not well
established. 

Auxiliary fits allowing a difference $\delta g$
between the $\rho^\pm$ and $\rho^0$ couplings, in order to
generate a floating $\delta \Gamma_\rho$, have been performed. 
One observed slightly more sensitivity to a free $\delta g$ than to a free
$\delta M_\rho$, but nothing conclusive enough 
to depart from $\delta g=0$  while increasing the number of fit parameters and their
correlations. Indeed, within the BHLS framework, the
$\tau$ data only play the role of an additional constraint and
their use is certainly not mandatory, except for
testing the ''$e^+e^-$ vs $\tau$ discrepancy'' which
has been shown to disappear~\cite{ExtMod3}.

Within the set of data samples which
are studied by means of the global fit framework provided
by BHLS, the single place where the charged $\rho$ meson
plays a noticeable role is the $\tau$ dipion spectrum.
Taking into account its relatively small statistical weight 
within this set of data samples, one does not expect 
to exhibit from global fits a noticeable sensitivity to mass 
and width differences with its neutral partner. 

\section{Confronting the Various  $e^+e^- \to \pi^+ \pi^-$ Data Sets}
\label{issue}
\subsection{The Issue}
\label{issue_dat}
\indent \indent
Although the BHLS Lagrangian should be able to describe more complicated
hadron production processes, in a first step one obviously has to focus
on low multiplicity states, primarily two particle production but also
the simplest three particle production channel $e^+e^- \to
\pi^+\pi^-\pi^0$. Four pion production, annihilation to $KK\pi$ \ldots are beyond 
the scope of the basic setup of the BHLS model. We expect that available data on the
lowest multiplicity channels provide a consistent database which
allows us to pin down all relevant parameters, such that our effective
resonance Lagrangian is able to {\it simultaneously} fit all possible
low multiplicity channels. In fact, what is considered are essentially
all relevant annihilation channels up to the $\phi$; in this energy range
 the missing channels ($4\pi,5\pi,6\pi,\eta\pi\pi, \omega\pi$) contribute
less than 0.3\% to $a_\mu^{\rm had}$  \cite{ExtMod3}.

Our previous study  \cite{ExtMod3}  has actually shown that the following groups of
complementary data samples and/or RPP \cite{RPP2010} accepted  particle properties
(mainly complementary branching fractions) support our global fit strategy~:

\begin{itemize}
\item{\bf (i)} All $e^+e^-$ annihilation data into the  $\pi^0\gamma$, $\eta\gamma$, $\pi^+\pi^-\pi^0$, 
$K^+K^-$, $K^0 \overline{K}^0$ final states admit a consistent simultaneous fit\footnote{
With -- possibly -- some minor tension between the $\pi^+\pi^-\pi^0$ data around the $\phi$ 
resonance and the 
dikaon data (see the discussions in \cite{ExtMod1,ExtMod3}). },
\item{\bf (ii)} The  $\tau^\pm \ra \nu_\tau \pi^\pm \pi^0$ dipion
spectra produced  by ALEPH \cite{Aleph}, CLEO \cite{Cleo} and BELLE \cite{Belle}, 
however  limited to the energy region where they are in reasonable 
accord with each other\footnote{The data sample from OPAL \cite{Opal}
behaves differently as can be seen in Fig. 3 in \cite{Ghozzi} or in Fig.1
from \cite{DavierHoecker} and, thus, it is not considered for simplicity.
 This behavior is generated by a (probably) too low measured cross section at the
 $\rho$ peak combined with the normalization of the spectrum to the
 (precisely known) total branching fraction;  this procedure enhances
 the distribution tails  and makes the OPAL distribution quite different
 from ALEPH, CLEO and Belle.},  {\it i.e.}$\sqrt{s} \le  1$ GeV, as can be
inferred from Fig. 10 in \cite{ExtMod3},
\item{\bf (iii)} Some additional partial width from the $P\gamma\gamma$ and
 $VP\gamma$ decays, which are independent of the annihilation channels listed
 just above, 
\item{\bf (iv)} Some information concerning the $\phi \ra \pi^-\pi^+$ 
decay, especially its accepted partial width $\Gamma(\phi \ra \pi^-\pi^+)$ \cite{RPP2010}.
This piece of information is supposed to partly counterbalance the lack of spectrum
for  the $e^+e^- \ra \pi^+\pi^- $ annihilation in the $\phi$ mass 
region\footnote{The BaBar data \cite{BaBar} allow, for the first time,
 to make a motivated statement concerning how this piece of information
  should be dealt with inside  the minimization code.
This is discussed below.}.
\item{\bf (v)} All the $e^+e^- \ra \pi^+\pi^- $ data sets (NSK) collected\footnote{
The data sets \cite{Barkov} collected by former detectors at Novosibirsk are
also considered.} by the scan experiments mounted at Novosibirsk, especially
CMD--2 \cite{CMD2-1995corr,CMD2-1998-1,CMD2-1998-2} and SND \cite{SND-1998}. 
\end{itemize}

They represent a complete reference collection of data samples and lead to fits 
which do not exhibit any visible tension between the BHLS model parametrization 
and the data
(see for instance Table 3 in \cite{ExtMod3}). This is worth being noted, as
we are dealing with  a large number of different 
data sets collected by different groups using different
detectors and different accelerators. The (statistical \& systematic)
 error covariance matrices
used within our fit procedure  are cautiously constructed following closely
the group claims and recommendations\footnote{
For what concerns all the $e^+ e^-$ data samples referred to just above,
the procedure, explicitly given in Section 6 of \cite{ExtMod1}, is
outlined in the header of Section \ref{globalfits} below; for
the $\tau$  spectra, the statistical \& systematic 
error covariance matrices
provided by the various Collaborations \cite{Aleph,Cleo,Belle}
 are added in order to
perform the fits \cite{ExtMod2,ExtMod3} .}.  
Therefore, the study in \cite{ExtMod3} leads to think that the $model$ 
correlations exhibited by  
BHLS reflect reasonably well the $physics$ correlations expected to 
exist between the various channels. 

However, beside the (NSK) $e^+e^- \ra \pi^+\pi^- $ data sets collected in scan 
mode, there exists now data sets collected using the Initial State Radiation (ISR) 
method by the KLOE  and BaBar  Collaborations. 
All recent studies (see \cite{DavierHoecker2,DavierHoecker}, for instance) 
report upon some "tension" between them. As
this issue has important consequences concerning the estimate of the muon anomalous
magnetic moment, it is worth examining if the origin of this  tension  can be identified
and, possibly, substantiated. Besides  scan and ISR data, it is also interesting 
to reexamine \cite{taupaper}
the pion form factor data   collected in the spacelike region  \cite{NA7,fermilab2}
within the BHLS framework; indeed, if valid, these data
provide strong constraints on the threshold behavior of 
the pion form factor and, therefore, an improved information on the muon $g-2$.

\subsection{The Analysis Method}
\label{method}

\indent \indent  The BHLS model has many parameters and a global fit has to be guided
by fitting those parameters to those channels to which they are the most
sensitive.  Obviously resonance parameters of a given resonance have to
be derived from a fit of the corresponding invariant mass  region. Similarly,
the anomalous type interaction responsible for $\pi^0 \to \gamma
\gamma$ or the $\pi\pi\pi$ final state \ldots $~$are sensitive to very 
specific channels only.

We also have to distinguish the gross features of the HLS model and
the chiral symmetry breaking imposed to it.
With this in mind, in our approach to comparing the various
$e^+e^- \to \pi^+ \pi^-$ data samples, the $\tau$ decay spectra play a key
role since the charged channel is much simpler than the neutral one
where $\gamma$, $\rho^0$, $\omega$ and $\phi$ are entangled by
substantial mixing of the amplitudes, which are not directly
observable.  In the low energy region, below the kaon pair thresholds and the
$\phi$ region, what comes into play is the $\rho^\pm$ form factor obtained from the
$\tau$ spectra.  Together with the isospin breaking due to
$\rho^0-\omega$  mixing -- characterized by the branching fractions
$Br(\omega \to \pi^+\pi^-)$ and $Br(\omega \to e^+e^-)$, which in a
first step can be taken from the RPP --  the $\rho^\pm$ form factor should 
provide a good prediction for the $e^+e^- \to
\pi^+\pi^-$ channel. Data from the latter can then be used to refine
the global fit. This will be our strategy in the following.

\vspace{0.5cm}
\indent \indent The annihilation channels referred to as {\bf (i)}
in the above Subsection  
as well as the decay information listed in {\bf (iii)} have little to do
with the $e^+ e^- \ra \pi^+ \pi^- $ annihilation channels, except for the physics
correlations implied by the BHLS model. On the other hand, as long as one limits
oneself to the region $(2 m_\pi, 1{\rm~ GeV})$, there is no noticeable
contradiction
between the various dipion spectra extracted from the 
$\tau^\pm \ra \nu_\tau \pi^\pm \pi^0$ decay by the various groups 
\cite{Aleph,Cleo,Belle}. Therefore, it is motivated to
examine  the behavior of each of the collected $e^+ e^- \ra \pi^+ \pi^-$ data 
sets, {\it independently of each other}, while keeping as 
$common$   reference the data corresponding to the
channels listed in {\bf (i)} -- {\bf (iii)}. Stated otherwise, the data
for the channels listed in {\bf (i)} -- {\bf (iii)},
 together with the BHLS model, represent a benchmark, able to examine 
critically any given $e^+ e^- \ra \pi^+ \pi^-$ data  sample.

It then only remains to account for isospin breaking effects specific of the
$e^+ e^- \ra \pi^+ \pi^-$ channel, in a clearly identified way.
{\it A priori}, IB effects specific
of the  $e^+ e^- \ra \pi^+ \pi^-$ annihilation are threefold and cover~:
\begin{itemize}
\item{\bf (j)} Information on the decay $\rho^0 \to e^+ e^-$,
\item{\bf (jj)} Information on the decay $\omg \to \pi^+ \pi^-$,
\item{\bf (jjj)} Information on the decay $\phi \to \pi^+ \pi^-$.
\end{itemize}
  
The importance of decay information on  $\rho^0 \to e^+ e^-$
to determine IB effects has been emphasized in only a few previous works
\cite{Fred11,ExtMod2,ExtMod3}. Within the BHLS model, the ratio
$f_{\rho \gamma}(s)/f_{\rho W}$
exhibits non--negligible IB effects  for this particular 
coupling (see Fig. 11 in  \cite{ExtMod3}). They amount to several percents in   
the threshold  region quite important for evaluating $g-2$.

There is certainly no piece of information in the data covered by 
the channels listed in {\bf (i)} -- {\bf (iii)} above concerning 
the decay information {\bf (jj)} 
or {\bf (jjj)}. In contrast, the vertex $\rho^0 e^+ e^-$ is certainly involved
in all the annihilation channels considered.
Imposing the RPP \cite{RPP2010} information  
$\Gamma(\rho^0 \to  e^+ e^-)=7.04 \pm 0.06$ keV  is, nevertheless, 
legitimate because
the channels {\bf (i)} -- {\bf (iii)} do not significantly 
constrain the decay width $\rho^0 \to e^+ e^-$. 

For the following discussion we define the
branching ratio products $F_\omega\doteq \mathrm{Br}(\omega
\to e^+e^-)\times\mathrm{Br}(\omega \to \pi^+\pi^-)$ and
 $F_\phi\doteq \mathrm{Br}(\phi \to e^+e^-)\times\mathrm{Br}(\phi \to
\pi^+\pi^-)$; these
pieces of information are much less model dependent
than their separate terms (see Subsection 13.3 in \cite{ExtMod1}).
The RPP accepted information for these products are
$F_\omega=(1.225 \pm 0.071)~10^{-6}$ and $F_\phi=(2.2 \pm 0.4)~10^{-8}$. 

Two alternative analysis strategies can be followed~:
\begin{itemize}
\item{\bf (k)} Use the accepted values \cite{RPP2010}
for the $\rho^0 \to e^+ e^-$, $\omg \to \pi^+ \pi^-$
and $\phi \to \pi^+ \pi^-$.
These are the least experiment dependent pieces of 
information\footnote{Nevertheless,  one should keep in mind that these
accepted values are highly influenced by the $e^+ e^- \ra \pi^+ \pi^-$ 
$scan$ data samples compared to others. Therefore, this choice could favor  
the CMD-2 and  SND data samples when fitting; 
however, as these accepted values are  
certainly not influenced by none of the BaBar or KLOE data samples,
the behavior of each of the various ISR data samples becomes a crucial piece
of information.}. 
We will be even more constraining by supplementing these
$\omg $ and $\phi$ branching ratios by phase information~:  The so--called
Orsay phase concerning the $\omg$ decay\footnote{We will use as input the value 
$104.7^\circ \pm 4.1^\circ$ found by \cite{ffVeryOld}, which is   
consistent with the results recently derived 
\cite{Maltman2009} while using an analogous (HLS) model
(see Tables VI--IX therein).}
and the reported phase\footnote{The single existing measurement $-34^\circ \pm 4^\circ$ 
is reported by the SND Collaboration \cite{SNDPhi}.} of the 
$\phi \to \pi^+ \pi^-$ amplitude relative to $\rho^0 \to \pi^+ \pi^-$.

\item{\bf (kk)} Use directly data when possible. Indeed, all relevant IB information
carried by $\rho^0 \to e^+ e^-$ and $\omg \to \pi^+ \pi^-$ can be numerically derived
within the BHLS model by the difference between the $\pi^+ \pi^-$ 
spectra and the dipion spectrum $\pi^\pm \pi^0$ in the $\tau$ decay;
more precisely, using the $\pi^+ \pi^-$ spectrum within the tiny
energy region $0.76 \div 0.82$ GeV should be enough to derive  the relevant
IB pieces of information in full consistency with our model.

As all scan (NSK) $e^+e^- \ra \pi^+\pi^- $ data samples
\cite{Barkov,CMD2-1995corr,CMD2-1998-1,CMD2-1998-2,SND-1998} and
both KLOE data sets (KLOE08 and KLOE10) stop below 1 GeV, 
the $\phi$ information should be taken from somewhere else, namely
from the RPP. Fortunately, the $\phi$  region
is now covered by the BaBar data set \cite{BaBar,BaBar2}.
Therefore, as soon as 
the consistency of the $\phi \to \pi^+ \pi^-$ information carried by
the BaBar data and by \cite{RPP2010,SNDPhi} is established, this part of 
the spectrum could supplement the scan and KLOE data 
sets\footnote{\label{bbr_phi} Actually, the few BaBar data points between, say,
1.0 and 1.05 GeV carry obviously more information than only the branching
ratio and the "Orsay" phase at the $\phi$ mass.}. 

Concerning the $\phi \to \pi^+ \pi^-$ information, this second strategy will 
be used by either taking the \cite{RPP2010,SNDPhi}
 information or the BaBar data points  between $1.0 $ and $1.05$ GeV.
 
\end{itemize}

\begin{figure}[!ht]
\begin{minipage}{\textwidth}
\begin{center}
\resizebox{\textwidth}{!}
{\includegraphics*{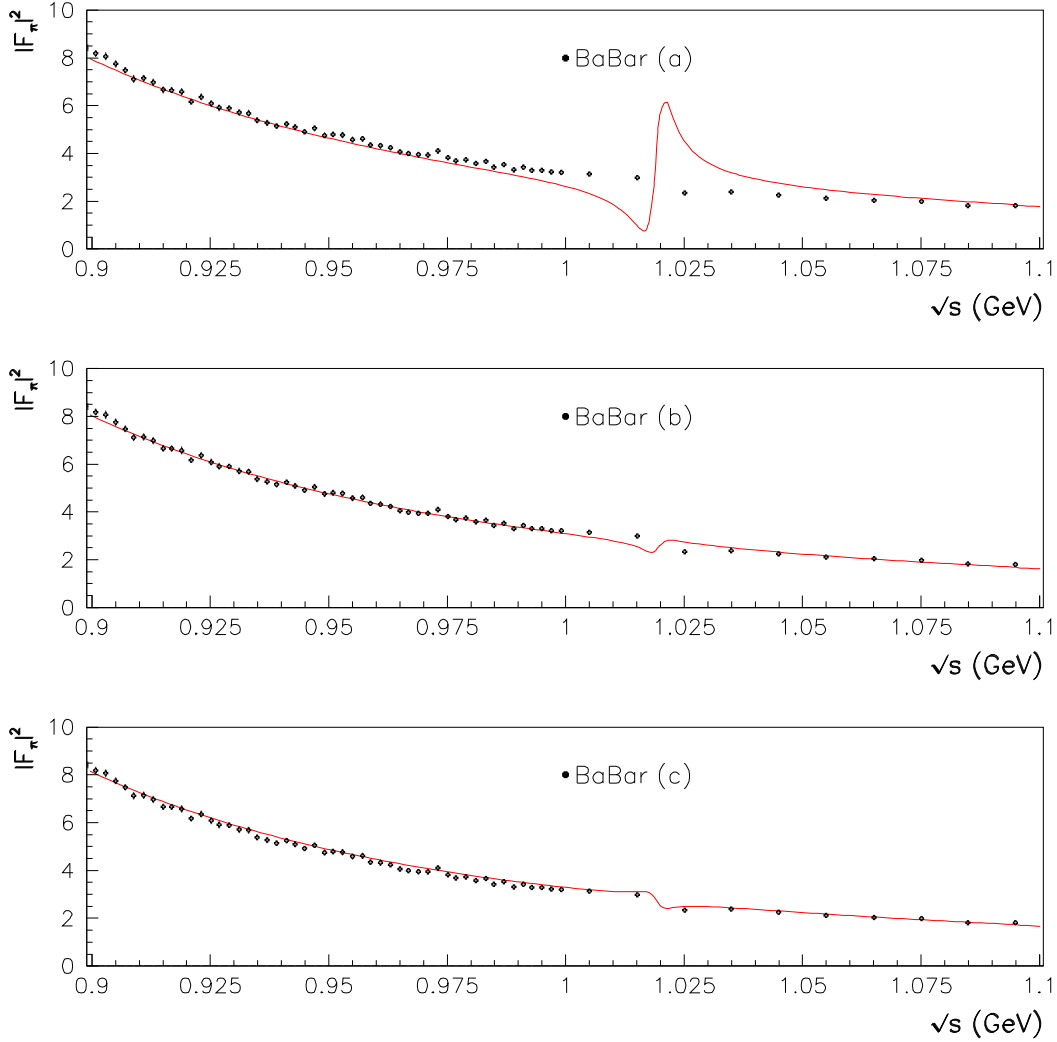}}
\end{center}
\end{minipage}
\begin{center}
\vspace{-0.3cm}
\caption{\label{Fig:modphi}
The  $e^+e^- \ra \pi^+ \pi^-$ cross section around the $\phi$ mass together
with BaBar data superimposed.
The curve in (a) displays the prediction using the RPP $\phi$ decay information
computed at the $\phi$  Higgs--Kibble mass; the curve in (b) displays the prediction using 
the PDG $\phi$ decay information computed at the experimental  $\phi$ mass.
In (c) the PDG $\phi$ decay information is replaced  by the five BaBar data points
located between 1. and 1.05 GeV.}
\end{center}
\end{figure}

\subsection{How to Implement  $\omg/\phi \to \pi^+ \pi^-$ PDG Information? }
\label{Vdecays}
\indent \indent 
The vector meson couplings to $\pi^+ \pi^-$ or $e^+ e^-$ depend
on the $s$--dependent "mixing angles" $\alpha(s)$, $\beta(s)$ and $\gamma(s)$.  
This does not give rise to any ambiguity as 
long as one deals with spectra; however,
when using the PDG information for vector meson decays, especially 
to  $\pi^+ \pi^-$ or $e^+ e^-$, one has to specify at which value
for $s$ each of the vector meson (model) coupling should be evaluated.

Within the HLS model, there are  {\it a priori} two legitimate choices for 
the mass of vector mesons; this can either be the Higgs--Kibble (HK) mass which
occurs in the Lagrangian after symmetry breaking or, especially  for the $\omg$ and $\phi$ mesons,
the experimental (accepted) mass as given in the RPP. Prior to
the availability of the BaBar data \cite{BaBar}, the published 
$e^+ e^- \ra \pi^+\pi^-$ cross section data did not include the $\phi$ mass
region
and, therefore, there was no criterion to check the quality of each possible 
choice in the  $\phi$ mass region\footnote{As the HK mass for the 
$\omg$ meson coincides almost exactly with its accepted RPP value,
 the problem actually arises only for the $\phi$ meson.}. 
The choice made in the previous studies
using the broken HLS model \cite{taupaper,ExtMod1,ExtMod2,ExtMod3} was the  
 $\phi$ HK mass.

As already noted, the broken HLS model, fed with the
data listed in {\bf i}--{\bf iv} (see Subsection \ref{issue_dat},
above), provides predictions for the pion form factor independently
of the measured $ e^+ e^- \to\pi^+ \pi^-$  data. This procedure is discussed 
in detail in the next section.
  Here we anticipate some results specific to the $\phi$ mass issue.

Fig. (\ref{Fig:modphi}a)
displays the prediction for the pion form factor in the $\phi$ region
using the HK mass to estimate the $\phi\pi^-\pi^+$ coupling constant with
the BaBar data superimposed (not fitted); it is clear that the prediction
is quite reasonable up to $\simeq 0.98$ GeV  as well as above $\simeq 1.05$ 
GeV. However, it is clearly unacceptable for the mass region in--between.
In contrast, using the $\phi$ mass as given in the RPP
to extract the $\phi\pi^-\pi^+$ coupling constant  from its accepted partial width 
\cite{RPP2010}
provides the spectrum shown in Fig. (\ref{Fig:modphi}b); this alternate choice
is certainly reasonable all along the mass region displayed. Therefore,
it is  motivated to update our former results \cite{ExtMod3} by performing
the change just emphasized\footnote{We show later on that choosing
the $\phi$ HK mass has produced some overestimate of  the prediction for
$a_\mu$ and, thus, some underestimate of the discrepancy with 
the BNL measurement  \cite{BNL,BNL2}.}. 
In order to be complete, it is worth mentioning here a fit result obtained
by exchanging the PDG/SND  $\phi$ decay information
with the BaBar pion form factor data$^{\ref{bbr_phi}}$ 
in the range ($1.0 \div 1.05$) GeV. The result,
given in Fig. (\ref{Fig:modphi}c),  shows that the lineshape 
of the BaBar pion form factor at the $\phi$ mass
can be satisfactorily accommodated. As the exact pole position of the $\phi$ 
meson is determined by a benchmark independent of the $e^+e^- \to \pi^-\pi^+$ 
process
(see Subsection \ref{issue}), the drop exhibited by  Fig. (\ref{Fig:modphi}c)
in the BaBar is perfectly consistent with an expected $\phi$  signal.

\section{$\tau$ Predictions of the Pion Form Factor}
\label{tauPred} 
\subsection{$\tau$+PDG Predictions}
\indent \indent 
        As mentioned before, the charged isovector $\tau^\pm \to
        \nu_\tau\pi^\pm \pi^0$ dipion spectra are not affected by
        $\gamma-\rho-\omega-\phi$ mixing and hence are of much simpler
        structure. Supplemented by the basic $\rho-\omega-\phi$ mixing
        effects  which derive from $SU(2)$ and $SU(3)$ flavor
        breaking, one has a good starting point to fix the parameters
        of the BHLS model to predict the process $e^+e^- \to \pi^+\pi^-$.
        Specifically, we are using the data including the channels listed
        in {\bf i-iv} of Section \ref{issue} together with RPP information
        relevant to fix the IB effects affecting the pion form factor.  
        This method is named, somewhat abusively\footnote{By abusively, we
	 mean,  first that the
	"Orsay" phases  for both the $\omg$ and $\phi$ mesons
	 have no entry in the RPP and, second, that the benchmark
	 represented by the processes listed in Subsection \ref{issue} 
	 within items {\bf (i)} to {\bf (iv)} have little to do
	 with $\tau$ or the RPP.}  $\tau$+PDG.

Specifically, the IB effects encoded in $\mathrm{Br}(\omega \to \pi^+\pi^-)$,
$F_\omega\doteq \mathrm{Br}(\omega \to e^+e^-)\times\mathrm{Br}(\omega \to
\pi^+\pi^-)$ and
        $F_\phi\doteq \mathrm{Br}(\phi \to e^+e^-)\times\mathrm{Br}(\phi \to
\pi^+\pi^-)$ are taken from the RPP. 
For the missing phase information
we adopt the result from the fit \cite{ffVeryOld} for the Orsay phase of 
the $\omega \to \pi^+\pi^-$
amplitude and the result from SND   \cite{SNDPhi} for the phase of the
$\phi \to \pi^+\pi^-$ amplitude\footnote{A preliminary version of the present 
  work was presented \cite{krakow} at the
Workshop on Meson Transition Form Factors held on May 29-30, 2012 in Krakow, Poland.
Some minor differences may occur with the present results due to the fact that  
the SND phase for  $\phi \to \pi^+ \pi^-$ was not imposed in the preliminary work.}. 
Following the discussion in
the preceding Subsection, the model branching ratios and
phases are computed at the vector boson masses accepted by the
RPP.

The fit returns a probability of 89.4\% with $\chi^2/n_{dof}=553.4/596$.
The fit quality ($\chi^2/n_{points}$) for each of the fitted channels 
is almost identical to our results in \cite{ExtMod3} (see the last column
in Table 3 therein). Each of the decay partial width 
extracted from \cite{RPP2010}  contributes
by $\simeq 1$ to the total $\chi^2$. It is also worth mentioning that
the dipion spectra from \cite{Aleph,Cleo,Belle} are nicely described
up to  $\sqrt{s}=1$ GeV and provide residual distributions indistinguishable 
from those shown in Figure 10 of \cite{ExtMod3}.
From this fit, one derives the
 ($\tau$+PDG) predictions for the pion form factor
 which can be compared with  the various existing $e^+e^- \ra \pi^+\pi^- $ 
 data samples.
 
 The overall view of the comparison is shown
in Fig. \ref{Fig:taupred_all}. This clearly indicates that the data associated
with the channels listed in {\bf i}--{\bf iv}, supplemented with a limited 
 PDG information is indeed able to provide already a satisfactory picture of 
the pion form factor as reported by all experiments having published
$e^+e^- \ra \pi^+\pi^- $ spectra. 

Let us stress that the predicted pion form factor relies on
the $\pi^\pm \pi^0$ spectra provided by ALEPH \cite{Aleph},
Belle \cite{Belle} and CLEO \cite{Cleo} only up to 1.0 GeV.
Therefore, the inset in  Fig. \ref{Fig:taupred_all} actually 
shows the $extrapolation$  of the prediction 
into the spacelike region with the NA7 data \cite{NA7} superimposed;
this clearly indicates that there is no 
{\it a priori} reason to discard the 
spacelike data from our data handling. One should also note
that the extrapolation of the prediction above the
$\phi$ mass is quite reasonable up to $\simeq 1.2\div 1.3$ GeV. This may 
indicate  that the influence of high mass vector mesons is negligible up to this energy region.

In order to make more precise statements, let us  
magnify piece wise the information carried by  
Fig. \ref{Fig:taupred_all}.  Thus, Fig. \ref{Fig:taupred_low} displays the behavior
of the various $e^+e^- \ra \pi^+\pi^- $ data samples in the
($0.3 \div 0.7$) GeV energy region. 
As a general statement the behavior expected
from the existing data samples looks well predicted by 
the $\tau$+PDG method. 
A closer inspection allows to infer that the CMD--2
and SND data points ({\it i.e.} NSK when used together)  are well
spread onto both sides of the predicted curve; this property is also
shared by the KLOE10 sample. Even if reasonably well described,
        the KLOE08 and BaBar data samples are lying slightly above
	the $\tau$+PDG expectations; this difference should vanish
	when including the $\pi^+\pi^-$ spectra inside
	the fit procedure.
 
Fig. \ref{Fig:taupred_high} displays the behavior
of the various $e^+e^- \ra \pi^+\pi^- $ data samples in the
($0.85 \div 1.2$) GeV energy region. Here also the predicted curve accounts well for
the data behavior. A closer inspection tells that the sparse NSK data 
are well described. The BaBar data are also well accounted for all along 
this energy interval except for the $\phi$ region.  As shown by Fig. 
(\ref{Fig:modphi}c) above, this can be cured  and one can
show that the difference is mostly due  to  the phase 
for the $\phi \to \pi^+ \pi^-$ amplitude which 
departs
significantly\footnote{\label{ref_phi}
This issue is examined in detail in Subsection \ref{PhiRegion}
below.} 
from those provided by SND \cite{SNDPhi}. One could also note that
both KLOE data samples look slightly below the $\tau$+PDG expectations in this region.

One may conclude from Figs. \ref{Fig:taupred_low} and \ref{Fig:taupred_high}
that our "$\tau$+PDG" predictions are in good agreement with the data
and that a fit using fully these data samples should provide marginal 
differences between all $\pi^+\pi^-$ data sets \footnote{
\label{errors} Precise information concerning fit qualities is ignored in this Section.}.

However, the picture becomes quite different in the medium energy region
($0.70 \div 0.85$) GeV as illustrated by Figure \ref{Fig:taupred_mid}. In this
region,  our  $\tau$+PDG  prediction  follows almost perfectly 
expectations$^{\ref{errors}}$
from both the KLOE08 and KLOE10 data and the detailed lineshape
of the $\omg-\rho$ interference region is strikingly reproduced. 
Paradoxically, the NSK data are slightly less favored -- especially around 0.8 GeV --
despite their influence on the PDG information used
in order to account for IB effects in the $\omg-\rho$ interference region; 
however, taking into account experimental uncertainties, we already know that
a global fit using NSK data is highly successful \cite{ExtMod3}. 

In contrast, the behavior of the BaBar data  
looks inconsistent with the $\tau$+PDG prediction, especially on the low
mass side of the interference region. Actually, the observed overestimate of
the BaBar spectrum affects the whole region from threshold to the $\omg$
mass but is more important in the range ($0.74\div 0.78$) GeV. At higher energies 
one observes a reasonable agreement with expectations as well as with both
KLOE data sets. 

One should also note that the $\omg$ mass and (total) width induced
by the data for the $\pi^0\gamma$, $\eta\gamma$, $\pi^+\pi^-\pi^0$
final states are in perfect agreement with all the examined data samples;
this indicates that the energy calibration around
the $\omg$ mass is good for all ISR data samples. Fig. (\ref{Fig:modphi}c)
has already shown that the BaBar energy calibration is also good in
the $\phi$ region.

\subsection{ Predictions Using the $\rho^0-\omg $ Interference Region From Data}
\label{localfit}
\indent \indent As stated in Subsection \ref{issue_dat}, one can replace
the PDG information for $\omg \to \pi^+ \pi^-$ and $\rho^0 \to e^+ e^-$
by any pion form factor spectrum limited to the region 
$(0.75\div 0.82$) GeV. In particular, this turns out to fit 
 $F_\omg={\rm Br}(\omg \to e^+ e^-) \times {\rm Br}(\omg \to \pi^+ \pi^-) $  and
 the Orsay phase as it comes out from each of the specified data
 sets. 
 
 Concerning the $\phi \to \pi^+ \pi^-$ mode, one can also check
 the BaBar data \cite{BaBar,BaBar2} versus the SND  datum \cite{SNDPhi}.
 This will be discussed below when reporting on fitting the whole spectra
 (see Subsection \ref{PhiRegion}). 
 
\begin{table}[ph]
\centering
\begin{tabular}{|| c  || c  | c | c | c||}
\hline
\hline
\hhhc
\hhhd  Data Sample   &\multicolumn{4}{|c|}{  \hhhv Local Fit}\\
\hhhd ~~~~ & \hhhv $F_\omega\,10^{6}$ 
   &  \hhhv Orsay Phase (degrees) &  \hhhv Prob. (\%) &  \hhhv $\chi^2/n_{\pi^+\pi^-}$\\
\hline
\hline
 \hhhv Reference values \cite{SND-1998,ffVeryOld}  & $1.225\pm 0.071$ & $104.70\pm 4.10$ 
 & --   & -- \\
\hline
  \hhhv $\tau$ + PDG & $1.157\pm 0.053 $ &$108.92\pm 2.36$  & 89.4\%  & -- \\
\hline
 \hhhv $\tau$ + NSK \cite{CMD2-1998-1,SND-1998}  &  $1.219\pm 0.043 $ &$106.71\pm 0.25$  
 & 90.0\% & 52/47 \\
\hline
 \hhhv $\tau$ + KLOE08 \cite{KLOE08}  &$1.076\pm 0.041 $ & $110.44\pm 1.24$&  87.8\% 
 & 18/11  \\
 \hline
 \hhhv $\tau$ + KLOE10 \cite{KLOE10}   & $0.973\pm 0.045 $&$113.62\pm 1.63$ & 92.7\% 
 & 11/11    \\
 \hline
 \hhhv $\tau$ + BaBar  \cite{BaBar} 
&$1.780\pm 0.011 $&  $107.50\pm 0.19$  &54.4\% &67/35\\
\hline
\hline
\end{tabular}
\caption {
\label{Table:T1} Fit results for
$F_\omg\doteq{\rm Br}(\omg \to e^+ e^-){\rm Br}(\omg \to \pi^+ \pi^-) $ 
and for the Orsay phase (in degrees) using only the  $(0.75\div 0.82$) GeV energy 
region of each $\pi^+\pi^-$ data sample. 
The fit is done following the (local)
procedure sketched in Subsection \ref{localfit}; we use
${\rm Br}(\omg \to \pi^+ \pi^-)=(1.53\pm 0.13)\%$ \cite{RPP2010}
and the Orsay phase from \cite{ffVeryOld} as input values for
the ``$\tau$ + PDG'' fit. The probabilities are those of the global fit;
the last data column shows the contribution of each $\pi^+\pi^-$ data set
to the total $\chi^2$ and gives the corresponding number of data points.
} 
\end{table}

 Using the RPP recommended value for  ${\rm Br}(\omg \to \pi^+ \pi^-)$
 and the Orsay phase information from \cite{ffVeryOld} yields a value
 for $F_\omega$ in good correspondence with expectations, as clear from  the 
 entry $\tau$ + PDG in Table \ref{Table:T1}.
 Using NSK data  or any of both KLOE samples, instead of
 the PDG information, does not lead to predicted curves substantially
different from their analogue already shown and commented upon in the previous
Subsection. Interesting parameter values have been extracted
from global fits using
only the $(0.75\div 0.82$) GeV region  from the NSK and KLOE spectra
for $e^+ e^- \to \pi^+ \pi^-$ and are reported in Table \ref{Table:T1}; 
they are  in reasonable agreement with the reported branching  
ratio product $F_\omg$ \cite{SND-1998} and the Orsay phase 
\cite{ffVeryOld,Maltman2009} as well. Indeed,
taking the RPP $F_\omg$  value as reference, our estimates using the 70 MeV
interval  surrounding the interference region are at $0.1 \sigma$,
$1.8  \sigma$ and  $3.0 \sigma$ for respectively the NSK 
\cite{CMD2-1998-1,SND-1998},
KLOE08 \cite{KLOE08} and KLOE10 \cite{KLOE10} data samples. 
The difference between the RPP recommended value for $F_\omg$ 
and our entry for NSK  also tells that the BHLS parametrization and
the more  standard form factor lineshape used by SND \cite{SND-1998} provide  
almost identical values for $F_\omg$.

As far as BaBar data are concerned, the situation looks different and the most relevant 
piece of information is provided in Figure \ref{Fig:tau_bbr_mid}. This proves that
the largest difference between BaBar data and the other analogous
data samples \cite{CMD2-1998-1,SND-1998,KLOE08,KLOE10} is the
$F_\omg$   information inherent to the BaBar data. Clearly, Table \ref{Table:T1}
shows that the BaBar value for  $F_\omg$ is off from its recommended value 
by $(7 \div 8) \sigma$.  This strong disagreement is substantiated by comparing
 Figures  \ref{Fig:taupred_mid} and \ref{Fig:tau_bbr_mid}. 

Table \ref{Table:T1} also reports the fit probability for each of the examined configurations.
With about 90\% probabilities, the "$\tau$+PDG" prediction and the NSK, KLOE08 and KLOE10 (global) 
fits exhibit a full consistency with the rest of our benchmark ({\it i.e.} all other
annihilation channel physics). The agreement of this with BaBar data, even limited to
such a tiny interval, is found much poorer  and exhibits a clear tension between
$F_\omg^{BaBar}$ and the rest of the (non--$\pi^+ \pi^-$) physics accessible to the HLS model. 

 Additional pieces of information are provided in the last data column of Table 
\ref{Table:T1} which complements the global fit probabilities. These are the
values for $\chi^2/n_{\pi^+\pi^-}$ for each of the various $\pi^+\pi^-$ data samples,
$n_{\pi^+\pi^-}$ being the number of data points included in the fitted energy range
({\it i.e.} $(0.75\div 0.82)$ GeV).

\subsection{Isospin Breaking Effects in the BHLS Model~: Comments}
\indent \indent It follows from the developments just above that the BHLS model
fed with a limited number of accepted values for some IB pieces of information is indeed able to provide
a quite satisfactory prediction for the $e^+ e^- \to \pi^+ \pi^-$ cross section
once the $\tau$ spectra are considered. This gives support to our breaking
model, especially to the $s$--dependent vector meson mixing mechanism.

The prediction is found in accord with the scan (NSK) data samples 
and with both KLOE data sets\footnote{Some issue with 
the uncertainties of the KLOE08 data sample will be discussed below.}.
 Indeed, the predicted lineshape  
strikingly follows  the central values from both KLOE data samples;
for the scan data, the prediction based on PDG information
is good but not as good as for the KLOE data. However, 
changing the PDG requested IB information by less
than 1 $\sigma$ -- as following from a mere comparison of the first and second lines in Table 
  \ref{Table:T1} -- leads to a perfect description of the NSK spectra 
 over the whole available energy range \cite{ExtMod3}. 
  In contrast, the RPP branching fraction product $F_\omg$ has to be changed by
  about   $ 7 \sigma$ in order to yield a comparable  description of the BaBar
 \cite{BaBar} data. 
 
 \vspace{0.5cm}
 \indent \indent 
  Basically, our approach is a $\tau$ based prediction of $\pi^+ \pi^-$ spectra; 
  it relies on the consistency of several different physics channels, the
   $\tau$ spectra  and on a model of isospin symmetry breaking (IB). {\it In fine},
   our breaking model does not carry IB parameter values plugged in from start, but
   yields the numerical IB effects in a data driven mode.
  It is thus interesting to examine the consequences 
 of this $\tau$ (and global) based approach on the muon $g-2$ estimated value. 
 For this purpose, it is worth stressing that the $\tau$ based estimates given 
 just below -- and later --
 are $not$ computed by integrating the $\tau$ spectra (and adding corrections
 like in \cite{Belle,DavierHoecker}, for instance), 
 but by integrating the $e^+e^-$ cross sections
 they allow to reconstruct through the global BHLS framework.
 
 This is what is shown in  Figure \ref{Fig:tau_gM2}. 
The first line displays the result derived from the ``$\tau$+PDG'' fit.
The four following lines are obtained by replacing the $\rho$ and $\omg$ IB 
information by the $0.76\div 0.82$ GeV region of the quoted data sets. The line named BNL
displays the experimental result \cite{BNL,BNL2} and 
the last line shows the $\tau$ based estimate
from \cite{DavierHoecker}. The last data column in  Figure \ref{Fig:tau_gM2}
gives the probability of the corresponding fits. 

It is clear  that all methods used to include IB effects within our $\tau$ based
approach give consistent results, all distant from the BNL measurement
at the $4 \sigma$ level. The associated probabilities indicate the 
quality of the   fits from where they are derived.

\section{Global Fits Using the  $e^+e^-\to \pi^+\pi^-$ Spectra}
\indent \indent
\label{globalfits}
As in our previous analysis \cite{ExtMod3}, we have performed
global fits using simultaneously all $e^+e^-$ annihilation data into the  
$\pi^0\gamma$, $\eta\gamma$, $\pi^+\pi^-\pi^0$, 
$K^+K^-$, $K^0 \overline{K}^0$ final states, the dipion spectra
collected in the decay of the $\tau$ lepton \cite{Aleph,Belle,Cleo}
and the decay information listed in Subsection \ref{issue_dat}. We also use
the RPP $\phi$ decay properties  in the (updated) way emphasized in Subsection \ref{Vdecays}.

For what concerns the $e^+e^-\to \pi^+\pi^-$ data included into the global 
fit procedure, 
we have performed fits using separately the NSK, KLOE08
and KLOE10 data samples. Global fits have also been performed for 
the BaBar dipion
spectrum restricted to the range of validity of our BHLS model.
Two options have been considered,
using BaBar data up to 1 GeV supplemented by
$\phi \to \pi^+\pi^-$ decay properties from the RPP or using the
BaBar data up to 1.05 GeV, thus including its $\phi$
region and avoiding the need of using the RPP information about the $\phi$.

In all cases, the errors (and the $\chi^2$) were constructed following the
information published/ recommended by the experimental groups who collected
these data. For instance, concerning  the scan data samples, the 
full covariance matrix is generally constructed by adding the systematic error
covariance matrix $V_{syst}$ to the (diagonal) statistical covariance 
$V_{stat}$ constructed from the tabulated uncertainties as described
in \cite{ExtMod1} (see Section 6 therein); $V_{syst}$ is constructed assuming
the reported systematic errors bin--to--bin correlated\footnote{When they
are reported, uncorrelated systematics are simply added in quadrature
to the statistical errors.}
. Nevertheless, for the
rather imprecise $e^+e^- \to (\pi^0/\eta) \gamma$ data, taking into account
the large magnitude of the reported systematics, the systematic and statistical
errors were simply added in quadrature. For the  $e^+e^- \to \pi^+ \pi^- \pi^0$
data, we dealt with depending on the level of precision of the relevant data 
sample (see Subsection 2.2.3 in \cite{ExtMod2} for more information).  

For the $\tau$ data, as both
 $V_{syst}$ and $V_{stat}$ are publicly available \cite{Aleph,Cleo,Belle},
 one only has to  add them up, as 
 already  performed for the study in \cite{ExtMod3}. 
 For  the BaBar sample,  the systematic uncertainties on the cross section
  are given (as a function of $\sqrt{s}$)   
in Table V from \cite{BaBar2} and are imposed to be 100 \% bin--to--bin 
correlated (following Section F in \cite{BaBar2}). Writing, for definiteness,
 each of these uncertainty functions as $f(s_i)$, for the data point $m_i$ 
 located at $\sqrt{s_i}$, the $(i,j)$ entry  of the    $V_{syst}$ matrix
 is given by the sum of the various $f(s_i)f(s_j)m_im_j$.
 
  For both the KLOE08 and KLOE10 data samples \cite{KLOE08,KLOE10}, the KLOE Collaboration 
 provides basically the same information as BaBar and
 therefore, we simply have to proceed  likewise.

Moreover, the various $\pi^+\pi^-$ data sets collected by CMD--2 and SND 
\cite{CMD2-1998-1,CMD2-1998-2,SND-1998} on the one hand and the older data samples
from  \cite{Barkov} on the other hand,
both carry common bin--to--bin $and$ sample--to--sample 
correlated uncertainties estimated resp. to 0.4\% and $\simeq 1$\%. As
in our previous studies
\cite{ExtMod2,ExtMod3}, this effect is accounted for in the minimization code.
These are the most important reported correlations of this type within the scan data
\cite{SimonPriv}.

\vspace{0.5cm}

We have also performed global fits using combinations of these individual 
$\pi^+\pi^-$ data samples. In this case,
the contributions of the NSK and KLOE10 data to the total $\chi^2$ were left
unweighted as their own $\chi^2/n$ contribution is always of the order 1
in fits using each of them in ``isolation''\footnote{\label{standalone} 
We remind that, in this work,
the wording  ``isolation''  or ``standalone'' is $always$ used to qualify the
fits performed using a specific $e^+e^-\to \pi^+\pi^-$ data sample. It is 
 understood that each of the  $\pi^+\pi^-$ data sets ``fitted in isolation''  
 is always fitted in conjunction with all the other data listed in Subsection \ref{issue_dat}.
When several of the $e^+e^-\to \pi^+\pi^-$ data samples are submitted to fit
(together with the other channels), we then refer to ``combined'' fits.
In order to warn the reader, we prefer keeping the word isolation
between quote marks.}. 
In contrast, in such combinations involving the KLOE08 
and/or BaBar data, the contribution of each of these
to the total $\chi^2$  was weighted by the ratio $f_M=n_{M}/\chi^2_{M}$
(M= KLOE08, BaBar) where $\chi^2_{M}$ is the $\chi^2$ of the M data set obtained
in the best fit using only M as $e^+e^-\to \pi^+\pi^-$ data set; $n_M$ is the 
corresponding number of data points. In fits involving 
the spacelike data \cite{NA7,fermilab2},   the corresponding weight was also 
used\footnote{In the fits referred to in \cite{krakow}, the spacelike data contributions 
to the total $\chi^2$ were left unweighted.}. 

For definiteness, when relevant, we have used
  $f_{NSK}=f_{KLOE10}=1$, $f_{KLOE08}=60/90\simeq 0.67$, 
$f_{BaBar}=270/346\simeq 0.78$ and $f_{space}=59/85\simeq 0.69$. 
These weights have been varied and it has been found that
 the sensitivity of the physics results to their precise value
 is marginal;  the main virtue of 
 these weights is to provide probabilities  not too much ridiculous. 
 On the other hand, as a matter of principle, when results are displayed
 which have been obtained using weights, it is quite generally for the reader's information. 
 We have preferred being  conclusive by only relying on the largest data 
 set combinations which do not call for any reweighting. Indeed, this
 simply reflects that global fit probabilities do not raise any objection
 to trusting the uncertainties  as they are reported together with each of 
 the used data sets. 
 
 This  method\footnote{This method is quite parent from the $S$--factor
 technique commonly used in the Review of Particle Properties to  account,
 while averaging, for marginal inconsistencies between the various reported 
 measurement/uncertainty of some physics quantity.},  turns out to consider 
 each data set as a global object, rather than defining local 
 ($s$--dependent) averages as
 done by others \cite{DavierHoecker3}. This method looks better
 adapted to the global fit method which provides a quality check
 reflecting the behavior of each $e^+e^-\to \pi^+\pi^-$ data set 
 within the global context of a large number of physics channels.
 Indeed, doing   local averages would prevent to detect discrepancies 
 originating from some given data set only. On the other hand, within
 a global framework as BHLS, such a method would lead to fit parameter
 values modified in a completely uncontrolled way. It is the reason why
 our final results will only rely on sample combinations which do not call for
 any reweighting ({\it i.e.}  necessarily going  beyond experimentally 
 reported uncertainty information).

For completeness,
it is also worth noting that the $\chi^2$ contributions of the -- more than 40 -- 
data sets associated with all the other channels were always left unweighted.

A feature common to all fits using the  $e^+e^-\to \pi^+\pi^-$ data
sets in ``isolation'' or combined is that the individual $\chi^2$ contributions
associated with the other channels ($\pi^0\gamma$, $\eta\gamma$, $\pi^+\pi^-\pi^0$, 
$K^+K^-$, $K^0 \overline{K}^0$, \ldots) were only marginally affected
by the specific choice of $e^+e^-\to \pi^+\pi^-$ data submitted to fit. 
Their typical values are almost identical to what can be found in the last 
data column of Table 3 in
\cite{ExtMod3}; more precisely, the $\chi^2$ value provided by each of these channels
never varies by more than a few  percents. Let us  remind that
the number of data points submitted to fit -- beside the
$e^+e^-\to \pi^+\pi^-$ data -- is $\simeq 600$ when working within\footnote{
The difference between configuration B and configuration A is that the former
excludes the use of the 3--pion data collected around the $\phi$ mass. } 
the configuration B  defined in \cite{ExtMod3} ($\simeq 675$ within configuration A). 

\subsection{Salient Features of the Various $e^+e^-\to \pi^+\pi^-$ Spectra}
\label{features}
\indent \indent We have performed several tens of 
fits of the various $e^+e^-\to \pi^+\pi^-$ spectra in ``standalone'' 
mode$^{\ref{standalone}}$  and/or
combined. It does not look  useful to report on each fit in detail. Instead
of overwhelming the reader with unnecessary information and plots, we have preferred
focusing on the salient features of  their behavior within the global fit 
context. Beside the fit properties of the full spectra (up to 1 GeV, generally),
 this covers  the muon anomalous magnetic moment value and the behavior at the 
 $\omg$ mass, more precisely the value for 
 $F_\omg={\rm Br}(\omg \to e^+ e^-) \times {\rm Br}(\omg \to \pi^+ \pi^-) $.
  The first of these topics will be addressed in a separate Section below.  
 Concerning the second topic, we remind that the value for  $F_\omg$ expected from the 
 RPP  is $(1.225 \pm 0.071)~ 10^{-6}$.  

On the other hand, we do not enter into much detail
 concerning the effects of the spacelike data, always used weighted in this paper;
 we limit ourselves to mentioning that they never modify   the fit
 qualities in a significant way.

\subsubsection{``Standalone'' Fits of the $e^+e^-\to \pi^+\pi^-$ Spectra}
\label{standaloneFits}
\indent \indent When using only$^{\ref{standalone}}$ the (unweighted) NSK data,
 the fit returns\footnote{ The agreement of this result with the corresponding one
given in Table 3 of  \cite{ExtMod3}  clearly proves that the
modification of the $\phi \to \pi^+\pi^-$ information is
marginal for the fits not containing the $\phi$ region. The
latter is covered by the BaBar data set only.} 
$\chi^2_{NSK}/N_{\pi^+\pi^-}= 128.30/127$ and a global fit probability of
96.3\%. The fit residuals are displayed in the top panel of 
Fig. \ref{Fig:nsk_res} and the fractional deviations from the fitting function 
in the bottom panel. Taking into account the uncertainties, 
both plots do not exhibit any significant departure from flatness.
The fit residual distributions of the ALEPH, CLEO and
Belle dipion spectra also submitted to fit in the present case
are rigorously given by Fig. 10 in \cite{ExtMod3}, where the fit 
also extends over the energy region
from threshold to 1 GeV. Therefore, one indeed observes
flat residual distributions and flat fractional deviations from the fitting
functions  simultaneously in the $e^+e^-$ and $\tau$ channels
as already claimed in  \cite{ExtMod3}.

Similarly, the KLOE10 data set \cite{KLOE10} returns 
$\chi^2_{KLOE10}/N_{\pi^+\pi^-}= 73.68/75$ and a global fit probability
of 87.7\%. The fit residual distribution  is shown in the top panel
of Figure \ref{Fig:kloe10_res}; the bottom panel in this Figure 
shows the fractional deviations from the fitting function. Both
distributions can be considered as reasonably flat. As in the above Fig. 
\ref{Fig:nsk_res},  the plotted errors are the square roots
of the diagonal elements of the (full) error covariance matrix.
Therefore, the value for $\chi^2_{KLOE10}$ and the flatness
of the residuals shown in Figure \ref{Fig:kloe10_res} illustrate
that the (reported in \cite{KLOE10}) full error covariance matrix looks
 correctly understood.

Therefore, within the global fit context, each of the NSK and KLOE10
data samples exhibits the same outstanding behavior \footnote{
\label{prob_qual} As emphasized
in our previous works \cite{ExtMod1,ExtMod2,ExtMod3}, the global
probabilities are enhanced thanks to the highly favorable
fit properties of the $e^+e^-\to (\pi^0/\eta)\gamma$ data;
indeed, $\chi^2/N$ for these are respectively $\simeq 0.76$ and 0.66.  }.
On the other hand, the fit parameters and error covariance matrix allows to derive,
using obvious notations,
$F_\omg^{NSK}= (1.205\pm 0.042)~10^{-6}$ close to the PDG value reported
at top of Table \ref{Table:T1}. One also gets $F_\omg^{KLOE10}= (1.074\pm 0.051)~10^{-6}$,
a $\simeq 2.5 \sigma$ difference with the central value for $F_\omg^{NSK}$.

Under the same conditions, the (unweighted) KLOE08 data set returns 
$\chi^2_{KLOE08}/N_{\pi^+\pi^-}= 96.55/60$ and a global fit probability
of 56\%, which is rather low. 
It is worth noting the remarkable flatness of the residual distributions displayed
by Figure \ref{Fig:kloe08_res} which
does not prevent to yield a relatively large value for $\chi^2_{KLOE08}$. 
The large $\chi^2_{KLOE08}$ and the flatness of the residual distribution
shown in Figure \ref{Fig:kloe08_res}, considered together,
might indicate an issue with the non--diagonal
part of the full error covariance matrix. Anyway, one can conclude  
that the poor KLOE08 fit probability reflects an issue with the
KLOE08 error  estimate rather than a distorted lineshape.

 Figure \ref{Fig:kloe08_res} should be compared with the similar distribution
derived formerly using a primitive version of the BHLS model (see Figure 3 in \cite{ExtMod1}).
The clear improvement substantiates the gain provided by the BHLS model in its present form.
On the other hand, one should also note that 
the value for $F_\omg^{KLOE08}= (1.117\pm 0.042)~10^{-6}$ is consistent with 
$F_\omg^{KLOE10}$.

The fit of the (unweighted) BaBar data leads to 
$\chi^2_{BaBar}/N_{\pi^+\pi^-}=343.08/270 = 1.27$ 
 when limited to 1 GeV (17\% probability) and to 
 $\chi^2_{BaBar}/N_{\pi^+\pi^-}=340.77/275=1.24$
when going up to 1.05 GeV (corresponding to a 22\% probability).
 The residual distribution yielded when fitting from threshold to 1.05 GeV is  
given in Figure \ref{Fig:babar_res_all} together with the fractional distribution.
The residual distribution derived when fitting from threshold to 
1 GeV, (not shown) is slightly flatter in the $\rho-\omg$ region,
which indicates that accomodating simultaneously
the BaBar $\omg$ and $\phi$ regions has some price.
This has some consequence on the BaBar estimate of the muon
anomalous moment, as will be emphasized in Subsection \ref{amuRef} below.

Nevertheless, these distributions look reasonable and are associated with  quite
reasonable values for $\chi^2_{BaBar}/N_{\pi^+\pi^-}$; the bottom
panel distribution in Figure \ref{Fig:babar_res_all}
is even quite similar to those derived by the BaBar Collaboration
fit shown in Fig. 47 of \cite{BaBar2} (with no quoted fit quality). 

Therefore, the real issue is not the description of the BaBar spectrum {\it stricto
sensu}, but its consistency  with  the rest of the data and physics involved 
in the global fit, especially the $\tau$ data. This is well reflected by
the poor global fit probabilities of 17\% or 22\%,
poorer than for the KLOE08 data sample. Qualitatively, this result could have 
been expected 
from comparing the $\tau$+PDG predictions with "$\tau$+$(\rho -\omg)$" 
(using this
region of the BaBar  spectrum instead of the PDG information)
already analyzed in Subsection \ref{localfit}.
  
Finally, it is worth noting that the fit outcome provides
 $F_\omg^{BaBar}= (1.628\pm 0.012)~10^{-6}$ (fit up to
1 GeV) or $F_\omg^{BaBar}= (1.575\pm 0.010)~10^{-6}$ (fit up to 1.05 GeV),
both being far from $F_\omg^{NSK}$, $F_\omg^{KLOE08/10}$ and from PDG
expectations \cite{RPP2010}.

\subsubsection{Fits Combining the $e^+e^-\to \pi^+\pi^-$ Spectra}
\label{CombFits}
\indent \indent In view of the ``standalone'' fits just reported, we have done
fits of different combinations of the scan (NSK) and ISR data samples.
In this case, the contributions of the KLOE08 and BaBar samples to the
(minimized) $\chi^2$ function  are $always$ weighted as already stated.
\begin{itemize}
\item
Combining the KLOE08 and KLOE10 data~: This returns a consistent picture
where $\chi^2_{KLOE10}$ and $\chi^2_{KLOE08}$ are almost unchanged compared
to their ``standalone'' values and the fit probability reaches 81.6 \%.
This, indeed, confirms that they share the same physics content. 
This is confirmed by the fit result $F_\omg^{KLOE08/10}= (1.121\pm 0.038)~10^{-6}$,
consistent with both of $F_\omg^{KLOE08}$ and $F_\omg^{KLOE10}$.

\item
Combining all the ISR data sets (KLOE08, KLOE10 and BaBar)~: This returns  
$\chi^2_{ISR}/N_{\pi^+\pi^-}=1.34$ when including the weights for
KLOE08 and BaBar and $\chi^2_{ISR}/N_{\pi^+\pi^-}=1.64$ when the weights 
are not included. The weighting used does not prevent the probability to
remain poor$^{\ref{prob_qual}}$ (1.8\%) reflecting 
the level of inconsistency of the BaBar and KLOE(08/10) data samples already noted.
In this case, one gets $F_\omg^{ISR}= (1.608\pm 0.010)~10^{-6}$, exhibiting a large 
distortion towards the BaBar lineshape, despite the weighting.

\item
Combining all $e^+e^-\to \pi^+\pi^-$ spectra~: Taking the weights into account
-- which is more favorable -- one gets $\chi^2_{Global}/N_{\pi^+\pi^-}=1.35$
and a fit probability$^{\ref{prob_qual}}$ of 1.3\%. Once
again, the lineshape of the fit function is highly influenced by the BaBar sample
in the $\rho-\omg$ interference region as shown by $F_\omg^{Global}=(1.582 \pm 0.089)~10^{-6}$.

\item
Combining the NSK and KLOE10 data~: In this case, there is no weight and
one gets $\chi^2_{NSK}/N_{\pi^+\pi^-}= 131.37/127$ and 
$\chi^2_{KLOE10}/N_{\pi^+\pi^-}= 72.90/75$ close to the 
``standalone''$^{\ref{standalone}}$ fit results
reported in Subsection \ref{standaloneFits} above, and thus, 
$\chi^2_{KLOE10+NSK}/N_{\pi^+\pi^-}= 1.01$, and the remarkable 
global fit probability of 96.9 \%, showing that 
the NSK and KLOE10 data samples are quite consistent with each 
other and with the rest of the BHLS physics as well. 
Figures \ref{Fig:andreas_centr} and \ref{Fig:andreas_lowhigh} display
the fit residuals of this common fit; in both Figures, the leftmost panels show
the usual residual distributions of the NSK and KLOE10 samples ({\it i.e.}
the  differences between each measurement and the corresponding
fitting function value), while the rightmost panels display the 
same information  corrected for the bin--to--bin 
correlated uncertainties (see the Appendix). 

In both the low and
medium energy regions, the data points look reasonably well distributed 
on both sides of the fitting function (the zero axis)  and, also, the 
corrected residuals look closer  to zero than the usual ones.
If the effect of residual corrections (of pure graphical concern) looks marginal
at low to medium energies, the rightmost panel of 
Figure \ref{Fig:andreas_lowhigh} clearly indicates the more
appropriate character of the corrected residuals to translate
the fit quality.

\begin{table}[ph]
\centering
\begin{tabular}{|| c  || c  | c | c | c|| c||}
\hline
\hline
\hhhc
\hhhv  Data Sample   &\multicolumn{4}{|c|}{  
 Local Probabilities (\%) ~(Regions in GeV)} & Overall\\
\hhhd NSK+KLOE10 & 
\hhhv $0.30\div 0.55$  
   &  \hhhv  $0.55\div 0.75$  &  
   \hhhv $0.75\div 0.85$  &  \hhhv $0.85\div 1.00$ &Probability \\
\hline
\hline
  \hhhv CMD--2 & $20.11$ &$83.24$  
  & $17.99$   &  $17.94$ & 20.55\\
\hline
  \hhhv SND & $91.35$ &$58.03$  
  & $86.36$   &  $34.36$ & 92.94\\
\hline
   \hhhv KLOE10 & $99.98$ &$1.69$  
  & $30.95$   &  $91.11$ &52.91\\
 \hline
\hline
\hhhv Combined& $98.69$ & $15.83$ 
&$37.03$ & $48.18$& 57.40\\
\hline
\hline
\end{tabular}
\caption {
\label{Table:T_Andreas} Probabilities associated with the
estimated $\chi^2/N_p$ contributions
of the $\pi^+ \pi^-$ data samples from CMD--2, SND and KLOE10 
for  the various energy regions and -- last column -- for each  $\pi^+ \pi^-$ data set. 
Last line is referring to what
has been  named NSK+KLOE10 in the text and yields
a global fit probability is 96.9\%;  see the last
paragraph in Subsection \ref{CombFits}) for more details.}
\end{table}

As the global fit including simultaneously the CMD--2 
\cite{CMD2-1995corr,CMD2-1998-1,CMD2-1998-2}, SND \cite{SND-1998}
and KLOE10 \cite{KLOE10}  $\pi^+ \pi^-$ data sets play a crucial
role in the present study, it looks worth to give
more information on their fit quality beyond the global properties
just emphasized.  This is the purpose of Table \ref{Table:T_Andreas}.
We choose here to present the results in terms of probabilities
of the $\chi^2_{\pi^+ \pi^- }$  contributions
to the total $\chi^2$ associated with the corresponding number of data points
\footnote{This does not correspond to numbers of degrees of freedom, impossible
to define in this circumstance.}
$N_{\pi^+ \pi^- }$. The last data column thus gives this probability
for the data sample(s) provided by the three experiments. From this
exercise one can conclude that the three data samples, each as a whole, behave 
normally with a remarquable fit quality of the SND data sample 
($\chi^2/N_{points} \simeq 0.7$), while KLOE10 ($\chi^2/N_{points} \simeq 1.0$)
and the (collection) of CMD--2 (($\chi^2/N_{points} \simeq 1.1$) data samples  
are quite reasonable. 

Even if more approximate, one can perform the
same exercise for various energy sub-regions\footnote{In this case, the notion of
partial $\chi^2$ should be handled with some care as the number of data points is 
small in several bins and the inter--region correlations are neglected.}. 
This is displayed in the
first four data columns of Table \ref{Table:T_Andreas}. The probabilities
of the various SND data subsamples ($N_{\pi^+ \pi^- }=45)$
look reasonably flat, pointing to a 
very good account by the model function
and one yields   $\chi^2/N_{points}$'s ranging between 0.5 and 1.1. The probabilities
of the CMD2 collection of data subsamples ($N_{\pi^+ \pi^- }=82$)
are less favorable (the $\chi^2/N_{points}$'s
range between 0.7 and 1.3) but remain quite acceptable. The results for KLOE10
($N_{\pi^+ \pi^- }=75$)
is more appealing as the probabilities are good for three regions (corresponding
to $\chi^2/N_{points}$ values of resp. 0.3, 1.2 and 0.5), while the fourth region
is much worse (1.7\% probability corresponding to a $\chi^2/N_{points}$ of
1.7). This could have been already inferred by inspecting the residuals
of the fit in "isolation" shown\footnote{See especially the 
$0.675\div 0.725$ GeV energy region.} in Figure (\ref{Fig:kloe10_res}). Nevertheless,
Figure (\ref{Fig:nsk_kloe10_abc})) shows a reasonable account of the fit pion form
factor even in this region. 

The last line in Table \ref{Table:T_Andreas}, which corresponds to
the combination NSK+KLOE10 exhibits good probabilities up to the maximum 
energy and, in this case, the number of data points per energy sub--region
(resp. 42, 53, 71, 36) is  large enough to limit the amplitude of possible
fluctuations in the $\pi^+\pi^-$ data samples.

The global fit combining  the NSK and KLOE10 data also yields
$$F_\omg^{NSK+KLOE10}= (1.166\pm 0.036)~10^{-6}~,$$ which
should supersede the present world average value \cite{RPP2010} because of the full
consistency it exhibits with the largest set of data ever fitted simultaneously
and with a splendid probability.

\end{itemize}

\subsubsection{Fits of the $e^+e^-\to \pi^+\pi^-$ Spectra : Concluding Remarks}
\indent \indent The results reported just above have allowed us to show
that the NSK data are
 in fair agreement with the physics represented by the
$\pi^0\gamma$, $\eta\gamma$, $\pi^+\pi^-\pi^0$, 
$K^+K^-$, $K^0 \overline{K}^0$ annihilation channels,  the dipion spectra
collected in the decay of the $\tau$ lepton and some more
decay information listed in Subsection \ref{issue_dat}. This is not
a really new result as this conclusion was already reached in our
previous analysis \cite{ExtMod3}.
The single difference with \cite{ExtMod3} is the new way to include the $\phi  \to
\pi^+\pi^- $ information (see Subsection \ref{Vdecays}), possible with the BaBar data.

The new information is that the KLOE10  data sample behaves likewise
and, more importantly, that the NSK and KLOE10 data sets are consistent
with each other as well as with the rest of the physics considered within the
BHLS model and the global fit context.

We have performed global fits which have shown that the data from  
KLOE08   and BaBar   have some difficulty to
accommodate the global fit context. 
For what concerns KLOE08, the problem appears to be related with
(underestimated?) systematic errors or, possibly, with their correlations. 
In the case of BaBar data, the problem looks more serious,
as it deals with the form factor lineshape itself
 in the $\rho-\omg$ interference region; this issue
 manifests itself in
the value for  $F_\omg$, much larger than expected.

For this purpose, it is interesting to see the behavior of the
most relevant fits of the pion form factor
in the $\rho-\omg$ resonance region. These are displayed
in Figure \ref{Fig:rho_omg}.  
 The best fit obtained   using only the NSK data  is shown in the top left
panel and is clearly quite satisfactory. The top right panel exhibits the
case when the NSK data are complemented with the KLOE10 sample; this fit is also
quite successful. 

Bottom left panel in Figure \ref{Fig:rho_omg} shows the behavior of 
the best fit function when only considering the ISR data 
(BaBar, KLOE08, KLOE10) and bottom right panel when taking into account
all existing scan and ISR data. Both are clearly less satisfactory
reflecting mostly the tension between BaBar and the KLOE data sets. 

As already stated, the largest favored  configuration is to include
simultaneously the NSK and KLOE10 samples within the  fit procedure.
Fig. \ref{Fig:nsk_kloe10_abc} shows the fractional deviations from
the fitting functions for the $e^+e^-$ and $\tau$ data derived simultaneously
from the global fit. The top panel is clearly consistent with a flat distribution.
The bottom panel shows that CLEO and Belle distributions are flat, but also
that ALEPH data may show a slight $s$--dependence starting around 
0.8 GeV (quite similar to the top panel in Fig. 55 from \cite{BaBar2}
or to Fig.12 in \cite{Belle}). This is a 
pure consequence of having introduced  the  KLOE10 data 
sample\footnote{Compare bottom panels in the present Fig. \ref{Fig:nsk_kloe10_abc}
and in Fig. 10 in \cite{ExtMod3}}. 

One may thus consider that the statistically favored  configurations
(the so--called NSK and NSK+KLOE10 global fit configurations) 
exhibit flat residual distributions simultaneously  in $e^+e^-$ and $\tau$ 
$\pi \pi$ channels.

The question is now whether the
KLOE08 and BaBar data samples can nevertheless help to improve some physics
information of important concern as the muon $g-2$. This will be discussed
in the forthcoming Section. Anyway, the KLOE10 data sample 
already allows to confirm
the results already derived using the NSK data 
and, even, helps in getting  improved results.
\subsection{Physics Information Derived From Global Fits}
\label{physicsInfo}
\subsubsection{The $\phi$ Region in the $e^+e^-\to \pi^+\pi^-$ Spectrum}
\label{PhiRegion}
\indent \indent Up to now, the $\phi$ pieces of information used
in our fits/predictions are the RPP value for the product
$F_\phi= {\rm Br}(\phi \to e^+ e^-){\rm Br}(\phi \to \pi^+ \pi^-)
=(2.2\pm 0.4)~10^{-8}$ and the "Orsay" phase for the $\phi\to \pi^+ \pi^-$ amplitude
provided by SND \cite{SNDPhi}, namely $-34.0\pm 4.0$ degrees. 
In order to avoid over interpreting the SND phase as an  Orsay phase ({\it i.e.}
identified with the phase of the product $F_{\omega \phi}^e(s)~ g_{\phi \pi \pi}(s)$
in Eq. (\ref{Eq14}) at $s=m_\phi^2$), we have found worth revisiting this assumption. 
Using the NSK and KLOE10 data sets as reference $\pi^+ \pi^-$ data sample,
we have performed several  global fits within the BHLS framework. 

For definiteness, and for an easy comparison, the first line in Table \ref{Table:T2} 
displays the values extracted from  \cite{RPP2010,SNDPhi} and used in  all 
fits referred to up to now.
The second line in this Table displays the corresponding values reconstructed 
using the final parameter values from the appropriate global fit. The goodness
of fit  (94.5\% probability) is  reflected by the $\chi^2$ values shown in the last two
data columns. The reconstructed value for $F_\phi$ is in very good accord
with the (input) RPP value, while the phase is shifted toward slightly
more negative values than the SND datum. 

The fit results summarized in the third line are obtained by withdrawing the SND phase
\cite{SNDPhi} from the minimization procedure. 
The fit quality is almost unchanged (95.3\% probability) 
and the expected $F_\phi$ is still well reproduced; on the other hand,
 the $\phi$ phase moves by about $3 \sigma$ but remains significantly negative.

In both cases, the pion form factor lineshape exhibits the behavior
shown in Figure (\ref{Fig:modphi}b), e.g. a tiny peak at the $\phi$ mass,
resembling what is shown in Figure 4 of \cite{SNDPhi}.

The next step has been to remove the reference values for $F_\phi$ and
for the $\phi$ phase and supplement the NSK and KLOE10 data with the
$\phi$ region spectrum from BaBar data ($\sqrt{s} \in [1.00,1.05]$ GeV). 
The important results are shown in the fourth line of Table \ref{Table:T2}.
The fit probability remains good (91.3\%), but the $\chi^2$ of the NSK+KLOE10
data is degraded by $\simeq 10$ units, which may indicate some tension
between these data sets and the coherent background beneath the $\phi$ peak. 
The fit value for $F_\phi$ increases
by 50\% but remains perfectly consistent with expectations \cite{RPP2010}. 
In contrast, the Orsay phase 
changes dramatically (from $-[30 \div 50]^\circ \to 150^\circ$). The lineshape
exhibited by the BaBar pion form factor is almost identical to those shown in
Figure (\ref{Fig:modphi}c).

\begin{table}[ph]
\hspace{-1.5cm}
\begin{tabular}{|| c  || c  | c | c  | c ||}
\hline
\hline
\hhhc
\hhhd $\phi$ Information Used & \hhhv $F_\phi\,10^{8}$ 
   &  \hhhv Orsay Phase &  \hhhv $[\chi^2_{\pi^+ \pi^-}/N_{points}]_{< 1.0~GeV}$ &  \hhhv $[\chi^2_{\pi^+ \pi^-}/N_{points}]_\phi$\\
\hline
\hline
 \hhhv Input \cite{RPP2010,SNDPhi}  & $2.2\pm 0.4$ & -$34.0\pm 4.0$ & -& - \\
\hline
 \hhhv Imposing Orsay phase \cite{SNDPhi} & $2.34\pm 0.42$ & -$41.45\pm 1.89$ & $204.3/202$ & $2.25/2$ \\
\hline
 \hhhv Releasing Orsay phase  & $2.23\pm 0.41$ & -$48.23\pm 1.88$ & $202.9/202$ & $0.04/1$ \\
\hline
\hline
 \hhhv BaBar $\phi$ Region Spectrum \cite{BaBar}  &  $3.31\pm 0.99 $ &$156.92\pm 2.76$   &$215.0/202$& $5.2/5$\\
\hline
\hline
 \hhhv   BaBar data \cite{BaBar} fit   &  $3.04\pm 0.60 $ &$153.41\pm 0.98$ & $(338.1/270)$ & $(2.69/5)$ \\
 \hhhw  Imposing BaBar $\phi$ parameters  &  $3.05\pm 0.50 $ &$153.37\pm 0.85$   &$208.0/202$ &$0.05/2$ \\
\hline
\end{tabular}
\caption {
\label{Table:T2} Fit results for
$F_\phi= {\rm Br}(\phi \to e^+ e^-){\rm Br}(\phi \to \pi^+ \pi^-) $ 
and for the "Orsay" phase (in degrees) for the $\phi$ amplitude.
See Subsection \ref{PhiRegion} for comments.} 
\end{table}

As stated above$^{\ref{bbr_phi}}$, the $[1.00,1.05]$ GeV region
of the BaBar spectrum carries information on the $\phi$ signal but
also on the underlying ($\rho$ coherent) background. In order to substantiate 
its effect, we have finally performed a fit using as input the $F_\phi$ and
$\phi$ phase values extracted from fitting the BaBar data up to 1.05 GeV 
(see Subsection \ref{standaloneFits}); the corresponding values are given
in the fifth data line of Table \ref{Table:T2} and, when imposed
to the fit procedure, these numbers lead to the results
shown in the last line. The fit probability improves to 94\%. 

As a summary, Table \ref{Table:T2} shows that some ambiguity
occurs about the pion form factor behavior in the $\phi$ region.
It  becomes interesting to have new data covering this region,
in order to decide which among Figures  (\ref{Fig:modphi}b) and
(\ref{Fig:modphi}c) reflects the right behavior. It is also interesting
to examine the consequences of this ambiguity on $g-2$ estimates.

Without going into details (addressed in the next Section),
we can give the values for $\Delta a_\mu= a_\mu^{exp}-a_\mu^{th}$ 
corresponding to the various cases shown in Table \ref{Table:T2}. 
In units of $10^{-10}$, the values for  $\Delta a_\mu$ 
are respectively $39.91\pm5.21$ (line \# 2), $39.35\pm5.23$ (line \# 3),
$37.18\pm5.16$ (line \# 4) and $39.32\pm5.20$ (line \# 6). The most likely effect this
implies on  $\Delta a_\mu$ is a possible $shift$ by $0.59 \times 10^{-10}$ (lines \# 2, 3, 6). 
A maximum possible shift is  $2.73 \times 10^{-10}$ (lines \# 2 and 4). Instead, the uncertainty
is practically unchanged. 
Taking into account the uncertainty on $a_\mu^{exp}$ \cite{BNL,BNL2},
the effect of this shift on the significance of  $\Delta a_\mu$
is quite marginal ($4.88\sigma \rightarrow 4.57\sigma$).
Nevertheless, this shift plays as a systematic effect to be accounted 
for in the final value which will be proposed for $a_\mu^{th}$.

\subsubsection{The HLS Favored Solution~: Fit Parameters and  Physics Quantities}
 \label{FitNumRes}                                                                                        
 \indent \indent 
 In our previous analysis  \cite{ExtMod3}, we encountered some tension in
the $e^+e^-$ annihilation data in the $\phi$ region  between
the $e^+e^- \to \pi^+\pi^-\pi^0$ channel and the $e^+e^- \to K^+K^-,
K^0 \overline{K^0}$ channels. Including all data in the global
fit defined what we called "configuration A". Alternatively,
we excluded the 3-pion data in the vicinity of the $\phi$ mass
from the fit. In this case, the corresponding collection of data included in
the fit  was called "configuration B", which
yields a better fit probability. As  \cite{ExtMod3}, the present analysis 
privileges configuration B as a basis. However, as the "observed" tension could 
not be abnormal, we also consider fits based on configuration A
for completeness, essentially in relation with the muon anomalous moment.

Tables \ref{Table:T3} and \ref{Table:T4} display the global
fit parameter values within configuration B, using the $e^+e^-$ data from  
only CMD--2, SND and KLOE10. This corresponds to the so--called NSK+KLOE10 case
which is favored by the BHLS model and breaking scheme.

The  top part of Table \ref{Table:T3} displays
information directly related with the standard parameters of the original HLS model.
The value for $a$ is found, as usual \cite{taupaper} significantly
larger than the standard VMD expectation ($a=2$) and the value for the universal 
vector coupling $g$ stays at accepted values \cite{Fred11}. The 
HLS parameters\footnote{We follow the conclusion
reached in \cite{ExtMod3} and assume $c_3-c_4=0$.} giving
the weight of the anomalous terms \cite{FKTUY,HLSRef} in the global HLS model
 $(c_3+c_4)/2$, $c_1-c_2$, get values close to our previous results \cite{ExtMod3}.
 The parameter $v$, related with the 't~Hooft determinant terms  \cite{tHooft}, 
 will be discussed below.
 
 The middle part of Table \ref{Table:T3} essentially displays
  the SU(3) breaking parameters  generated by the original BKY mechanism \cite{BKY};
 their role is crucial in the $ e^+e^- \to K \overline{K}$ cross sections
 and in the $VP \gamma$ couplings (see Appendix E in \cite{ExtMod3}).
The values for $z_A$, $z_V$ and $z_T$ show that the breaking of
the SU(3) symmetry generates departures from 1 (unbroken case) by 30\% to 60\%.

The bottom part of Table  \ref{Table:T3} shows specifically
the values for the BKY parameters generating the (direct) isospin breaking
mechanism  \cite{Hashimoto,ExtMod3} defined in Subsection \ref{direct_brk}~:
 $\Delta_V$, $\Delta_A$ and $\Sigma_V$ which vanish when Isospin  
 Symmetry is not broken. Numerically,
 they are at the level of a few percents, {\it  i.e.} much
 smaller than those corresponding to the breaking of the SU(3) symmetry, 
 as expected. The value for $g_{\omg \to \pi \pi}^{direct}$ which only depends 
 only on the direct IB breaking mechanism  is discussed just below.

 The fit results shown in Table \ref{Table:T4} deal with the subtraction
 polynomials coming with the various loops (see Section 6 in  \cite{ExtMod3}
 or, more deeply, Section 12 in \cite{taupaper} and especially Table 3 therein). 
 The bottom part of Table \ref{Table:T4} gives the subtraction polynomials
 of the functions $\varepsilon_1(s)$ and  $\varepsilon_2(s)$ which define the vector 
 meson mixing as reminded in  Subsection \ref{mix1}.
  They  are given for sake of completeness  
 as, generally,  they do  not carry an obvious intuitive meaning; they mostly play 
  a role in the definition of the vector meson  mixing \cite{ExtMod3}. Nevertheless,
  as clear from Eqs. \ref{Eq9}, via the mixing angles $\alpha(s)$   and
  $\beta(s)$,  the functions $\varepsilon_1(s)$ and  $\varepsilon_2(s)$ play
  a fundamental role in reconstructing  $F_\omg$ and $F_\phi$
  discussed  at several places above and they have been shown to produce
  the expected effect either isolatedly ($F_\phi$ only depends on vector
  meson mixing) or combined with the direct IB  \cite{Hashimoto,ExtMod3}
  discussed above ($F_\omg$ depends on both).

 \hspace{0.5cm}
 
 Table \ref{Table:T5} display physics parameters which can be compared with
 corresponding information derived by others~:   different theoretical
 frameworks or measured quantities not included in our fit procedure; 
 this may give confidence in the validity of the BHLS model and  its structure.
 
 If the universal coupling $g$ is indeed returned by fits at usual values,
 the reconstructed value for  $g_{\rho\pi\pi}$ compares reasonably well
 with its value derived from fits using the Gounaris--Sakurai 
 parametrization \cite{Fred11} $g_{\rho\pi\pi} =6.1559$. 
 
  In their mini-review  \cite{RPP2010},
 Rosner and Stone conclude that the most likely estimate for $f_K/f_\pi$ 
 can be summarized by
   $f_K/f_\pi=1.197 \pm 0.002_{exp} \pm 0.002_{CKM} \pm 0.001_{rad.~cor.}$ which compares
   reasonably well with our own estimate reported in Table \ref{Table:T5}.  This also
   indicates that this proposed value could be included  in our fit procedure
   as an additional  constraint.
  
  The following datum is our estimate for $g_{\omg \to \pi \pi}^{direct}$. 
  A piece
  of information recently studied in \cite{Maltman2009} is related with
  the ratio $G=g_{\omg \to \pi \pi}^{direct}/g_{\rho\pi\pi}$. Our previous 
  estimate \cite{ExtMod3},
  using only the CMD--2 and SND data within our global framework, was 
  $(3.47 \pm 0.64)\%$ smaller than the findings
  of \cite{Maltman2009}; from Table  \ref{Table:T5}, this ratio is
  now estimated at $(6.27 \pm 0.94)\%$, just 
  in between the two estimates proposed by Maltman and Wolfe\footnote{
  $(7.3 \pm 3.2)\%$  and $(4.4 \pm 0.4)\%$, depending on the
  data sets used in their derivation.} 
  \cite{Maltman2009}, showing the influence of the KLOE10 data.

 Our estimate for the 't~Hooft parameter $\lambda$ \cite{tHooft}, which governs
the magnitude of nonet symmetry breaking in the pseudoscalar sector
(cf. Eq. (7) in \cite{ExtMod3}), can be derived from the values for $v$
and $z_A$ (cf. Eq. (22) in \cite{ExtMod3}). We confirm that 
$\lambda$ is in the range of $ 10 \%$ with a substantial
uncertainty (30\%); we are not aware of published external 
estimate of this quantity.

\hspace{0.5cm}
 
The following part of Table \ref{Table:T5} deals with information concerning mixing
properties in the pseudoscalar sector. We have adopted 
as parametrization of the $\pi^0-\eta-\eta^\prime$ mixing
those proposed in \cite{leutw96}. Within this parametrization,
the  $\pi^0-\eta$ mixing is governed by $\epsilon$,
the  $\pi^0-\eta^\prime$ mixing is governed by $\epsilon^\prime$.
Both parameters are expected to carry values at the few percent level,
and one may expect $\epsilon^\prime<\epsilon$. These expectations
are satisfied  by our estimates.
Moreover,  $\epsilon$ and $\epsilon^\prime$ can be expressed \cite{ExtMod3} in terms of 
the pseudoscalar singlet--octet mixing angle $\theta_{PS}$ and 
of some $\epsilon_0$, function of the quark masses \cite{leutw96}.
Recent estimates yield
$\epsilon_0=(1.31\pm0.24)\%$  \cite{Dominguez}
and
$\epsilon_0=(1.16\pm0.13)\%$  \cite{Colangelo},
significantly smaller than our former result ($\epsilon_0=(3.16\pm0.23)\%$).
Our present evaluation  (see Table \ref{Table:T3}), is much closer
to these expectations.

\hspace{0.5cm}
 
The pseudoscalar mixing angles are not basic parameters in our
fit procedure but depend on $\lambda$ and $z_A$ already discussed
(cf. Eqs. (26) in \cite{ExtMod3}). An early study \cite{WZWChPT}
concluded that the mixing angle $\theta_0$ defined in the mixing
scheme developed in \cite{leutw,leutwb} was consistent with zero.
This has been confirmed in \cite{ExtMod3} 
($\theta_0 = -1.11^\circ\pm 0.39^\circ$);  quite recently,   
 Chen {\it et al.} \cite{mixingNew} yield
$\theta_{0}=(-2.5\pm 8.2)^\circ$ from a study of radiative decays.
This motivates imposing the condition $\theta_0 \equiv 0$ to our
fit procedure. From the expressions for $\theta_8$ and
$\theta_{PS}$  (in terms of $z_A$ and $\lambda$, or, equivalently, $v$),
we derive 
$\theta_8 = -24.61^\circ \pm 0.21^\circ$, quite
consistent with the recent $\theta_{8}=-21.1^\circ\pm 6.0^\circ$
\cite{mixingNew} and much more precise.

Our  evaluations of $\theta_{PS}=-13.54^\circ\pm 0.15^\circ $ 
 compares favorably  with its KLOE 
measurement\footnote{The value published is actually
expressed in the flavor basis; for an easy comparison,
we convert to the singlet--octet mixing angle using
 $\theta_{PS}=\phi_P-\arctan{\sqrt{2}}$. }
\cite{KLOEetap} ($-13.3^\circ\pm 0.3_{stat}^\circ\pm 0.7_{syst}^\circ\pm 0.6_{th}^\circ$),
recently confirmed by GAMS ($-13.4^\circ \pm 1.8^\circ$).

To end up with this topic, we should mention that
the fit values for the partial widths
$\pi^0 \gamma \gamma$, $\eta \gamma \gamma$ and
$\eta^\prime \gamma \gamma$ are at resp. $0.3 \sigma$,
$0.5 \sigma$ and $0.2 \sigma$ from RPP expectations \cite{RPP2010}.

\hspace{0.5cm}
 
Our evaluation for the $\omg$
and $\phi$ mass and width are also shown in Table  \ref{Table:T5}.
These correspond to the usual definition of the inverse propagator
with fixed width~:
\be
D_V^{-1}(s)=s-m_V^2 + i m_V \Gamma_V\;.
\label{Eq18qq} 
\ee

They are found in fair consistency with RPP expectations and 
our global fit estimates compare favorably
with the corresponding data extracted from fits to the individual measured
annihilation spectra reported in \cite{RPP2010}.

The last line in Table  \ref{Table:T5} displays information on the
$\rho$ meson mass and width. Concerning objects as broad as the $\rho$ 
meson, mass and width values are highly definition--dependent 
\cite{taupaper} and, depending upon the choice, their spread can easily amount
to about 20\%. In Table  \ref{Table:T5}, we give  the location $s_\rho$ of the 
$\rho^0$ propagator pole located in the unphysical sheet; it has been derived
using the parameter values and their error covariance matrix within
a Mathematica code. Defining $s_\rho=M_\rho^2 -iM_\rho \Gamma_\rho$,
one can derive  mass and width values, also given in  Table  \ref{Table:T5}.
As far as we know, one can only compare with \cite{Bernicha1994} (see
also \cite{Bernicha1995} for a hint on the parametrization dependence
of the pole location), who performed fits by approximating the $\rho$
propagator by a Laurent series; this Reference reports $M_\rho=757.5\pm 1.5$ MeV
and $\Gamma=142.5 \pm 3.5$ MeV relying on fitting only the data reported in
\cite{Barkov}, close to our results. The previous estimate of these within
the HLS model \cite{taupaper} where larger (resp. 760.4 and 144.6)
with comparable uncertainties; however, the data from Belle and KLOE were not
available at that time. Because of our approximations (see Subsection 
\ref{IB_effects}), the present study is not sensitive to differences between 
the charged and the neutral $\rho$ mesons.

\begin{table}[!htb]
\begin{center}
\begin{tabular}{|| c  | c  | c | c  | c ||}
\hline
\hline
\hhhu $g$ (HLS)  &  $a$ (HLS) &  $(c_3+c_4)/2$   &  $c_1-c_2$   & $v$  \\ 
\hline
\hhhb   $5.578 \pm 0.001$ & $2.398 \pm 0.001$ &  $0.920 \pm 0.004$ & $1.226\pm 0.026$& $0.030 \pm 0.012$   \\
\hline
\hline 
\hhhv $z_A$  &$z_V$  &$z_T$ &$\epsilon_0$ &--\\
\hline
\hhhb $1.608 \pm 0.006$ & $1.319 \pm 0.001$&$1.409 \pm 0.062$ &$0.026 \pm 0.003$ &--\\
\hline
\hline
$\Delta_A$ \hhhv & $\Sigma_A$ & $\Delta_V$ &$\Sigma_V$&$h_V$\\
\hline
\hhhb $0.048 \pm 0.007$ & {\bf 0}  & $-0.028 \pm 0.003$ & 
$-0.034 \pm 0.001$&  $2.853 \pm 0.291$\\
\hline
\hline
\end{tabular}
 \end{center}
\caption{
\label{Table:T3}
Parameter values from the global fit  using the CMD--2, SND and KLOE10 data. The piece of
information written boldface was not allowed to vary within the fit procedure.}
 
\end{table}

\begin{table}[!htb]
\begin{center}
\begin{tabular}{|| c  || c || c ||}
\hline
\hline
\hhhu Parameter  & Sub. Pol. $\Pi_{\pi\pi}^\rho(s)$ \hhhb 
& Sub. Pol.  $\Pi_{\pi\pi}^{W/\gamma}(s) $  \\
\hline
$C_1$ (GeV$^{-2}$) \hhhb &\bf{0}& $0.671 \pm 0.041$ \\
\hline
$C_2 $\hhhb & $-0.473 \pm 0.001$   & $0.730 \pm 0.063$  \\
\hline
\hline
\hhhv   & Sub. Pol.  $\varepsilon_2(s)$ &
Sub. Pol.  $\varepsilon_1(s)$   \\
\hline
\hline
$C_1$  (GeV$^{-2}$)\hhhb & $-0.075 \pm 0.006$ & $-0.015 \pm 0.002$ \\
\hline
$C_2$\hhhb & ~~~$0.034 \pm 0.005 $ &  ~~~$0.015\pm 0.002$\\
\hline
\hline
\end{tabular}
\end{center}
 
\caption{
\label{Table:T4}
Parameter values from the global fit  using the CMD--2, SND and KLOE10 data (cont'd).
The boldface parameter is not allowed to vary.
Each subtraction polynomial is supposed to be written in the form $C_1
s+ C_2 s^2$. The functions $\varepsilon_1(s)$ and $\varepsilon_2(s)$
are combinations of the kaon loops which govern the neutral vector
meson mixing (see Subsections \ref{mix1} and \ref{FitNumRes}). 
}
\end{table}

 \begin{table}[!htb]
\begin{center}
\begin{tabular}{|| c | c| c | c ||}
\hline
\hline
\hhhv $g_{\rho \pi \pi}$ \hhhb & $f_K/f_\pi$   & $g_{\omg \to \pi \pi}^{direct}$ & $\lambda$   \\ 
\hline
\hhhb  $ 6.505\pm 0.003$ &$ 1.279\pm 0.010$  & $ 0.408\pm 0.061$  & $(7.64\pm 3.19)~10^{-2}$\\
\hline
\hline
$\epsilon$  \hhhv & $\epsilon^\prime$ &  $\theta_{8}$ (deg) &$\theta_{PS}$  (deg)\\
\hline
\hhhb  $(4.027\pm 0.474)~10^{-2}$  &  $(0.975\pm 0.122)~10^{-2}$  & $(-24.61\pm 0.21)^\circ$  &$(-13.54\pm 0.15)^\circ$  \\
\hline
\hline
\hhhv $m_\omg$ (MeV)& $\Gamma_\omg$ (MeV)&$m_\phi$  (MeV)&$\Gamma_\phi$ (MeV)\\
\hline
\hhhb $ 782.52\pm 0.03$ &  $8.66 \pm 0.04$&  $ 1019.25 \pm 0.26$ &   $4.18\pm 0.02$  \\
\hline
\hline
\hhhv $m_\rho$ (MeV) & $\Gamma_\rho$ (MeV)& $ {\cal R}e (s_\rho)$ (GeV$^2$)& 
$ {\cal I}m (s_\rho)$ (GeV$^2$)  \\
\hline
\hhhb $ 753.8\pm 0.5$ &   $ 138.10\pm 0.5$ &  $ 0.5682\pm 0.0007$ &  $ 0.1041\pm 0.0004$   \\
\hline
\hline
\end{tabular}
\end{center}
 
\caption{
\label{Table:T5}
Physics parameters extracted from the BHLS favored fit.
}
\end{table}

\section{Hadronic Contribution to the Muon $g-2$ Estimates}
 \label{gM2_studies}                                                                                        
 \indent \indent In order to estimate the hadronic contribution to the muon 
 anomalous magnetic moment $a_\mu$, the method followed in the present work
 is identical to the one used in \cite{ExtMod3}. In this Section, we examine
 the global fit solutions provided by the data samples listed in Subsection \ref{issue_dat};
 the various cases correspond to varying the $e^+e^-\to \pi^+\pi^-$ data
 sample combination submitted to the global fit procedure. We mostly work under 
 the configuration B which has been reminded in the header of Subsection \ref{FitNumRes}
 just above.

 \subsection{$a_\mu(\pi\pi)$ Contribution from the Reference $m_{\pi \pi}$ Region}  
\label{amuRef}
  \indent \indent  
In order to emphasize what is going on, it is worth examining the
contribution to $a_\mu(\pi\pi)$ provided by the invariant mass interval
${\cal A} \equiv \left[ 0.630,~0.958 \right ]$ GeV. As for this reference region, 
all experimental groups have published the numerical integration leading to their
estimate of $a_\mu(\pi\pi,\cal{A})$ (using Eq. (\ref{Eq1}) with the experimental spectra), 
one can compare our fit outcome with these.
When combining data sets within the fit procedure, we compare to the usual 
weighted average of the $a_\mu(\pi\pi,\cal{A})$ values provided
by each experimental group.

The results for $a_\mu(\pi\pi,\cal{A})$ are collected in Fig. \ref{Fig:amu_ref} 
and will be discussed from top to bottom. The first point has no experimental
partner, as it comes out from the $\tau$+PDG fit described in Section \ref{tauPred}.
The dashed--dotted line is drawn through the central value of this prediction,
the dotted lines show the corresponding $\pm 1 \sigma$ band.

It should then be noted that most experimental estimates happen to be inside
the $1 \sigma$ band of the $\tau$+PDG prediction. KLOE10  is at the border,
while BaBar is rather larger by about $2 \sigma$. One should also note
that KLOE08 as well as the combination CMD2+SND+KLOE10 (our preferred combination
because of its global fit properties) provide central values which coincide almost exactly
with the $\tau$+PDG expectation; even if shifting the  $\tau$+PDG prediction by
$(1\div 2)~10^{-10}$  should be performed (see Subsection \ref{IBsystematics} below),
 the overall agreement 
remains noticeable. This looks to us an important property taking
into account the long--standing discussion about the $\tau - e^+e^-$ issue. Our 
results tend to prove that this issue vanishes once the breaking mechanism
is appropriate.

The next point shows the global fit estimate (square symbol) while 
using$^{\ref{standalone}}$ 
``solely'' the NSK data
(especially those from \cite{CMD2-1998-1,CMD2-1998-2,SND-1998}, but also those
from \cite{Barkov,CMD2-1995corr}).  The present fit value is found smaller
than those published in  \cite{ExtMod3}  by about $1 \times 10^{-10}$. This  shows that 
the effect of our updating the $\phi$ data decay information extends over the whole spectrum. 
The uncertainty is unchanged and
amounts to a 40\% gain compared with the experimental estimate.
This gain is a pure effect of the global fit procedure where the channels others
than $\pi^+ \pi^-$ allow to improve $also$ the $\pi^+ \pi^-$ contribution importantly
because of the underlying  VMD physics correlations.

The following point is derived using$^{\ref{standalone}}$ 
``solely'' the KLOE08 data sample. As while
fitting ``solely'' the NSK data, the fit outcome also differs by $1 \times 10^{-10}$ 
from the experimental central value.  The fit result associated with the KLOE10 data sample 
is also found in good agreement with the experimental expectation and one 
should note that both differ  by $1.5 \times 10^{-10}$ in opposite directions.
In contrast, the fit result for the mixed KLOE10+(weighted) KLOE08 data sample 
is found to agree very well with expectation (a $ 0.6 \times 10^{-10}$
difference only).

The data point associated with the CMD2+SND+KLOE10 combined data sample is found in
much better accord with the average of the experimental data; 
in this case, the error is also improved
by about a factor 2 compared with the experimental value, but,
interestingly, first, the central reconstructed value does not
exhibit any significant systematic shift ($ \simeq 0.1 \times 10^{-10}$) compared to the
experimental expectation, second, the central reconstructed value
is in perfect accord with the $\tau$+PDG prediction, as already noted.

The next pair of points shows the case for the BaBar data. 
While using the whole spectrum up to 1 GeV, the fitting function 
provides an estimate for $a_\mu(\pi\pi,\cal{A})$ in good agreement with
the expected experimentally integrated value.
 However, extending the fitting region up to 
1.05 GeV, increases the difference between the fit outcome and the
experimental expectation from $1.5 \times 10^{-10}$ to $4.4 \times 10^{-10}$,
which illustrates some issue in the $\omg$ region
when also fitting  the BaBar $\phi$ region.

The last two points (ISR  and scan + ISR data, resp.) correspond to fitting the whole spectra
up to  1 GeV while weighting the KLOE08 and BaBar contributions to the $\chi^2$. 
The results look in good correspondence with the weighted averages of the 
direct integration results, and one remains within the $1 \sigma$ band
of the $\tau$+PDG expectation.

\vspace{0.5cm}
\indent \indent The contribution to $a_\mu(\pi\pi)$ 
from the invariant mass interval
${\cal A} \equiv \left[ 0.630,~0.958 \right ]$ GeV and the global fit properties
(see Section \ref{globalfits}) allow to draw a few important conclusions~:
\begin{itemize}

\item{\bf (l)} There is no mismatch between the $\tau$+PDG expectations and
the estimates derived from the fitting functions  or those derived from the  numerical
integration of the measured $e^+e^- \to \pi^+\pi^-$ cross sections. To be more precise, when some
departure is observed, it is always closely associated with poor global fit qualities
of the corresponding data sample.

\item{\bf (ll)} One can take as reference $e^+e^- \to \pi^+\pi^-$ data  the NSK
(CMD-2 \& SND) $and$ KLOE10 data samples. Separately and together, they are found
in perfect accord with all annihilation data considered ($\pi^0\gamma$, $\eta\gamma$, $\pi^+\pi^-\pi^0$, 
$K^+K^-$, $K^0 \overline{K}^0$ final states) as well as with the published dipion spectra in the decay
of the $\tau$ lepton and the few additional decay information introduced within the global fit
BHLS framework. The uncertainty is improved by a factor of about 2.

\item{\bf (lll)} Comparing the central value for $a_\mu(\pi\pi,\cal{A})$ derived from the fitting
function -- using the  NSK and KLOE10 data samples within the global framework -- with 
its direct estimate indicates that the estimate derived from the fit
is almost unbiased ($\simeq 0.1 \times 10^{-10}$). In contrast,
the fitting functions derived from the fits to each of the NSK and KLOE10 data 
separately may exhibit a small bias (a shift of $\simeq  1\times 10^{-10}$).

\item{\bf (lv)} One may remark that the global fit using all existing $e^+e^- \to \pi^+\pi^-$ 
scan \& ISR data also provides reasonable values for $a_\mu(\pi\pi,\cal{A})$ without
a significant bias and with a much improved uncertainty. This reduced uncertainty
may look too optimistic, taking into account the global fits properties
of the corresponding sample combinations. Nevertheless, one may note
that the last two entries in Fig. \ref{Fig:amu_ref}, derived by weighting
the KLOE08 and BaBar samples with respect to all others, do not
depart from the general trend.

\end{itemize}

 \subsection{The Full Muon  Anomalous Magnetic Moment $a_\mu$}  
\label{gM2_all}
\indent \indent  Here we are interested in the theoretical prediction of the
muon anomalous magnetic moment and its comparison with the
experimental result \cite{BNL,BNL2}. Our BHLS based global fit results provides
the dominant contribution to the hadronic vacuum polarization contribution.

Several global fits have been performed using different groups of
 $e^+e^- \to \pi^+\pi^-$ data samples. The resulting fitting functions and the 
 parameter error covariance  matrices have allowed to derive the leading order
 hadronic contribution (LO--HVP) to $a_\mu$ up to 1.05 GeV in various
 cases. The full HVP is obtained
 by combining our present results with the other hadronic contributions
 listed in Table 9 of our previous work  \cite{ExtMod3}. The various evaluations
 of the muon $g-2$ are calculated by summing up  $a_\mu^{LO-HVP}$ with the
 Higher order HVP, the QED and electroweak (EW) contributions and the so--called 
 light--by--light (LBL) contribution; for this purpose, the values given in
 Table 10 of  \cite{ExtMod3} have been updated (see Table \ref{Table:T6} below).

 In Figure \ref{Fig:gM2_final},  we display
results for the deviation $\Delta a_\mu=a_\mu^{\rm
exp}-a_\mu^{\rm th}$ between experiment and theoretical
predictions in units of $10^{-10}$, for various fit results
which only differ by the indicated $e^+e^-\to \pi^+\pi^-$ data
sample considered within the fit procedure.

Specifically, the fit results included in Fig. \ref{Fig:gM2_final}  are
all based on the ``configuration B'' data set introduced in \cite{ExtMod3}.
Furthermore, always included in the
fits are all channels listed in Subsection \ref{issue_dat},
especially the $\tau$ data.

 At top of the Figure, we provide the $\tau$+PDG 
 entry ({\it i.e.}, without using any   $e^+e^- \to \pi^+\pi^-$ data set in the fit
 procedure). The twin points drawn correspond to $\Delta a_\mu = a_\mu^{exp}-a_\mu^{th}$
 in units of $10^{-10}$; the lower point is obtained using the spacelike data
 \cite{NA7,fermilab2}, the upper one by excluding them from the fit. The number following
 the point is the estimate for $\Delta a_\mu$ from fits excluding the spacelike data.
 The significance of this difference is written [$x/y ~\sigma$], $x\sigma$ being 
 the significance obtained when excluding
 the spacelike data, $y\sigma$ when they are included into the fit procedure.  The last column 
 in this Figure provides the $\chi^2/n_{points}$ for the $e^+e^- \to \pi^+\pi^-$
 data; this number is given only for (combination of) samples used without
 any weight.
 
 The first remark which can be drawn is that including the spacelike data does not
 change noticeably the value for $\Delta a_\mu$ nor its statistical 
 significance. The second important point is that all the features noted
 while discussing $a_\mu(\pi\pi,\left[ 0.630,~0.958 \right ] ~{\rm GeV})$ 
in the previous Subsection survive when dealing with the
whole provided spectra (see ({\bf l})--({\bf lv}) just above).

It was noted in \cite{ExtMod3} that the use of the ISR data  should not 
dramatically improve the uncertainty of the theoretical estimate~: going 
from scan data only (the NSK entry) to scan+ISR data (the "Global Fit" entry) improves
the uncertainty by only $0.3 \times 10^{-10}$; this is the major gain of working in a global
fit context \cite{ExtMod3}. In contrast, the central value 
exhibits slightly more sensitivity, as can be seen by comparing the various entries.

Anyway, the present analysis allows to conclude that the discrepancy
between the measurement for $a_\mu$ \cite{BNL,BNL2} and the Standard Model
prediction is certainly in the range  of $4.5 \sigma$ or slightly higher. Taking into account
the already reported fit qualities, the most favored value for $a_\mu^{LO-HVP}$ is~:
\be
a_\mu^{LO-HVP}= (681.23 \pm   4.51)~10^{-10}.
\label{Eqr}
\ee
 This value is derived by introducing 
the $e^+e^-$ NSK and KLOE10 data samples within our global fit procedure; systematics
on this number will be discussed with the final result (Subsection \ref{favoredEstimate}).

  Finally, we have redone several of our fits within the configuration A. These confirm
  the results derived within the configuration B which are the body of the present
  paper. One also confirms the 'tension' reported in \cite{ExtMod3} between
  the 3--pion spectra in the $\phi$ region and the data collected for both
  $e^+e^- \to K \overline{K}$ cross sections. Whether this is an experimental issue,
  or if this indicates that our breaking model should be extended\footnote{One
  might have to revisit the possibility that nonet symmetry should also
  be broken in the vector meson sector as it is already done in the pseudoscalar
  sector.}   is an open issue. Nevertheless, some physics results derived
  within the configuration A will be presented just below for completeness.
  
 \subsection{Comparison with Other Estimates of $a_\mu$}  
\label{comparison}
\indent \indent
Fig.  \ref{Fig:gM2_comp} reports on some recent estimates of the muon anomalous magnetic moment
$a_\mu$ together with the BNL average value \cite{BNL,BNL2}.  
In this Figure, our (favored) results -- derived using only the CMD--2, SND and KLOE10
data samples -- are given under the entry tags BHLS::A (for configuration A)
and BHLS::B (for configuration B). Some evaluations proposed by other authors are also 
shown; they are extracted from \cite{DavierHoecker3} 
(DHMZ10), \cite{Fred11} (JS11), \cite{DavierHoecker} (DHea09) and \cite{Teubner} (HLMNT11).

\begin{table}[ph]
\hspace{-2.cm}
\begin{tabular}{|| c || c | c  || c ||c ||}
\hline
\hline
\hhhv $10^{10} a_\mu$  &    \multicolumn{2}{|c|}{Values (incl. $\tau$)}  &  
 \multicolumn{2}{|c|}{Values (excl. $\tau$)}\\
\hhhv ~~~   &   scan only &   scan $\oplus$ ISR  &   scan  $\oplus$ ISR & scan only \\
\hline
\hhhv LO hadronic   &    $685.66 \pm 4.54$   &   $688.60 \pm 4.24$& $687.53 \pm 4.34$
& $684.25 \pm 5.15$   \\
\hline
\hhhv HO hadronic &   \multicolumn{4}{|c|}{ $-9.97  \pm 0.09 $} \\
\hline
\hhhv LBL &   \multicolumn{4}{|c|}{ $10.5 \pm 2.6$ }   \\
\hline
\hhhv QED &   \multicolumn{4}{|c|}{$11~658~471.8851 \pm 0.0036$}   \\
\hline 
\hhhv EW &   \multicolumn{4}{|c|}{$15.40\pm
0.10_{\rm had} \pm 0.03_{\rm Higgs,top,3-loop}$ }   \\
\hline  
\hline 
\hhhv Total Theor. & $11~659~173.48 \pm  5.23 $ &  $11~659~176.42 \pm  4.89 $
& $11~659~175.35 \pm 5.06 $& $11~659~171.89 \pm 5.77 $ \\
\hline 
\hline 
\hhhv Exper. Aver. &  \multicolumn{4}{|c|}{$11~659~208.9 \pm 6.3 $ } \\
\hline 
\hline 
\hhhv $\Delta a_\mu$ & $35.43\pm 8.14 $ & $32.48\pm 8.03 $ & $33.56 \pm 8.08 $
& $36.89 \pm 8.54$
 \\
\hline
\hhhv Significance ($n \sigma$) & $4.33 \sigma $ & $4.05 \sigma
$ & $4.16 \sigma $ & $4.32 \sigma $    \\
\hline
\hline
\end{tabular}
\caption{\label{Table:T6} 
\footnotesize The various contributions to $10^{10} a_\mu$. 
$\Delta a_\mu= a_\mu^{exp}-a_\mu^{th}$ is given in units of $10^{-10}$.
These results have been derived using $all$ existing annihilation data 
samples (configuration A).The naming "scan" and "ISR"  in the subtitles refers
only to the $\pi^+ \pi^-$ channel. For the measured value $a_\mu^{exp}$,
 we have adopted  the value reported in the RPP which uses the updated value for
$\lambda=\mu_\mu/\mu_p$ recommended by the CODATA group~\cite{CODATA2012}.
}
\end{table}

The analyses reported in \cite{DavierHoecker3},  \cite{Teubner} and \cite{Fred11}
provide $g-2$ evaluations based on $all$ existing data samples. In particular,
 all $\pi^+\pi^-$ data samples are used (ISR and scan) and all other existing 
 annihilation data samples, especially  the 3--pion data samples. These evaluations
 of $g-2$ may also include or not the $\tau$ data. Therefore, it is also instructive to give the
results derived from BHLS under the same conditions, {\it i.e.} working in the so--called
configuration A \cite{ExtMod3} and using $all$ available
ISR data samples beside the scan (NSK) data, whatever are the global fit probabilities.
 The corresponding results are reported at the "Full BHLS" entries in Fig.  \ref{Fig:gM2_comp} 
More precise information is also provided  in Table \ref{Table:T6}. 
However, one should  note that, while the fit probabilities are good when using only the
scan (NSK) data (90.6 \% or 72.7 \% respectively), they are below the $10^{-4}$ 
level$^{\ref{prob_qual}}$
when all the ISR data samples are included. This reflects the statistical inconsistency
between both KLOE data samples on the one hand and the BaBar data sample on the other hand.

We have explored  this issue in cases including the $\tau$ data samples. 
Without going into much details, it looks
worth to comment about fit qualities. The just reported probabilities bounded by  $10^{-4}$
are worrying, taking into account that the problematic data samples have been weighted.
For this purpose, we have first performed a fit discarding the BaBar data sample, to 
check the level of statistical inconsistency of NSK, KLOE10 and (weighted) KLOE08; the
obtained fit probability is 61.6 \% and the numerical results are almost identical
to those already derived using only NSK and KLOE10. In particular, the scan data
yield $\chi^2(NSK)/n=136/127=1.07$ and $F_\omg=(1.11 \pm 0.03)~10^{-6}$,
{\it i.e.} a limited distortion of the scan data description. 
The same exercise performed
using NSK and (weighted) BaBar yield 12.4\% probability but $\chi^2(NSK)/n=171/127=1.35$ and
$F_\omg=(1.52 \pm 0.04)~10^{-6}$; {\it i.e.}  a severely degraded
description of the scan data.

It should be noted that, when using all $\pi\pi$ data samples, the value 
returned by the BHLS model is close to JS11  \cite{Fred11}. An interesting point here
is that  \cite{Fred11} uses a mixing procedure ($\gamma-\rho-\omg$) parent to ours
($\rho-\omg -\phi$), while the IB effects are introduced differently in order to
get the DHMZ10 result \cite{DavierHoecker3}. Comparing the Full BHLS values derived
when using or not the $\tau$ spectra indicates that BHLS is much less sensitive
to the use of the $\tau$ spectra that the method from \cite{DavierHoecker3} (compare
the  DHMZ10 entries). Finally, comparing DHea09 \cite{DavierHoecker} and
BHLS when no ISR data are considered, one observes a good consistency; this seems to
indicate that the problem lays in the amount of isospin breaking compared
with $\tau$.  

Comparing now all the BHLS results (BHLS::A,BHLS::B and Full BHLS) indicates 
a remarkable stability for $\Delta a_\mu= a_\mu^{exp}-a_\mu^{th}$ under various 
fit conditions. One should also note that the $\tau$+PDG prediction 
is always in quite nice agreement
with using any kind of configuration for  $\pi\pi$ data 
samples (BHLS::A,BHLS::B and Full BHLS),
including those discarding the $\tau$ spectra and the ISR data (see
also Table \ref{Table:T6}).

Nevertheless, one should stress that, because of the poor fit quality
yielded when using the KLOE08 and BaBar  data samples beside the NSK and KLOE10 ones,
we do not consider the results shown in Table \ref{Table:T6} and in Fig.  \ref{Fig:gM2_comp}
under the entry tag "Full BHLS" as reliable. 
As a general statement, we prefer relying on estimates using only
the NSK and KLOE10 data samples for which the probabilities
are good ($\simeq 90\% \div 98\%$ for BHLS::B, 
$\simeq 56\% \div 73\%$ for BHLS::A). Numerically, the BHLS::B
and BHLS::A estimates happen to be always close to each other.
Nevertheless, we prefer favoring the BHLS::B result over BHLS::A
for reasons already explained.

The statistical significance for $\Delta a_\mu= a_\mu^{exp}-a_\mu^{th}$
is displayed on the right--hand  side of the Figure for each of the reported analyses.
The updating of the $\phi \to \pi \pi$ treatment within our minimization
code turns out to increase the significance for $\Delta a_\mu$ by about $0.5 \sigma$
without any significant change in the fit qualities. The present study, however, 
has shown that this update is highly favored by the $\phi$ information carried
by the BaBar spectrum.

 On the other hand, one should note  that using $all$
ISR data samples only leads to a $\simeq 4 \%$ improvement for the uncertainty
on $a_\mu^{th}$ compared with discarding all of them. This is actually, as reminded in
the Introduction, a specific consequence of the global fit procedure.

Another remark is worth being expressed~: Whatever is the configuration
(A or B), whatever is the set (or subset) of $\pi^+ \pi^-$ data samples
 considered, Table \ref{Table:T6} and  Fig.  \ref{Fig:gM2_comp} 
show that the role of the $\tau$ data
is not limited to returning improved uncertainties on $a_\mu^{th}$.
Indeed, one should note that the global fits using the $\tau$ data 
always returns $a_\mu^{th}$ from fits  larger than when 
discarding them; the difference amounts to $1 \div 2$ units.

Before closing this Section, it is worth remarking that the central values 
for $a_\mu^{LO-HVP}$ derived using scan data only or scan $\oplus$ ISR are
very close to each other. This is due to a numerical conspiracy of the BaBar and KLOE data 
samples~: The former data tend to increase  $a_\mu^{LO-HVP}$ and the latter to
decrease it, by a quite comparable amount. Taking into account the above reported
fit probabilities, the increase of $a_\mu^{LO-HVP}$ is not justified on statistical
grounds; in contrast the decrease of $a_\mu^{LO-HVP}$ produced by KLOE10 (and
also KLOE08) is supported by quite good probabilities of simultaneous fits
of scan and KLOE data.

 \subsection{The HLS Favored Solution~: Properties Of Its Estimate For $a_\mu$}  
\label{favoredEstimate}
\indent \indent It follows from the present study that a consistent data sample
can be defined which allows to yield a reliable estimation of  $a_\mu^{th}$.
This sample contains all the collected data samples covering the $\pi^0 \gamma$,
$\eta \gamma$, $K^+ K^-$ and $K_L K_S$ annihilation channels and the dipion $\tau$
spectra up to 1 GeV. It also contains the $\pi^0\pi^+ \pi^-$ annihilation channel.
For this last process, we prefer working within the so--called configuration
B by excluding the data sets collected around the $\phi$ mass. However, as clear
from  Figure  \ref{Fig:gM2_comp}, taking them into account (configuration A) does not
lead to substantial differences. Concerning the fundamental $\pi^+ \pi^-$ annihilation
channel, the statistical properties exhibited by the fits lead us to rely on using only
the scan (NSK) and the KLOE10 data samples; indeed, for this set of data samples 
one does not observe any mismatch between the claimed uncertainties and the 
statistical properties of global fits.

With this respect, three cases reported in Fig. \ref{Fig:gM2_comp}  deserve special
attention. Working in units of $10^{-10}$ and using obvious notations, 
the first one is the so--called ``$\tau$+PDG'' estimate, 
$\Delta a_\mu(\tau+{\rm PDG}) =38.10 \pm 6.80$, the second is the one obtained using
NSK, KLOE10 and the $\tau$ data, $\Delta a_\mu({\rm NSK+KLOE10}+\tau)=39.91\pm 5.21$ and
the last one where  the $\tau$ data are excluded, 
$\Delta a_\mu({\rm NSK+KLOE10}+\cancel{\tau})=41.91\pm 5.45$.

\subsubsection{Systematics Due To $\tau$ Data}
\label{IBsystematics}

\indent \indent
Here we come back to the issue of IB, discussed in 
Subsection \ref{IB_effects} before, and how the value for
$a_\mu^{th}$ can be affected by the approximations performed on the
$\tau$ spectrum versus $e^+e^-$, namely $\delta M_\rho$ and $\delta
\Gamma_\rho$~?

In the standard data-driven approach, while dealing $solely$ with their spectrum, 
the Belle \cite{Belle} Collaboration
has to sum up a list of IB corrections on the 
muon anomalous moment $a_\mu^\tau[\pi\pi]$ to derive the
physically relevant 
$a_\mu^{ee}[\pi\pi]=a_\mu^\tau[\pi\pi] + \delta a_\mu^{\rm had,LO}[\pi\pi,\tau]$.
 The contributions given in Table XII from \cite{Belle},
 relevant for our purpose
are $\delta a_\mu(\delta M_\rho)= (0 \pm 2)~10^{-10}$
and
$\delta a_\mu(\delta \Gamma_\rho)= (-1.4 \pm 1.4)~10^{-10}$ 
which sum up to $\delta a_\mu(tot)=(-1.4 \pm 2.44)~10^{-10}$.
In a slightly more recent study \cite{DavierHoecker}, it
is argued that the value should be slightly different~:
 $\delta a_\mu(tot)=(-0.62 \pm 1.11)~10^{-10}$. 

So, in total, if one assumes that the traditional evaluations
of the corrections to $\tau$ estimates apply without any thinking
to the BHLS estimate $a_\mu^{(\tau ),HLS}$, one may have to consider
a shift  $\delta a_\mu^{\rm had,LO}[\pi\pi,\tau] \equiv \delta a_\mu(tot) \simeq - 1.4 ~10^{-10}$, 
if one relies
on Belle \cite{Belle} estimates.

In the ``$\tau$+PDG'' fit configuration, there is no data on the 
$ e^+e^- \to \pi^+ \pi^-$ annihilation process. Therefore, the sensitivity
to the  $\rho^0$ meson lineshape is certainly quite marginal, while the
use of $\tau$ data introduces a strong influence of the $\rho^\pm$ meson.
 Hence, the value derived for  $a_\mu^{th}$ is likely to be corrected
 for the ($\delta M_\rho$, $\delta \Gamma_\rho$) effects discussed in 
 Subsection \ref{IB_effects} and just reminded.
 
In the QFT--driven BHLS parametrization, which is similar to what we
know from the electroweak Standard Model, the corresponding checks show 
that the effects of shifting the width by radiative effects leads to much 
more stable results ({\it i.e.} smaller effects). More precisely, using only
the Belle data corrected for IB effects as sketched above $and$ the $e^+e^-$
data from CMD--2, SND and KLOE10 (the ${\rm NSK+KLOE10}+\tau$ case), the shift for
$a_\mu$ does not exceed $\simeq  0.5 ~10^{-10}$.

If one accepts the evaluation method described above, a first estimate 
of the ($\delta M_\rho$, $\delta \Gamma_\rho$) effects   on  $a_\mu^{th}$ is 
$\delta a_\mu^{\rm had,LO}[\pi\pi,\tau]=\Delta a_\mu(\tau+{\rm PDG})-\Delta a_\mu({\rm NSK+KLOE10}+\tau)=-1.81$,
 not that far from the Belle estimate 
 ($\delta a_\mu^{\rm had,LO}[\pi\pi,\tau]\simeq -1.4$).

The difference 
$\Delta a_\mu(\tau+{\rm PDG})-\Delta a_\mu({\rm NSK+KLOE10}+\cancel{\tau})$,
obviously carries the same information as $\tau+{\rm PDG}$ does not
depend on $e^+e^-$ data and ${\rm NSK+KLOE10}+\cancel{\tau}$ does not depend
on $\tau$ data  and, thus, does not call for any correction. This evaluation yields 
 $-3.81$, about twice the preceding estimate and in the same direction. 

How can one interpret the issue? We get $\chi^2_{NSK+KLOE10}/N_{\pi^+  \pi^-}=204.27/202$
by fitting ${\rm NSK+KLOE10}+\tau$;
correspondingly, $\chi^2_{NSK+KLOE10}/N_{\pi^+  \pi^-}=201.59/202$
when fitting ${\rm NSK+KLOE10}+\cancel{\tau}$. So, the fitted 
$e^+e^- \to \pi^+  \pi^-$ becomes closer to the $\pi^+  \pi^-$ data, 
but the ``improvement'' looks so marginal, that the effect can
also be interpreted as a statistical fluctuation. 
  
One may prefer favoring the most constrained result, {\it i.e.}
$\Delta a_\mu(NSK+KLOE10+\tau)$. Indeed, in this case, 
the estimation of IB effects is  certainly much more constrained 
 by having simultaneously  data strongly affected by
IB effects ($e^+ e^-$) and data marginally depending on them ($\tau$). 
The single motivated uncertainty this provides on 
 $a_\mu^{th.}(NSK+KLOE10+\tau)$ is a (possible) negative shift ($-2~10^{-10}$). 

\subsubsection{Other Sources Of Systematics}

\indent \indent 
The discussion above leads us to conclude that the most favored value for  
 $\Delta a_\mu=a_\mu^{exp}- a_\mu^{th}$
is $39.91 \pm 5.21$ (in units of $10^{-10}$) derived from using ${\rm NSK+KLOE10}+\tau$.
The comparison with data performed in
Subsection \ref{amuRef} tends to indicate that this value should be almost unbiased. 

However, the ambiguity about the choice of the 
Orsay phase discussed in Subsection \ref{PhiRegion} may also indicate a possible systematic shift,
likely of $0.59 \times 10^{-10}$, but certainly limited to  $2.73 \times 10^{-10}$.
This upper bound  supposes that the BaBar spectrum behavior in the $\phi$ region
should be trusted\footnote{One may remark that the pion form factor
fitting function used in \cite{BaBar2} does not include a $\phi$ term. }.
Only a devoted measurement of the pion form factor in the $\phi$ region may allow to conclude about
this possible systematic shift. 

Looking at the differences between BHLS::A and BHLS::B, one also observes that
the most significant difference between both solutions is concentrated -- as
could be expected -- in the contribution of the 3--pion data around the $\phi$ peak.
This may lessen   $\Delta a_\mu$ by $1.26 \times 10^{-10}$.

On the other hand, it is worth noting the behavior of the predicted pion form factor
in the region $m_{\pi\pi} < 0.5$ GeV down to the negative $s$ region. 
 Figure \ref{Fig:na7} shows that the pion form factor derived from fits describes quite well
the (highly constraining) spacelike data down to about $-0.15$ GeV$^2$. Whether spacelike data are included
within the data sample submitted to fit (full curve) or not (dashed curve) does not
make any difference (both curves coincide within the thickness of the lines).
This is a noticeable property as, following Section 6 in \cite{ExtMod1}
and reminded in the Appendix, the NA7
data should be rescaled. The rescaling factor $1-\lambda$  is such that $\lambda$
depends on the spacelike data, on their scale uncertainties (0.8 \% for NA7 \cite{NA7}) and on the 
fitting function. Therefore, by ascertaining the threshold behavior of the pion form factor,
Figure \ref{Fig:na7} proves that the contribution of the $\pi \pi$ threshold region to $g-2$
predicted using the NSK and KLOE10 data is reliable and confirmed by the existing spacelike data.
Indeed, these are well accounted for without any regard to their being included
within the data sample submitted to fit (see Figure (\ref{Fig:na7}c)).

\subsubsection{The HLS Favored Estimate for $a_\mu^{th}$}

\indent \indent 

From the discussion above, it follows that the preferred solution is those derived
using NSK+KLOE10+$\tau$. Its estimate for $\Delta a_\mu$ carries a  statistical significance 
of $4.88 \sigma$. On the other hand, Figure  \ref{Fig:gM2_comp} has shown that using 
or not the $\tau$ spectra does not significantly modify
the significance for  $\Delta a_\mu$ ($5.0 \sigma \to 4.9 \sigma$). In contrast,
Figure  \ref{Fig:gM2_comp} also shows that using the KLOE10 data sample increases
the significance by about $ 0.5 \sigma$. 

In view of all these considerations, especially the issues encountered with the $\phi$
region physics information, the most motivated estimate we can propose is~:
 \be
\left \{
\begin{array}{lr}
a_\mu^{th}=(11~659~169.55 + \left[^{+1.26}_{-0.59} \right]_\phi  
+ \left[^{+0.00}_{-2.00} \right]_\tau
\pm 5.21_{th})~10^{-10}~~,&\\[0.5cm]
\Delta a_\mu= a_\mu^{exp}-a_\mu^{th}= (39.35 + \left[^{+0.59}_{-1.26}\right]_\phi 
+ \left[^{+2.00}_{-0.00} \right]_\tau
\pm   5.21_{th} 
\pm   6.3_{exp} )~10^{-10}~~,&\\[0.5cm]
a_\mu^{LO-HVP}=(681.82+ \left[^{+1.26}_{-0.59} \right]_\phi + \left[^{+0.00}_{-2.00} \right]_\tau 
\pm 4.51_{th} )~10^{-10}~~.
\end{array}
\right.
\label{Eqqq}
\ee

The quoted $[\delta a_\mu^{th}]_\phi$ is an estimate of the possible uncertainties affecting the 
$\phi$ region  and emphasized in Subsection \ref{PhiRegion} and also just above; 
the central value is derived from a fit within configuration B and
avoiding the use the SND phase constraint.
 $[\delta a_\mu^{th}]_\phi$  should not be added in quadrature to the theoretical
error but linearly to the central value. Accounting for this possible shift slightly
lessens the significance for $\Delta a_\mu$ from $4.81 \sigma \to 4.67 \sigma$.
The uncertainty attributed to $\tau$ effects, is also as a possible shift
and plays in the opposite direction.

Therefore, Figure  \ref{Fig:gM2_comp} illustrates that the global fit approach
allows for a gain in the significance for $\Delta a_\mu$ of about $1 \sigma$ or more
compared to the traditional methods 
based on numerical integration of the experimental cross sections
\cite{DavierHoecker,DavierHoecker3,Fred11,Teubner}. 
Moreover, 
our approach  has allowed us to define the largest set of data samples which exhibit 
full consistency with each other. 

The results involving the fit
of the scan $\oplus$ (all/selected) ISR data, do not lead to a noticeably smaller uncertainty for $\Delta a_\mu$
compared to using only scan data, as already inferred in \cite{ExtMod3}. Indeed, Figure  \ref{Fig:gM2_comp}
indicates that in  changing the fit
conditions  [scan $\oplus \tau ] \to$ [scan $\oplus \tau \oplus$ KLOE10], the uncertainty
marginally improves ($5.28  \to 5.21$), while the central value moves by 3 units 
($36.88  \to 39.91$), increasing the significance ($4.5 \sigma \to 4.9  \sigma$).

\section{Conclusion and Perspectives}
\label{Conclusion}
\indent \indent
The $e^+e^- \to \pi^+\pi^-$ annihilation channel has been widely studied and
one is faced with several data sets collected by different groups under various conditions.  
Beside the scan experiments  performed by the CMD-2 and SND collaborations which
have produced valuable data samples 
\cite{CMD2-1995corr,CMD2-1998-1,CMD2-1998-2,SND-1998}, experiments using the
Initial  State Radiation (ISR) method have also been performed by the KLOE
and BaBar collaborations and have produced (much) higher statistics data samples
\cite{KLOE08,KLOE10,BaBar}. The noticeable gain in statistics -- for what concerns
the $\pi^+\pi^-$ intermediate state -- is partly balanced by the
issues raised by the dominance of systematic uncertainties.
Indeed, comparing the properties of these data samples reveals 
inconsistencies which leads to somewhat contradictory conclusions
concerning the predicted value for the muon anomalous moment.

It thus becomes a relevant challenge to find a tool able to examine critically the
properties of the various available samples and substantiate their differences.
The  present study proves that the existing data for~:
\begin{itemize}
\item The $e^+e^-$ annihilation to the
$\pi^0\gamma$, $\eta\gamma$, $\pi^+\pi^-\pi^0$, $K^+K^-$, $K^0 \overline{K}^0$ final states,
 \item The dipion spectra in the $\tau^\pm \ra \pi^\pm \pi^0 \nu$ decay,
 \item Some radiative decays (namely, 
 $\pi^0/\eta/\eta^\prime \to \gamma \gamma$, $\rho^\pm \to \pi^\pm \gamma$, $K^* \to K  \gamma$,
 $\eta^\prime \to \omg \gamma$, $\phi \to \eta^\prime \gamma$) as given in the RPP,
\end{itemize}
supplemented with some limited isospin breaking (IB) information, provide a benchmark
able to reconstruct with a noticeable precision the pion form factor measured in the
$e^+e^- \to \pi^+\pi^-$ annihilation. 
As could have been expected, the requested IB pieces should cover the $\omg/\phi \to \pi^+\pi^-$ decays,   
but also the $\rho^0 \to  e^+e^-$ decay, rarely stressed explicitly \cite{Fred11}.

The tool for this prediction is the broken HLS model (BHLS) defined and studied in \cite{ExtMod3}.
In this last reference, the scan data for the $e^+e^- \to \pi^+\pi^-$ annihilation channel
were used. However, the present study shows that replacing these data by only the
$\rho^0/\omg/\phi$ decay data just referred to,  allows BHLS  to
pin down all the parameters of the model and provide strikingly precise predictions for  
$e^+e^- \to \pi^+\pi^-$. These have been named "$\tau$+PDG" predictions of the pion form factor.
As these predictions are clearly a new way to formulate $\tau$ based predictions 
for the HVP, BHLS provides also a new tool to explore
the long reported $\tau$ versus $e^+e^-$ discrepancy between the HVP evaluations.

Then, the present study leads to the following conclusions~:
 \begin{itemize}
\item {\bf 1/} There is no mismatch between   $\tau$ based predictions and direct $e^+e^-$ evaluation
of the HVP if one relies on BHLS and, especially, on its isospin breaking scheme. This is essentially
implemented in two steps~: direct breaking at the HLS Lagrangian level followed by vector meson mixing.
 The vector meson mixing, which generates physical vector fields,
 is unavoidable because, at first order in breaking parameters, the ideal
 vector fields are no longer mass eigenstates. This simply confirms the study in \cite{ExtMod3}.

 \item {\bf 2/} The $\tau$+PDG  predictions compare astonishingly to the KLOE data samples
 all along the spectra. Because of their influence on the PDG information, it is somewhat
 paradoxical to observe a difference between the NSK data and the $\tau$+PDG predictions,
 more marked than with both KLOE data samples.
 However, a  $1 \sigma$ modification of the width for $\omg  \to \pi^+\pi^-$ allows
 to recover a good agreement between the NSK data and the $\tau$+PDG predictions.
  In contrast, the predicted pion form factor reveals an important disagreement with BaBar data,
 essentially concentrated  in the $[0.74, 0.78]$ GeV region.
 
 \item {\bf 3/} By varying the fit conditions, it has been found that the KLOE10 and
 the scan (NSK) data samples together lead to global fits with outstanding statistical properties. 
 Even if the KLOE08  data sample has  properties similar to KLOE10
 (see Figures \ref{Fig:kloe08_res} and  \ref{Fig:kloe10_res}), it looks premature 
 to include it within the global fit procedure\footnote{
 A reanalysis of the KLOE08 data has been published after completion of this work
 \cite{KLOE12} which might change the picture. Using this revisited sample,
  however, requires additional work which will be done in due time.
  }.   
 \end{itemize} 
 
 From our analysis, one also  gets~~:  
$$F_\omg \equiv {\rm Br}(\omg \to e^+ e^-) \times {\rm Br}(\omg \to \pi^+ \pi^-)= (1.166\pm 0.036)~10^{-6}~,$$
close to the presently accepted \cite{RPP2010} value and twice more precise, with a corresponding
Orsay phase of $104.73^\circ\pm 0.63^\circ$. This value is in accord with the values 
derived from separate fits to scan, KLOE10 and KLOE08 data samples, reflecting that the
lineshape of their pion form factors  are consistent with each other. In contrast,
the $F_\omg$ value derived from (global) fit to the BaBar data sample is different
by about $(7\div 8) \sigma$.

Concerning the $\phi$ meson region, our analysis
of BaBar data provides 
 ${\rm Br}(\phi \to e^+ e^-) \times {\rm Br}(\phi \to \pi^+ \pi^-)= (3.31\pm 0.99)~10^{-8}$
 in good agreement with the accepted value  \cite{RPP2010}. The corresponding Orsay phase, however,
 seems to disagree with expectations \cite{SNDPhi}; our fits tend  to favor
 $-48.23^\circ\pm 1.88^\circ$, much closer to the SND phase.

\vspace{0.7cm}

Using the KLOE10 and the scan data samples leads to  the
most probable value for the muon anomalous moment~:

$$a_\mu^{th}=(11~659~169.55 + \left[^{+1.26}_{-0.59} \right]_\phi  + 
\left[^{+0.00}_{-2.00} \right]_\tau \pm 5.21_{th})~10^{-10} ~,~$$
which exhibits a significance for $\Delta a_\mu= a_\mu^{exp}-a_\mu^{th}$  at a $(4.7 \div 4.9)\sigma$ 
level, significantly larger than the results fully derived by direct numerical integration 
of the experimental cross sections.  This estimate is
free of any reweighting going beyond the reported uncertainties affecting the
data samples involved in the derivation.

Some additional remarks are worth being made concerning this result and our approach~:

\begin{itemize}
\item {\bf 4/}  The cross sections which should be integrated within the BHLS model
to evaluate $a_\mu^{th}$ can be considered as
an  "effective field theory induced interpolation" between  data points. By using a
relatively fine energy binning, the interpolation uncertainties are certainly
minimized in all regions, and for all cross sections, which exhibit sharp energy 
variations. This is certainly an advantage over the standard method which
should rely on trapezoidal estimation between relatively distant measured data points, possibly
improved by taking somewhat into account the local curvatures.

\item {\bf 5/}  One has certainly noted that the value for  $a_\mu^{th}$ we privilege
is smaller than all estimates involving $all$ ISR data. As explained above, this
is because the effects of the KLOE sample is not balanced by the BaBar data.
However, as noted several times above, the value for $F_\omg$ is a criterion
which leads us to conclude that NSK and KLOE10 only have homogeneous properties
which justifies to privilege a global simultaneous fit of these.

\item {\bf 6/}  The description of the
 BaBar data within the global fit framework does not look worse than 
 in the really standalone fit published by this Collaboration.
Actually, the main issue is not that much the description
of the BaBar data sample {\it stricto sensu} than its consistency 
with the underlying physics correlations with other
channels implied by the global approach within the same energy range.
As, it is unlikely that
BHLS may have some inherent reason to exhibit some tropism towards the scan or KLOE
data, we indeed consider the   $a_\mu^{th}$ value we propose as the most motivated
one on statistical grounds.  Of course, new ISR data one can expect from Belle
are clearly valuable.
\end{itemize}

Nevertheless, the results derived using all $\pi^+\pi^-$ ISR data samples have been examined
(see Figure \ref{Fig:gM2_final}) and shown to agree with the favored estimate 
just quoted. Because of the properties expected from the (BHLS) global framework recalled
in the Introduction, using only the KLOE10  and NSK data samples  provides already as
precise results as those derived using all available 
ISR data samples and the traditional evaluation method of $a_\mu^{LO-HVP}$. This
allows to avoid arguing on the relevance of (re)weighting procedures, which are
always delicate matters.

 Our work tells that the significance for $\Delta a_\mu$ starts to be close
 to the $5\sigma$ level. How close to this value it could be, should be confirmed
 by more precise annihilation data in the $[0.95~,1.05]$ GeV mass region, especially
 in the $\pi^+\pi^-$, $\pi^+\pi^-\pi^0$ final states. Finally, new measurements for the
 experimental value of the muon $g-2$ are planned \cite{LeeRoberts,Iinuma}; they
 are important so as to confirm the central value for $a_\mu^{exp}$, and also
 to lessen the experimental uncertainty which starts now to be dominant when
 estimating the significance for $\Delta a_\mu$. 
 
\section*{Acknowledgments}
 F.J. thanks for the support by the EC Program {\it Transnational Access 
 to Research 
Infrastructure} (TARI) INFN - LNF, HadronPhysics3 - 
Integrating Activity, Contract No. 283286.
\clearpage
\newpage
\section*{\Large{Appendices}}
\appendix
\section{Graphical Account of Correlated Uncertainties}
\label{Appendix}
\label{AA}
\indent \indent
Most of the $\pi \pi$ data we are dealing with are subject to systematic
bin--to--bin correlated uncertainties. In the most simple case, this 
is only a constant (not $s$--dependent) scale uncertainty with
a supposed standard deviation $\sigma$. In this case, the $\chi^2$ to be
minimized can be written~:
\be
\chi^2= \displaystyle 
\left [ m- M -A \lambda\right ]^T V^{-1} \left [ m-  M -A \lambda\right ]
+\frac{\lambda^2}{\sigma^2}
\label{seq1}
\ee
where $m$ is the vector of measurements, $M$ is the corresponding vector of model 
values and $V$ is the statistical error covariance matrix, which generally 
absorbs the parts of systematics which are not bin--to--bin correlated. 
An appropriate choice is $A=m$. $\lambda$ is the parameter which reflects the
scale uncertainty; it can be considered as a random variable and each particular
experiment can be viewed as a particular sampling of this. Assuming $\lambda$
independent of the measurements and of the model parameters, its particular value is 
fixed by solving $\partial\chi^2/\partial  \lambda =0$ which yields~:
 \be
\displaystyle \lambda =
\frac{A^T  V^{-1} \left[m-M \right]} 
{A^T V^{-1} A +\displaystyle \frac{1}{\sigma^2}}
\label{seq2}
\ee
\indent \indent
Substituting this into Eq. (\ref{seq1}) leads to the standard $\chi^2$ 
to be minimized when dealing with a constant bin--to--bin correlated 
uncertainty (global scale error)~:
 \be
\begin{array}{ll}
\chi^2= \displaystyle 
\left [ m- M \right ]^T W^{-1}(\sigma^2) \left [ m- M \right ]~~~,~~&
 \displaystyle 
\left ( W(\sigma^2) = [ V + \sigma^2 A A^T  ]\right )
\end{array}
\label{seq3}
\ee

When plotting superimposed the data and the model function,
or their difference (standard residual distribution), the drawing
may be in $visual$ contradiction with the final $\chi^2$ obtained
when minimizing Eq. (\ref{seq2}). This generally happens when the
scale uncertainty is not very small.
However, Eq. (\ref{seq1}) clearly
tells\footnote{Indeed, formally, if one assumes that 
the value for the sampling $\lambda$ of the random variable is known
for sure, it is clear that  $m- \lambda A-M$ makes  sense better
than $m-M$, from the point of view of the minimizer.
} that it is not $m$ which should compared to  $M$
but $m- \lambda A$; this has been accounted for while plotting 
 the spacelike
data \cite{NA7,fermilab2} in Figure (\ref{Fig:na7}). A similar
issue is encountered with the (standard) residuals $m_i- M_i$ 
and is solved by plotting the  "corrected residuals"
$m_i- M_i- \lambda A_i$ rather than $m_i- M_i$.

With the advent of ISR data, the structure of systematic uncertainties
has become much more complicated~: there is generally a large number
of independent bin--to--bin correlated uncertainties\footnote{Several constant
bin--to--bin correlated uncertainties can be summed up to only one 
scale uncertainty.} 
(10 for BaBar \cite{BaBar},
5 for KLOE08 \cite{KLOE08}, 13 for KLOE10 \cite{KLOE10} \ldots). In all cases 
\cite{BaBar,KLOE08,KLOE10}
most of these are additionally $s$--dependent. 

So, let us assume for definiteness
that some data sample -- represented by its measurement vector
 $m$ of length $n_i$ and its $n_i \times n_i$ statistical error covariance matrix $V$ --
 is subject to $n_\alpha$ independent
 bin--to--bin and $s$--dependent correlated uncertainties. These are
 generally considered
 as independent random variables of zero means and are  supposed 
  to carry  $s$--dependent standard deviations 
 ($\sigma_\alpha(s)$,   $\alpha=1, \cdots n_\alpha$).  In this case, Eq. (\ref{seq1})
should be generalized to\footnote{This generalization becomes obvious if
one performs in Eq. (\ref{seq1}) the change of variable $\lambda=\sigma \mu$). }~:
\be
\chi^2= \displaystyle 
\left [ m- M -B_\alpha \lambda_\alpha
\right ]^T V^{-1} \left [ m- M -B_\beta \lambda_\beta
\right ]
+\lambda_\alpha \lambda_\beta \delta_{\alpha \beta}
\label{seq4}
\ee
(summation over repeated greek indices is understood), defining
the $n_\alpha$ vectors $B_\alpha$ by their components 
$[B_\alpha]_i=\sigma_\alpha(s_i) m_i$ and using otherwise obvious notations.

In this case, the $\lambda_\alpha$ parameters can be considered as
independent random variables each of zero mean and of unit standard deviation.
Eq. (\ref{seq4}) shows that $n_\alpha$ scale parameters are to be fixed,
one for each of the bin--to--bin correlated uncertainty functions.
A trivial mathematical recurrence, using in sequence the solution for
$\partial\chi^2/\partial  \lambda_\alpha=0$ (cf. Eq. (\ref{seq2})),
allows to prove that the $\chi^2$ to be minimized becomes~:
\be
\begin{array}{ll}
\chi^2= \displaystyle 
\left [ m- M \right ]^T W^{-1}  \left [ m- M \right ]~~~,~~&
\displaystyle 
\left ( W  =  [ V +  \sum_\alpha B_\alpha B_\alpha^T  ]\right ).
\end{array}
\label{seq5}
\ee

Then, one simply gets $ W_{ij}=[V_{ij} + 
\sum_\alpha\sigma_\alpha(s_i) \sigma_\alpha(s_j)  m_i m_j]$ as expected
for independent bin--to--bin correlated uncertainties. No explicit
knowledge of the values for the various $\lambda_\alpha$ is needed
for this derivation.

For graphical purpose however, one may be interested in getting numerically
the various $\lambda_\alpha$ values. One can easily prove that they are
solutions of the linear system~:
\be
\begin{array}{lll}
\displaystyle 
 \sum_\beta\left[\delta_{\alpha \beta}+G_{\alpha \beta}\right ] \lambda_\beta= H_\alpha
~~~{\rm with}~
& G_{\alpha \beta}=B_\alpha^T V^{-1} B_\beta ~~~{\rm and}~
& H_\alpha=B_\alpha^T V^{-1}\left [  m-M \right ]
\end{array}
\label{seq6}
\ee
which is certainly always regular for systematics small compared with the data.

In this case, plotting together $m-\sum_\alpha\lambda_\alpha B_\alpha$ and $M$
 gives a more appropriate representation than  plotting $m$ and $M$; on the other hand, 
 $m-\sum_\alpha\lambda_\alpha B_\alpha -M$ better reflects the  
 $\chi^2$ value derived by minimizing Eq. (\ref{seq5}) than $m-M$.
 
 This
 does not exhaust all issues due to bin--to--bin correlations in the graphical 
 representation of data and fits; indeed, at least the spread of the model
 function (when derived from fits) adds a usefull piece of information. Anyway,
 as all other correlations (especially those due the statistical errors) are 
 not considered, plotting  $m-\sum_\alpha\lambda_\alpha B_\alpha$ cannot
 exhaust the full issue  but it certainly goes closer to facts.

\newpage
\begin{figure}[!ht]
\begin{minipage}{\textwidth}
\begin{center}
\resizebox{\textwidth}{!}
{\includegraphics*{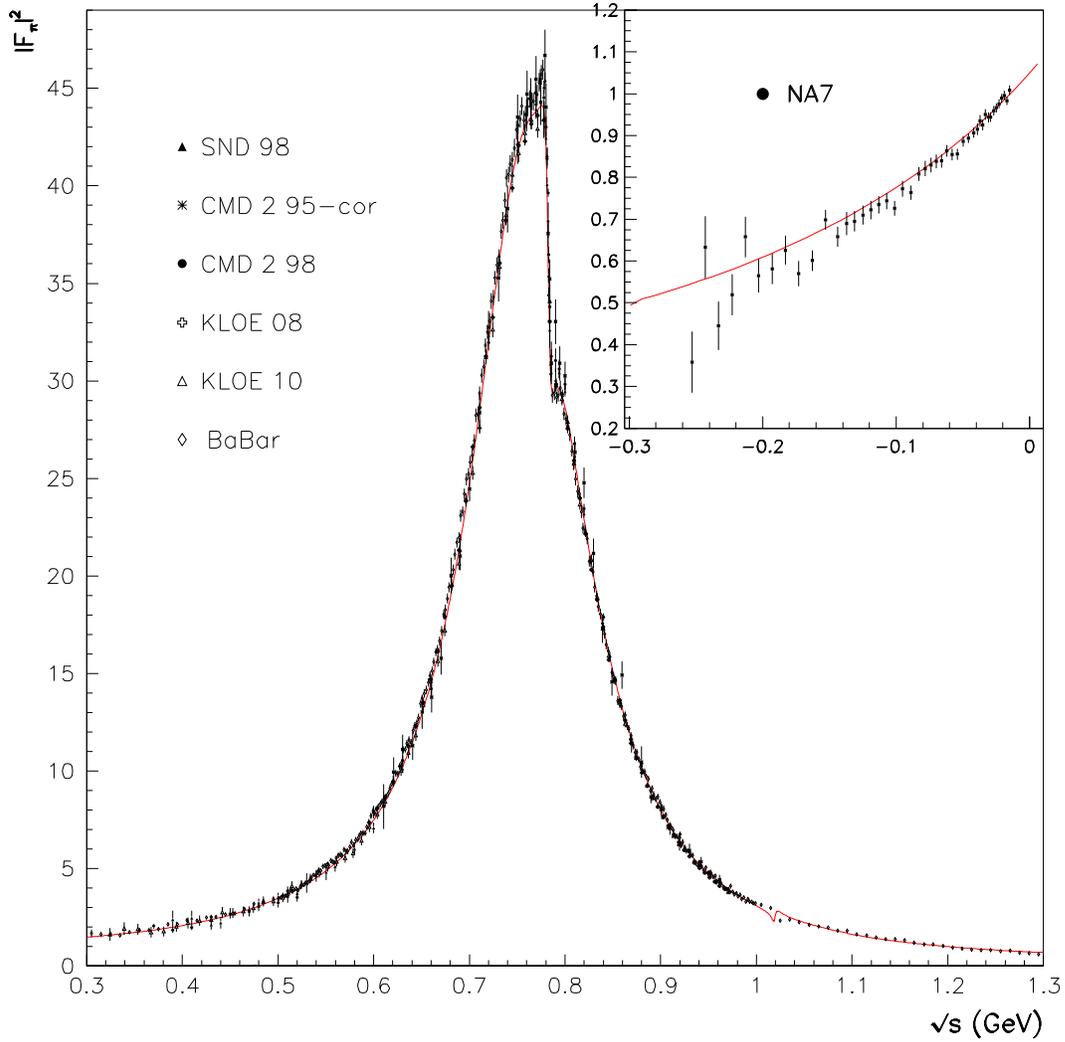}}
\end{center}
\end{minipage}
\begin{center}
\vspace{-0.3cm}
\caption{\label{Fig:taupred_all}
The  Pion Form Factor $prediction$ based on $\tau$ data and PDG information.
The most important experimental data are superimposed; they do not influence
the predicted curve.}
\end{center}
\end{figure}

\begin{figure}[!ht]
\begin{minipage}{\textwidth}
\begin{center}
\resizebox{\textwidth}{!}
{\includegraphics*{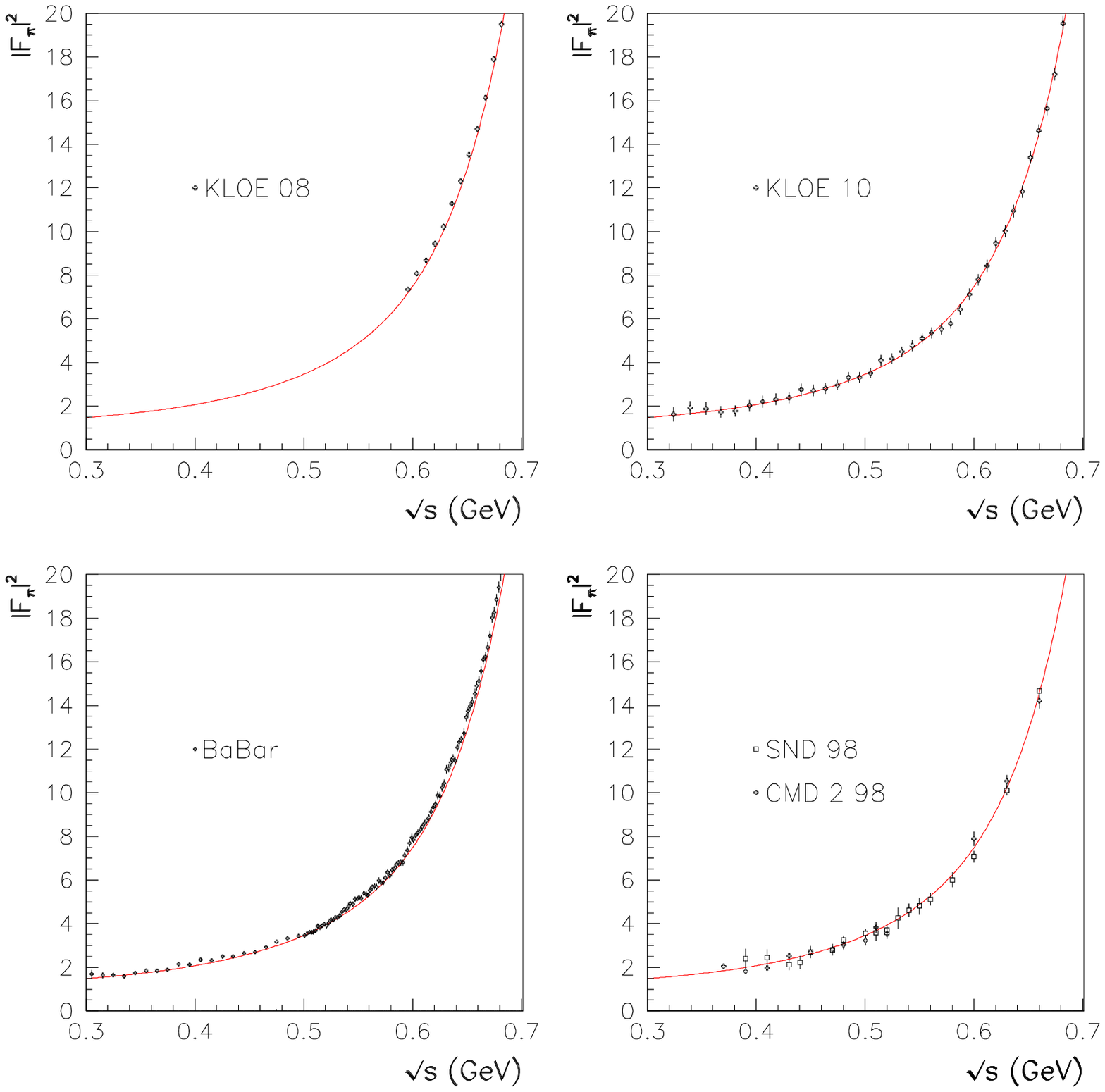}}
\end{center}
\end{minipage}
\begin{center}
\vspace{-0.3cm}
\caption{\label{Fig:taupred_low}
Magnified view of
the  Pion Form Factor $prediction$ based on $\tau$ data and PDG information;
the ($0.3 \div 0.7$) GeV region is shown with the indicated data superimposed. 
}
\end{center}
\end{figure}

\begin{figure}[!ht]
\begin{minipage}{\textwidth}
\begin{center}
\resizebox{\textwidth}{!}
{\includegraphics*{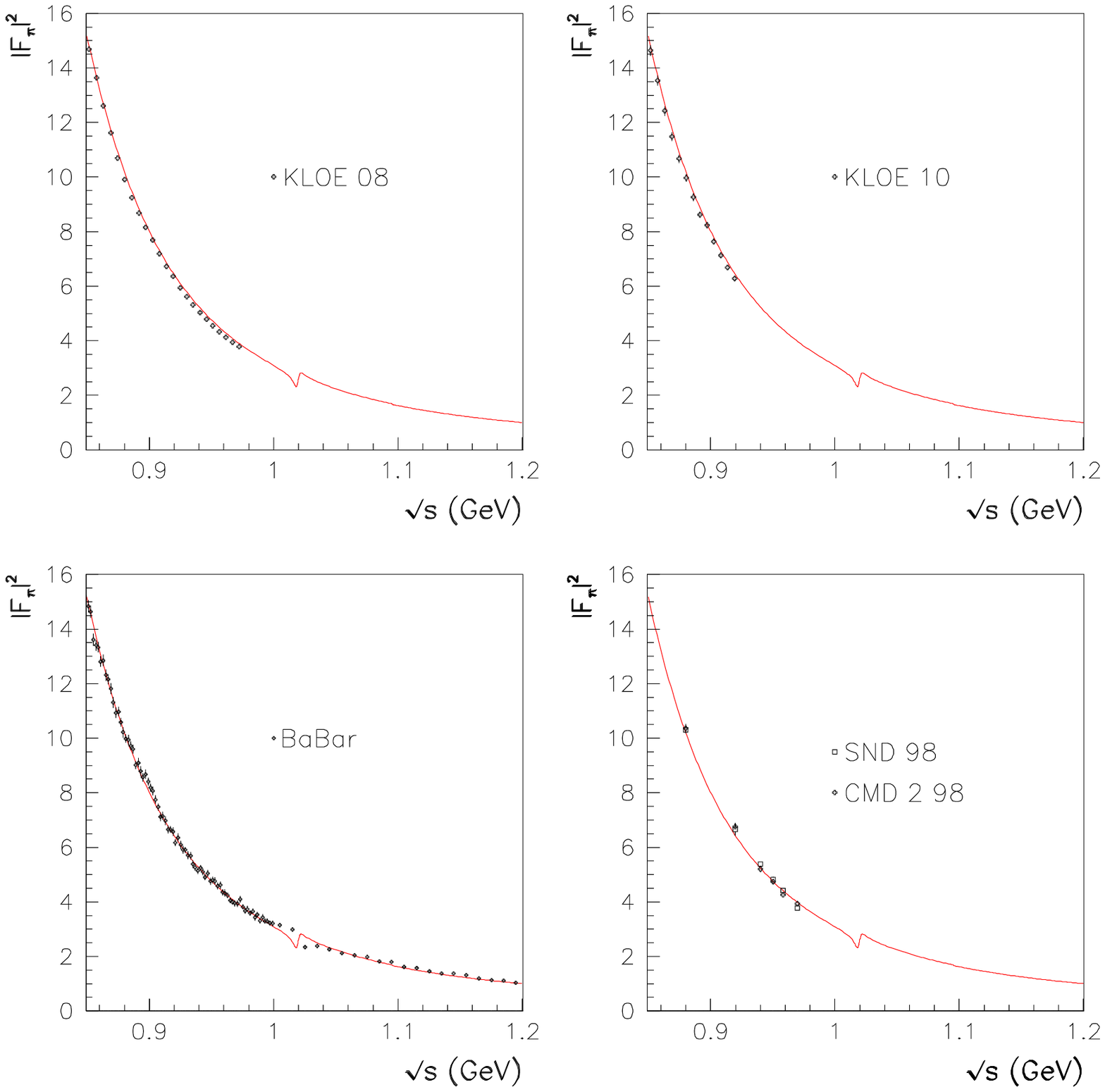}}
\end{center}
\end{minipage}
\begin{center}
\vspace{-0.3cm}
\caption{\label{Fig:taupred_high}
Magnified view of
the  Pion Form Factor $prediction$ based on $\tau$ data and PDG information;
the ($0.85 \div 1.2$) GeV region is shown with the indicated data superimposed. 
}
\end{center}
\end{figure}

\begin{figure}[!ht]
\begin{minipage}{\textwidth}
\begin{center}
\resizebox{\textwidth}{!}
{\includegraphics*{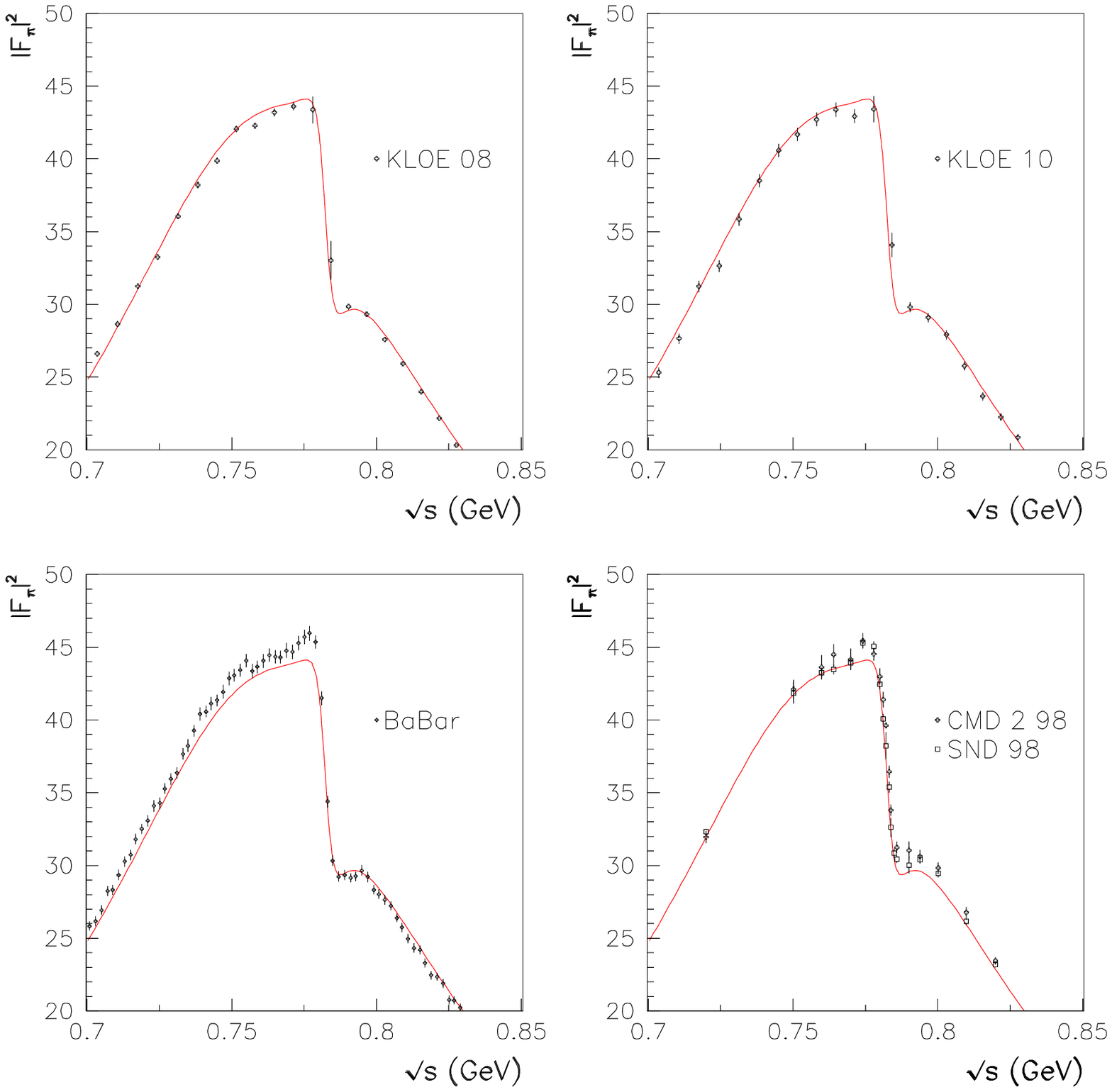}}
\end{center}
\end{minipage}
\begin{center}
\vspace{-0.3cm}
\caption{\label{Fig:taupred_mid}
Magnified view of
the  Pion Form Factor $prediction$ based on $\tau$ data and PDG information;
the ($0.7 \div 0.85$) GeV region is shown with the indicated data superimposed. 
}
\end{center}
\end{figure}

\begin{figure}[!ht]
\begin{minipage}{\textwidth}
\begin{center}
\resizebox{\textwidth}{!}
{\includegraphics*{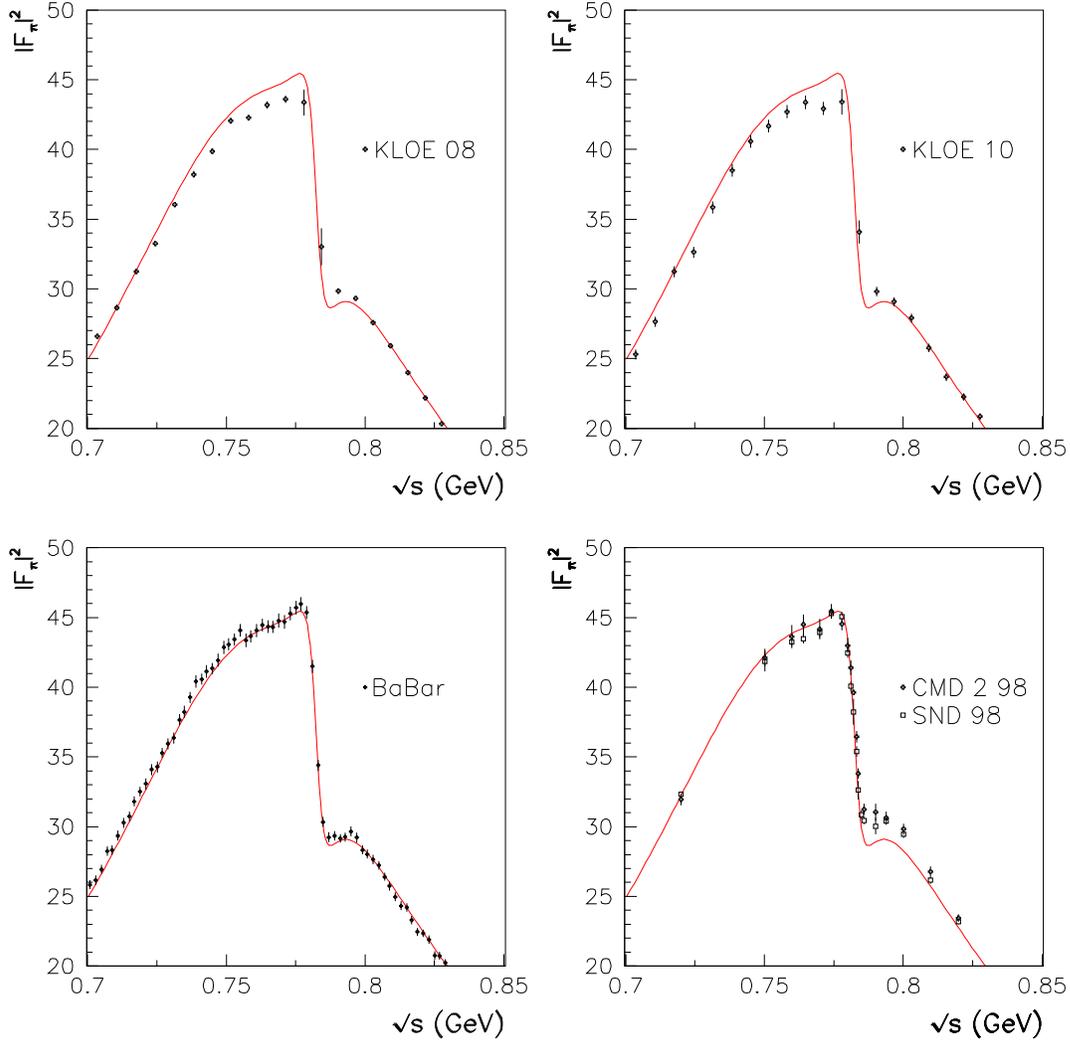}}
\end{center}
\end{minipage}
\begin{center}
\vspace{-0.3cm}
\caption{\label{Fig:tau_bbr_mid}
The  Pion Form Factor $prediction$ based on $\tau$ data  
and the ($0.76 \div 0.82$) GeV region of the BaBar spectrum \cite{BaBar};
the ($0.7 \div 0.85$) GeV region is shown with the indicated data superimposed. 
}
\end{center}
\end{figure}

\begin{figure}[!ht]
\begin{minipage}{\textwidth}
\begin{center}
\resizebox{\textwidth}{!}
{\includegraphics*{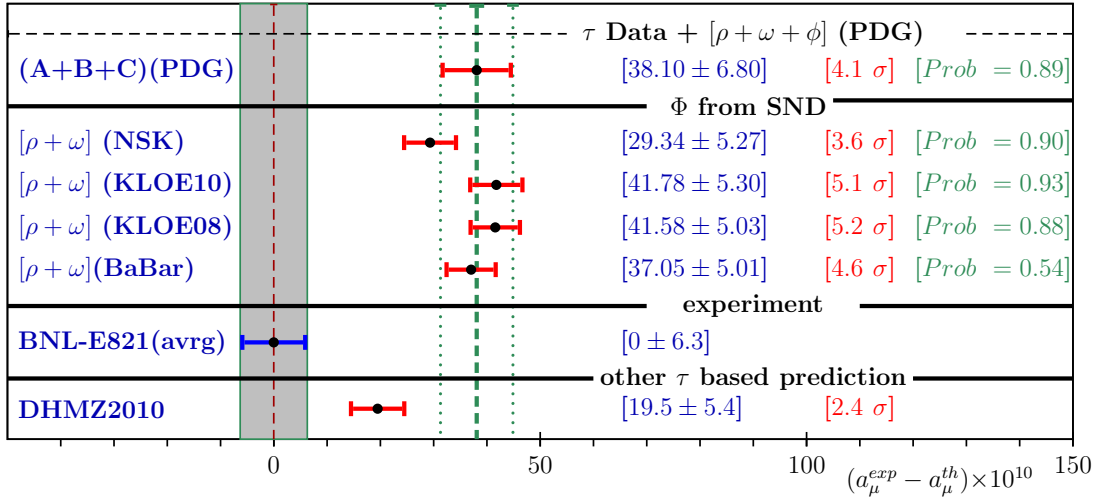}}
\end{center}
\end{minipage}
\begin{center}
\vspace{-0.3cm}
\caption{\label{Fig:tau_gM2}
$\tau$ based estimates using a global fit of the BHLS model (see Section \ref{tauPred}).
The first two numbers in each line display resp. the central value and the r.m.s. of $\Delta a_\mu=a_\mu^{exp}-a_\mu^{th}$.
The last two numbers give resp. the distance to the BNL measurement and the
fit probability -- essentially dominated by the non $\pi^+\pi^-$ data. The  
lower probability found when using the BaBar data, even in the limited energy region 
involved ($[0.76\div0.82]$ GeV), exhibits the tension of this data sample relative 
to the rest of the physics considered.
}
\end{center}
\end{figure}

\begin{figure}[!ht]
\begin{minipage}{\textwidth}
\begin{center}
\resizebox{\textwidth}{!}
{\includegraphics*{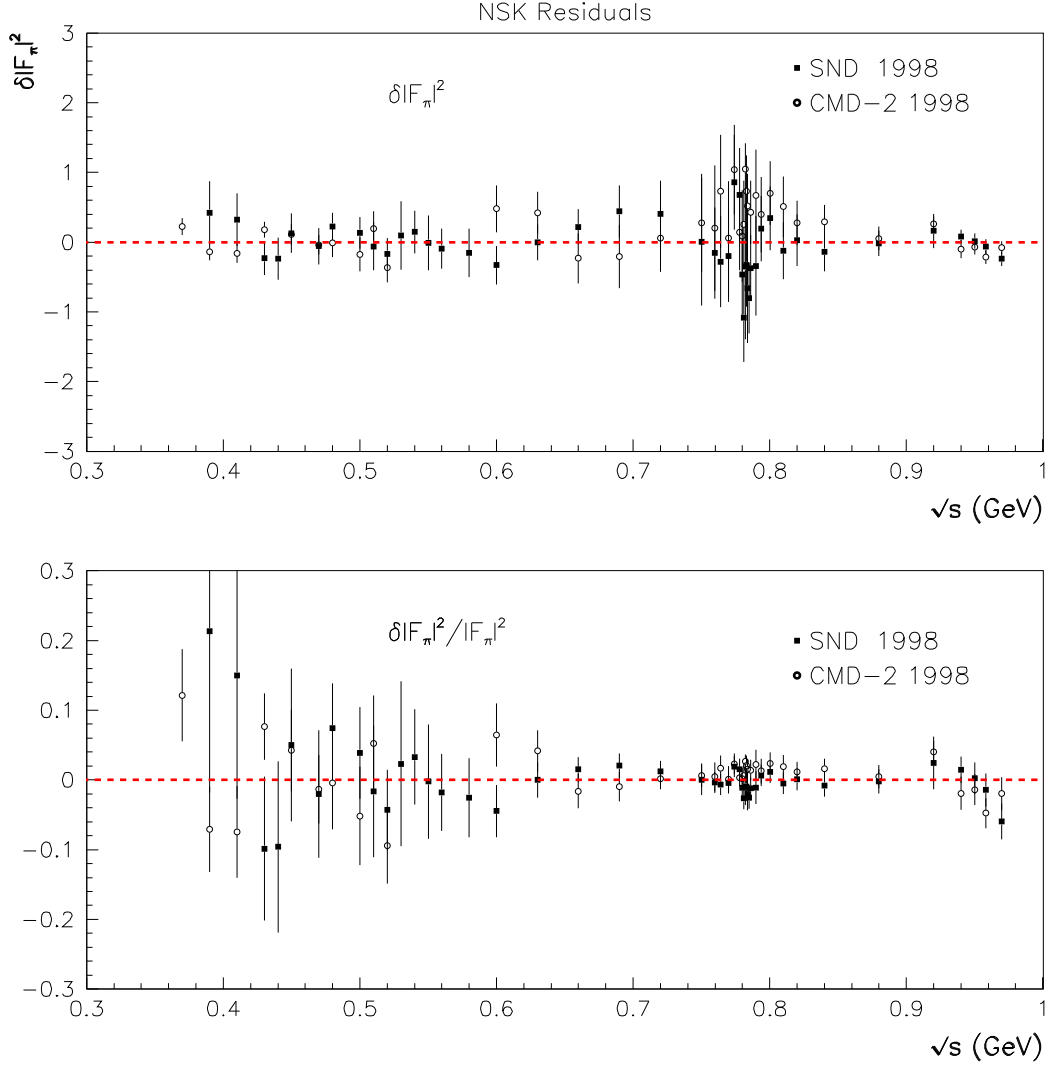}}
\end{center}
\end{minipage}
\begin{center}
\vspace{-0.3cm}
\caption{\label{Fig:nsk_res}
Fit residuals for the CMD--2 and SND  data samples  (NSK)
in isolation$^{\ref{standalone}}$.
The top panel shows the function 
$\delta |F_\pi(s)|^2=|F_\pi^{NSK}(s)|^2-|F_\pi^{fit}(s)|^2$; the bottom
panel displays the distribution $\delta |F_\pi(s)|^2/|F_\pi^{fit}(s)|^2$.}
\end{center}
\end{figure}

\begin{figure}[!ht]
\begin{minipage}{\textwidth}
\begin{center}
\resizebox{\textwidth}{!}
{\includegraphics*{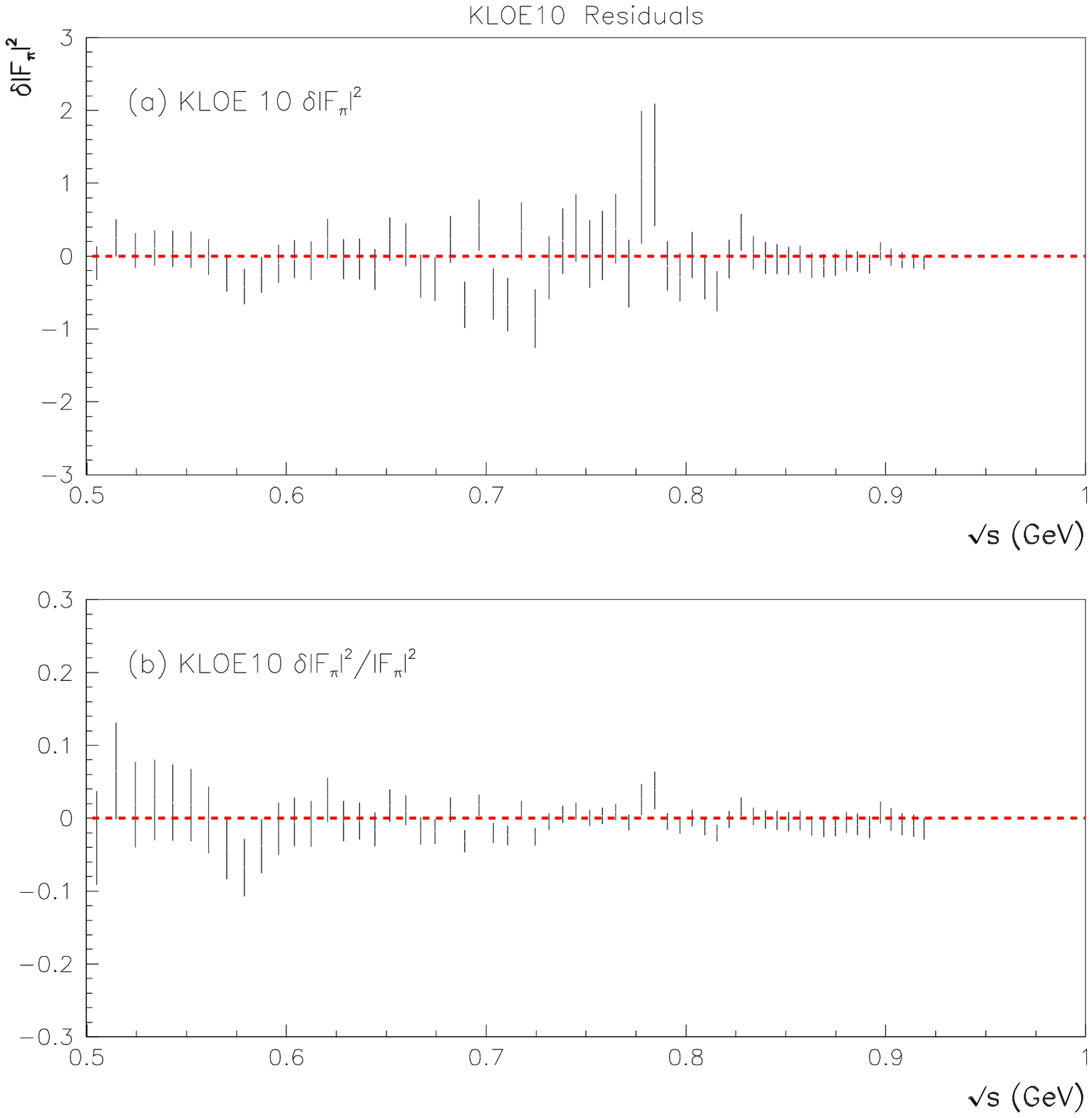}}
\end{center}
\end{minipage}
\begin{center}
\vspace{-0.3cm}
\caption{\label{Fig:kloe10_res}
Fit residuals for the KLOE10 data in isolation$^{\ref{standalone}}$.
The top panel shows the function 
$\delta |F_\pi(s)|^2=|F_\pi^{KLOE10}(s)|^2-|F_\pi^{fit}(s)|^2$; the bottom
panel displays the distribution $\delta |F_\pi(s)|^2/|F_\pi^{fit}(s)|^2$.}
\end{center}
\end{figure}

\begin{figure}[!ht]
\begin{minipage}{\textwidth}
\begin{center}
\resizebox{\textwidth}{!}
{\includegraphics*{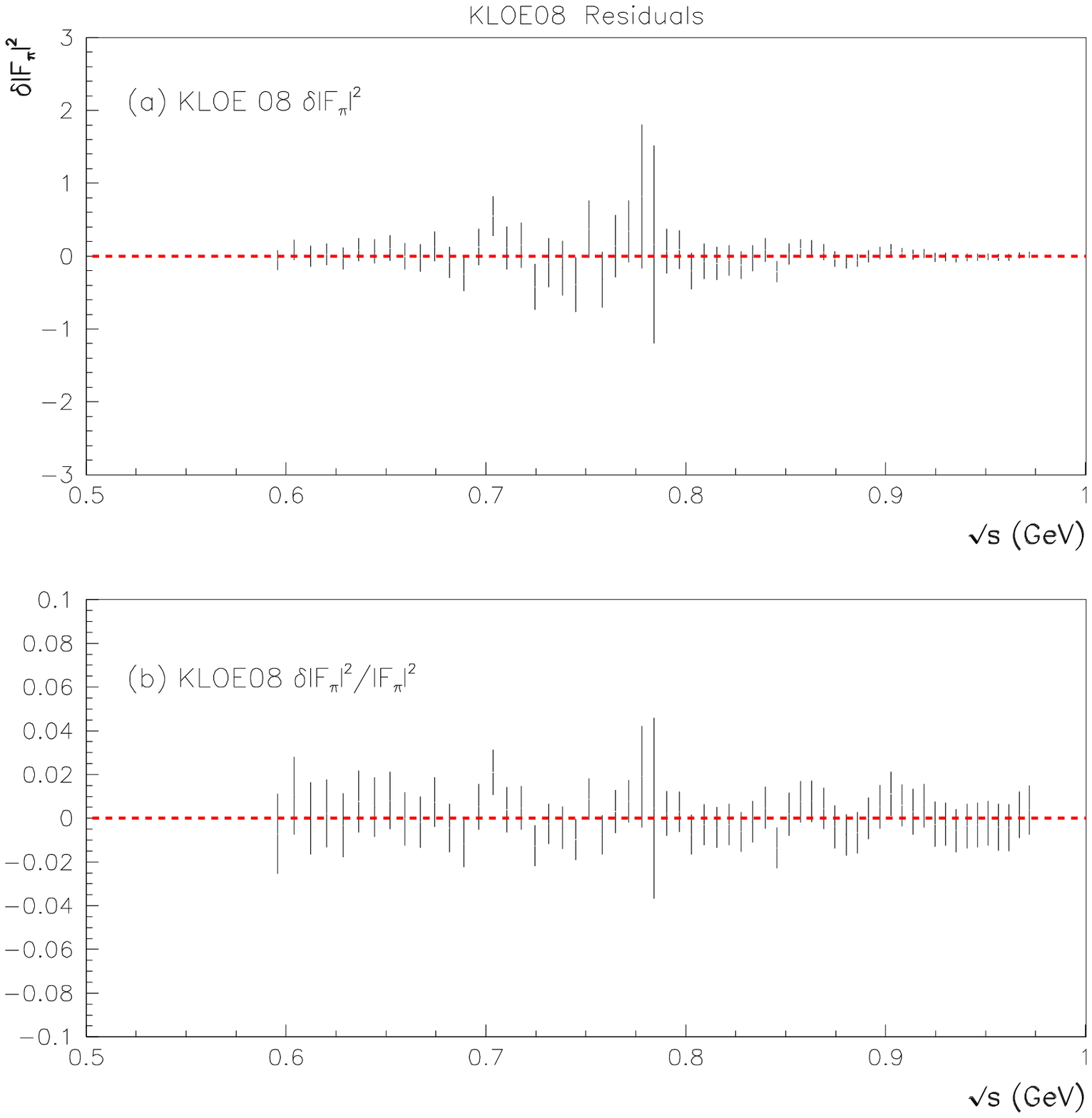}}
\end{center}
\end{minipage}
\begin{center}
\vspace{-0.3cm}
\caption{\label{Fig:kloe08_res}
Fit residuals for the KLOE08 data in isolation$^{\ref{standalone}}$.
The top panel shows the function 
$\delta |F_\pi(s)|^2=|F_\pi^{KLOE08}(s)|^2-|F_\pi^{fit}(s)|^2$; the bottom
panel displays the distribution $\delta |F_\pi(s)|^2/|F_\pi^{fit}(s)|^2$.}
\end{center}
\end{figure}


\begin{figure}[!ht]
\begin{minipage}{\textwidth}
\begin{center}
\resizebox{\textwidth}{!}
{\includegraphics*{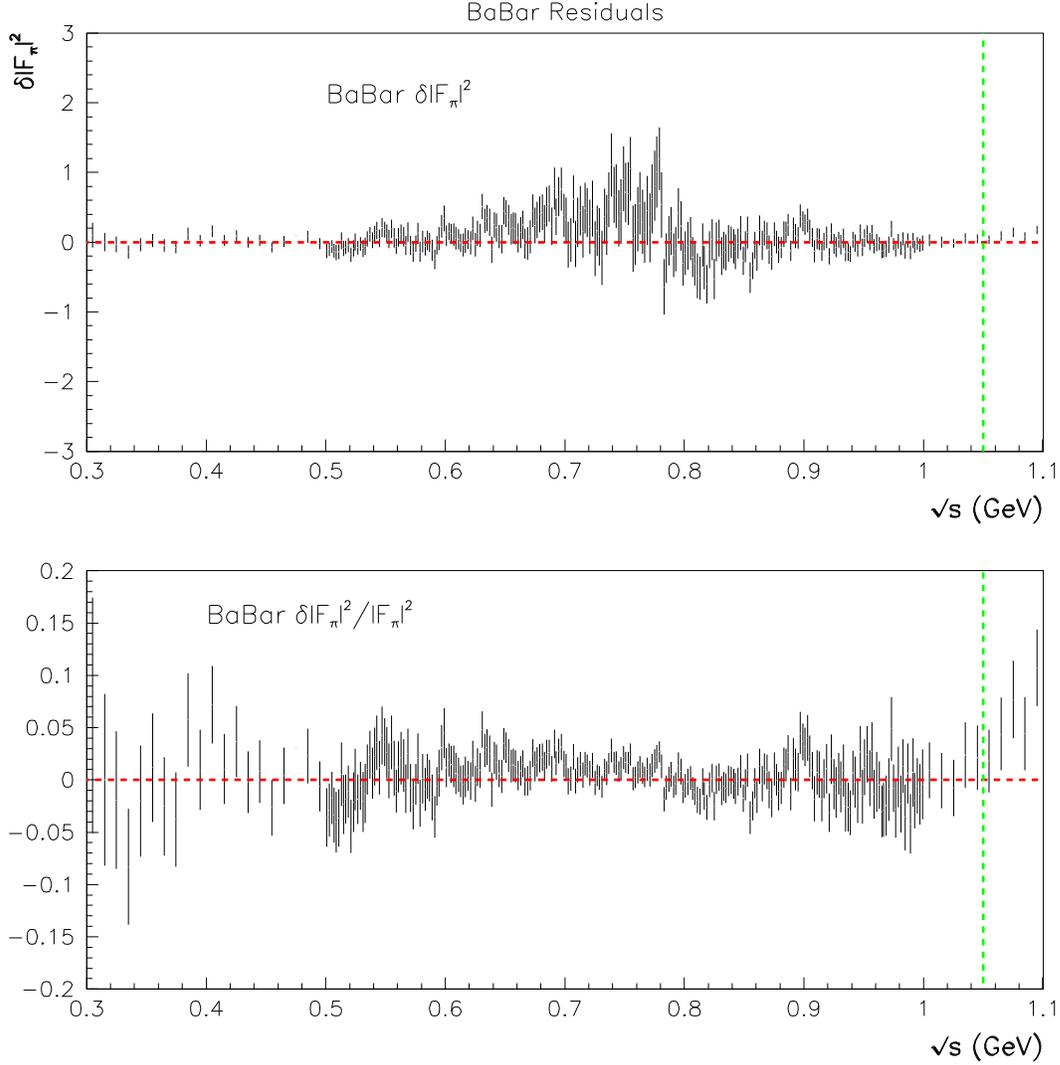}}
\end{center}
\end{minipage}
\begin{center}
\vspace{-0.3cm}
\caption{\label{Fig:babar_res_all}
Fit residuals for the BaBar data sample in isolation$^{\ref{standalone}}$
over the energy range from threshold to 1.05 GeV.
The top panel shows the function 
$\delta |F_\pi(s)|^2=|F_\pi^{BaBar}(s)|^2-|F_\pi^{fit}(s)|^2$; the bottom
panel displays the distribution $\delta |F_\pi(s)|^2/|F_\pi^{fit}(s)|^2$.
The vertical dashed lines indicate the upper end of the fitted spectrum. 
}
\end{center}
\end{figure}

\begin{figure}[!ht]
\begin{minipage}{\textwidth}
\begin{center}
\resizebox{\textwidth}{!}
{\includegraphics*{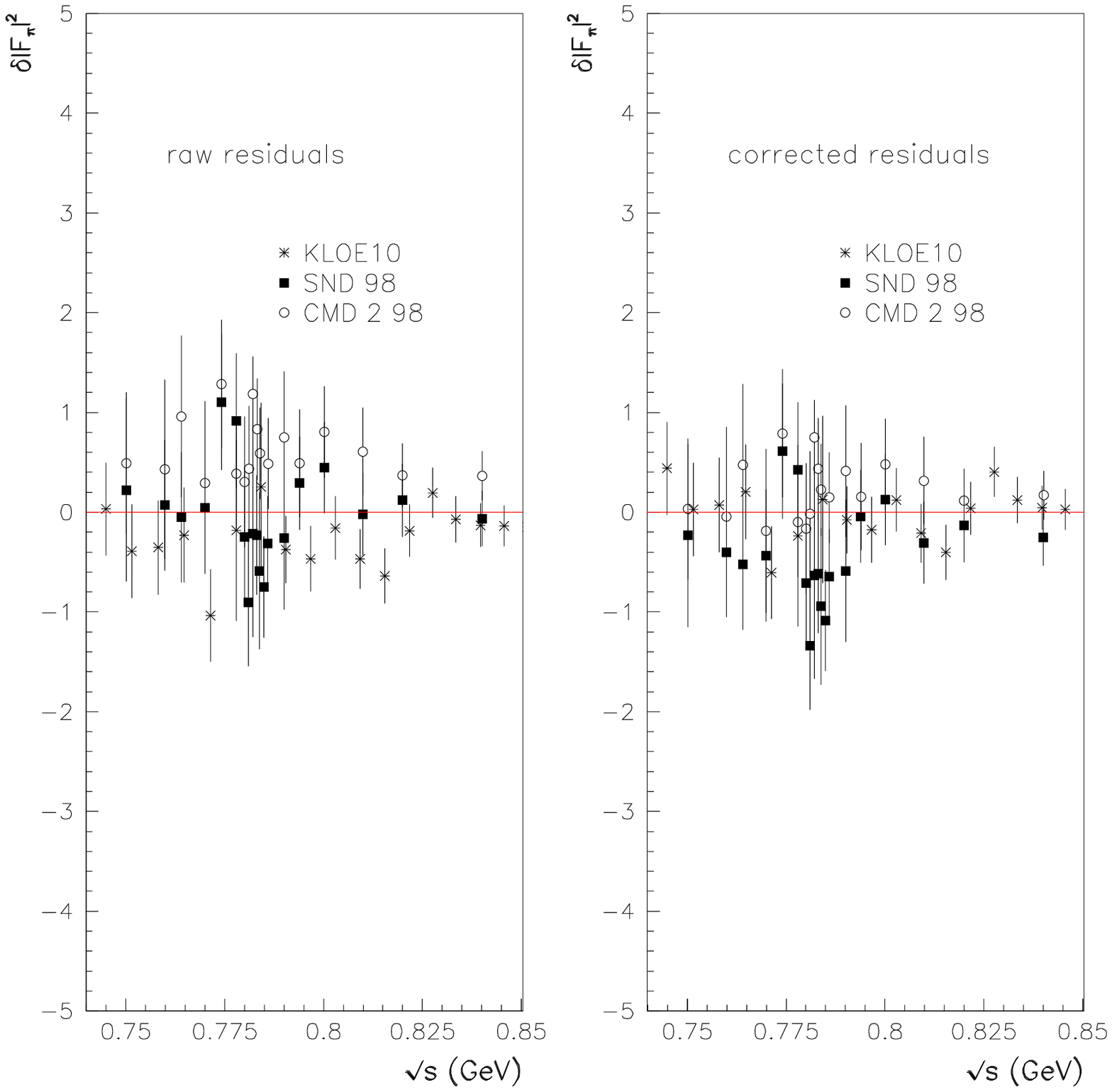}}
\end{center}
\end{minipage}
\begin{center}
\vspace{-0.3cm}
\caption{\label{Fig:andreas_centr}
Fit residuals of the global fit combining NSK and KLOE10.
The leftmost panel displays superimposed the residual distributions
$\delta |F_\pi(s)|^2=|F_\pi^{EXP}(s)|^2-|F_\pi^{fit}(s)|^2$
for each data sample in the central energy region. The rightmost
panel shows the corresponding corrected residual distributions as defined
in the Appendix.
}
\end{center}
\end{figure}

\begin{figure}[!ht]
\begin{minipage}{\textwidth}
\begin{center}
\resizebox{\textwidth}{!}
{\includegraphics*{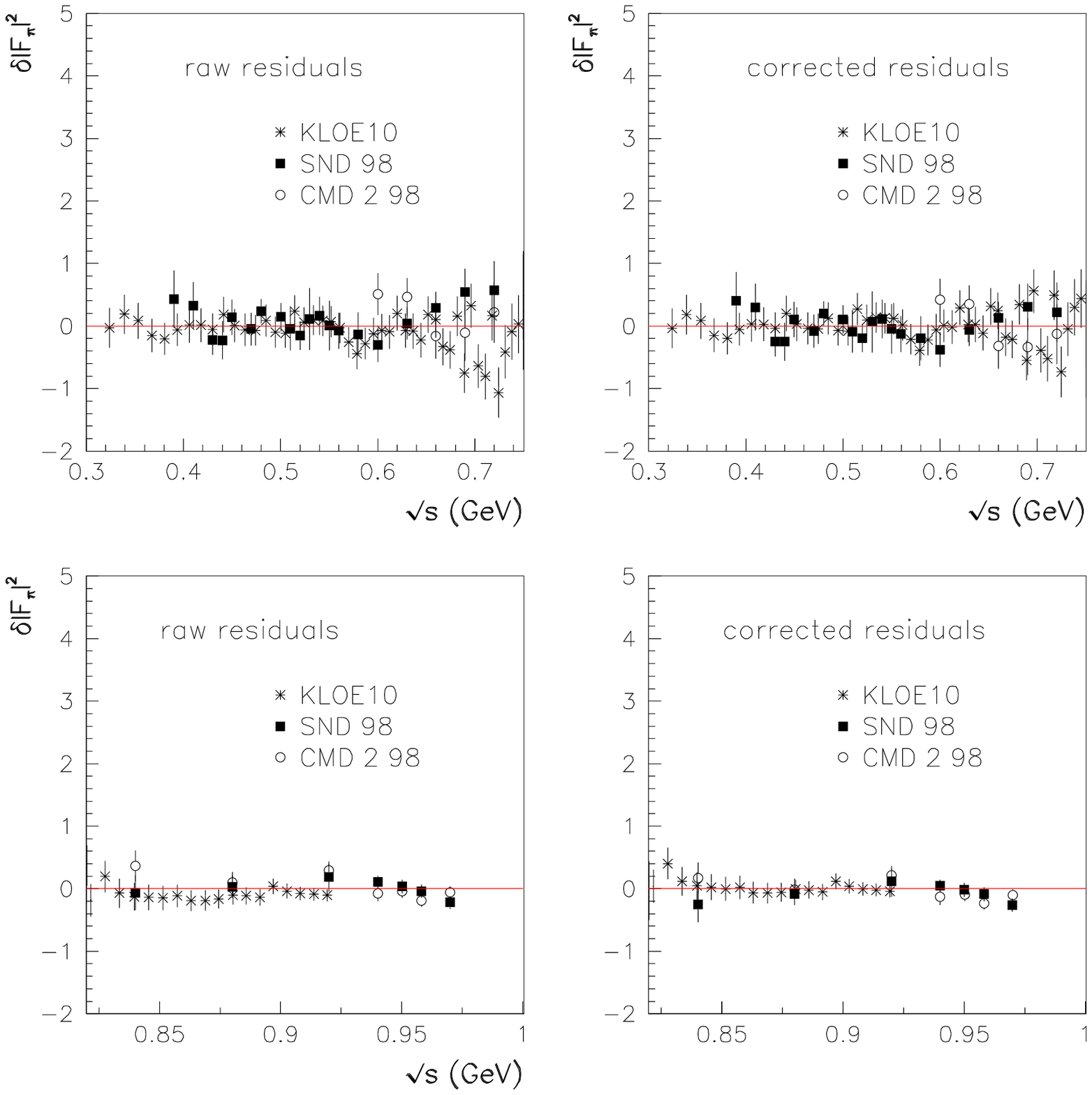}}
\end{center}
\end{minipage}
\begin{center}
\vspace{-0.3cm}
\caption{\label{Fig:andreas_lowhigh}
Fit residuals of the global fit combining NSK and KLOE10.
The leftmost panels display  (superimposed) the residual distributions
$\delta |F_\pi(s)|^2=|F_\pi^{EXP}(s)|^2-|F_\pi^{fit}(s)|^2$
for each data sample in the side energy regions. The rightmost
panels show  the corresponding corrected residual distributions as defined
in the Appendix.
}
\end{center}
\end{figure}

\begin{figure}[!ht]
\begin{minipage}{\textwidth}
\begin{center}
\resizebox{\textwidth}{!}
{\includegraphics*{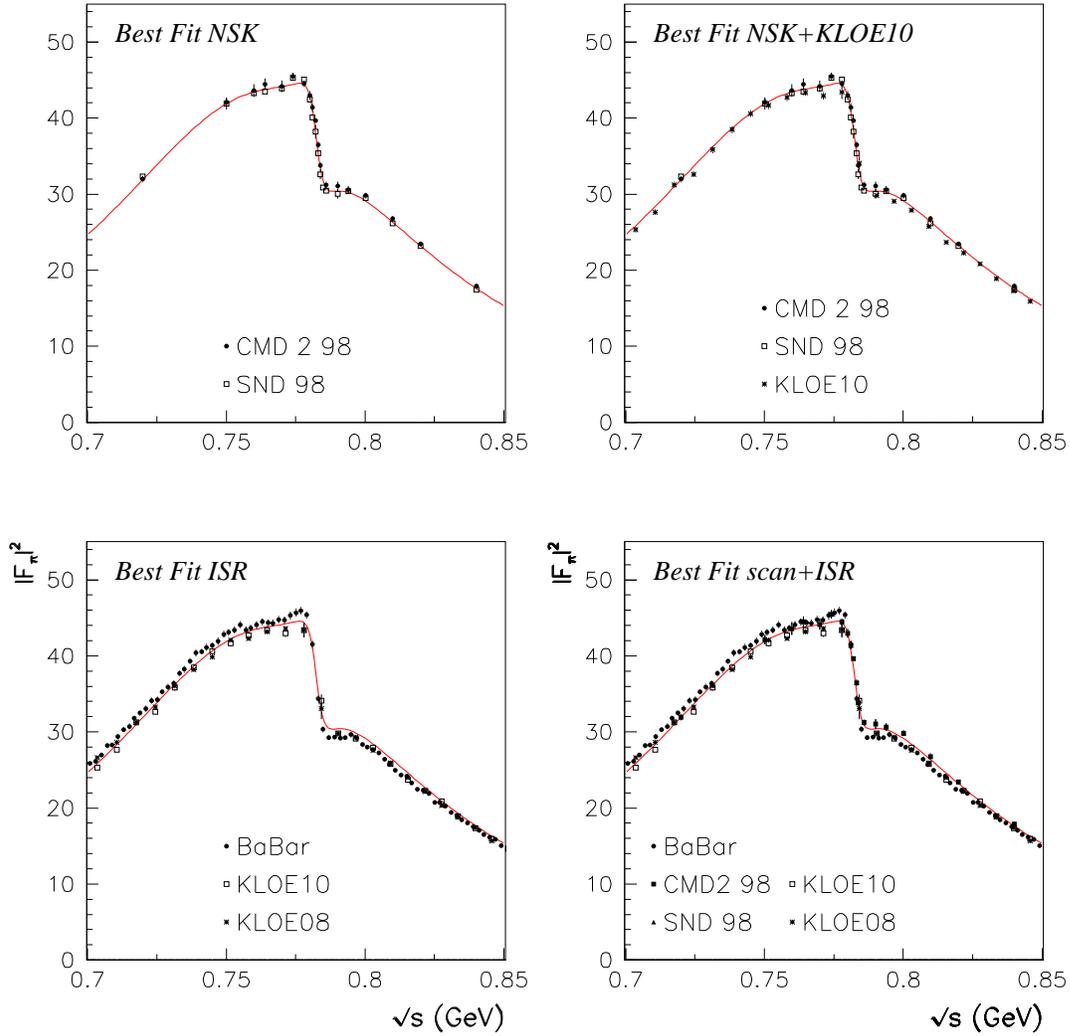}}
\end{center}
\end{minipage}
\begin{center}
\vspace{-0.3cm}
\caption{\label{Fig:rho_omg}
The  Pion Form Factor in the $\rho-\omg$ region from various global fit configurations.
Top left panel displays the best fit using only the CMD-2 and SND data as
$e^+e^-\to \pi^+\pi^-$ spectra. Top right panel shows the case when
the CMD-2, SND and KLOE10 data samples are fitted simultaneously.
Bottom left panel shows the best fits when the data samples considered are those
from BaBar, KLOE08 and KLOE10. Bottom right panel shows the best fit using simultaneously
 the CMD-2, SND, KLOE08,  KLOE10 and BaBar data samples.
}
\end{center}
\end{figure}

\begin{figure}[!ht]
\begin{minipage}{\textwidth}
\begin{center}
\resizebox{\textwidth}{!}
{\includegraphics*{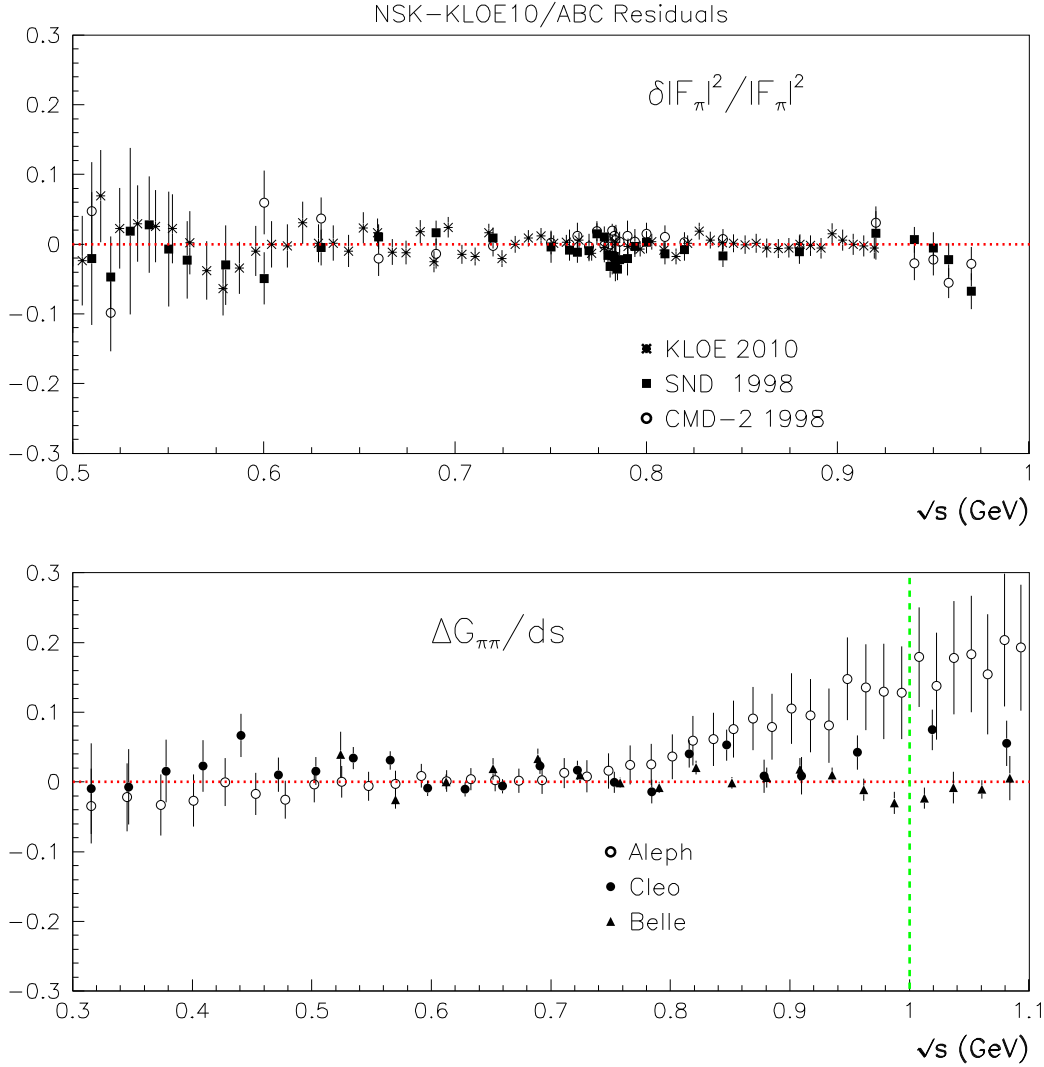}}
\end{center}
\end{minipage}
\begin{center}
\vspace{-0.3cm}
\caption{\label{Fig:nsk_kloe10_abc}
Fractional deviations from the fitting functions for the statistically favored
configuration (NSK + KLOE10).   Top panel shows the case for the
pion form factor in $e^+e^-$ annihilations, bottom 
panel exhibits the case for the $\tau$ dipion spectra.
The corrected   residuals $\delta |F_\pi|^2$, as described in the Appendix, have been 
used for the top panel; as the spectrum in the energy region from 0.3 GeV to 
0.5 GeV does not carry any interesting information, it has been removed from
the Figure.
The vertical line in the bottom panel indicates the end of the fitted region.
}
\end{center}
\end{figure}

\begin{figure}[!ht]
\begin{minipage}{\textwidth}
\begin{center}
\resizebox{\textwidth}{!}
{\includegraphics*{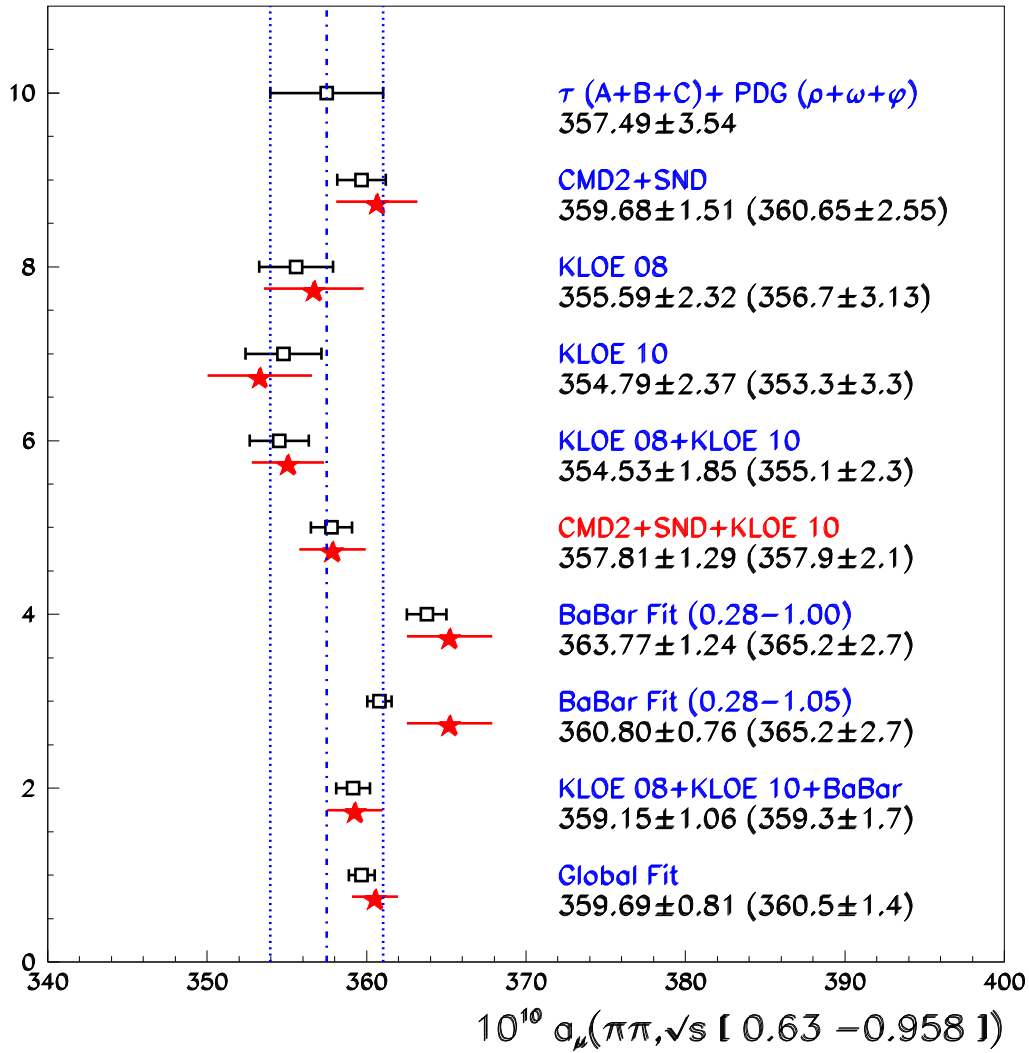}}
\end{center}
\end{minipage}
\begin{center}
\vspace{-0.3cm}
\caption{\label{Fig:amu_ref}
The muon anomalous magnetic moment.
The numbers gives the $\pi \pi$ contribution  
from the $(0.630,~0.958)$ GeV 
invariant mass region. The results from fits are shown with  
empty squares, the experimental estimates by stars.  Fit values are given
followed by the experimental estimate within brackets. Global fit properties favor
the result from the CMD2+SND+KLOE10 combined sample.}
\end{center}
\end{figure}

\begin{figure}[!ht]
\begin{minipage}{\textwidth}
\begin{center}
\resizebox{\textwidth}{!}
{\includegraphics*{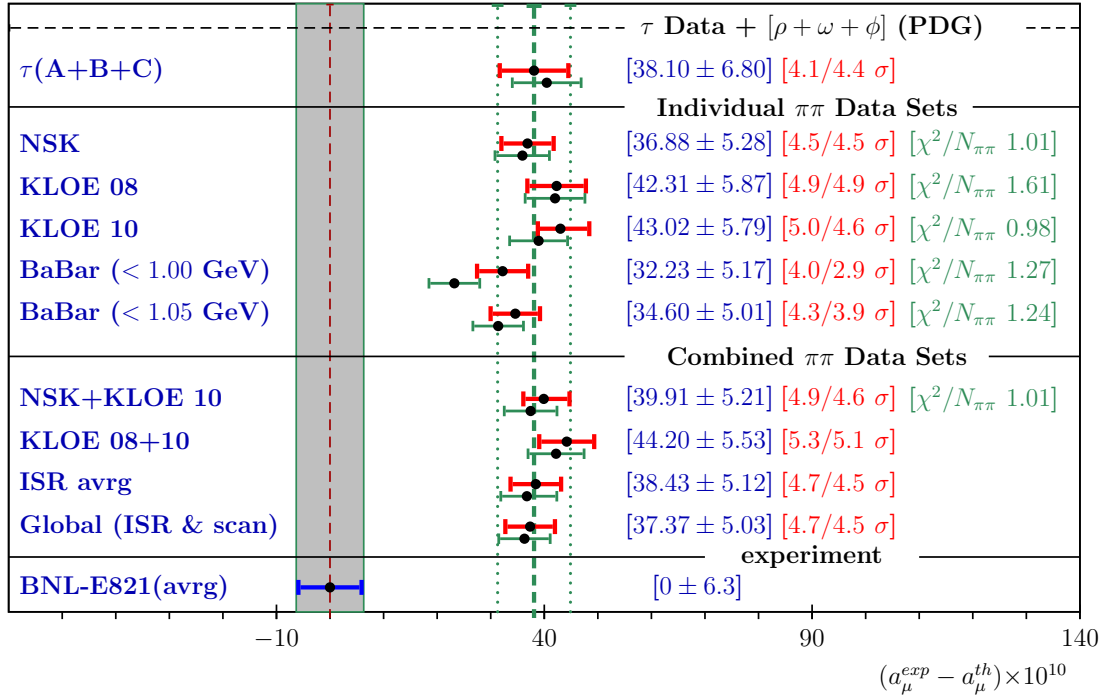}}
\end{center}
\end{minipage}
\begin{center}
\vspace{-0.3cm}
\caption{\label{Fig:gM2_final}
The deviation $\Delta a_\mu \times 10^{10}$ between experiment and theory for 
the anomalous magnetic moment of the muon. The leading hadronic vacuum polarization
contribution has been estimated via global BHLS model fits for different
$e^+e^- \to \pi^+\pi^-$ data samples.The $\tau$ predictions -- using or not 
the spacelike data -- are given in the top pair lines, followed by the fit results 
using each scan or ISR data sample in isolation 
or combined; see text for comments.}
\end{center}
\end{figure}

\begin{figure}[!ht]
\begin{minipage}{\textwidth}
\begin{center}
\resizebox{\textwidth}{!}
{\includegraphics*{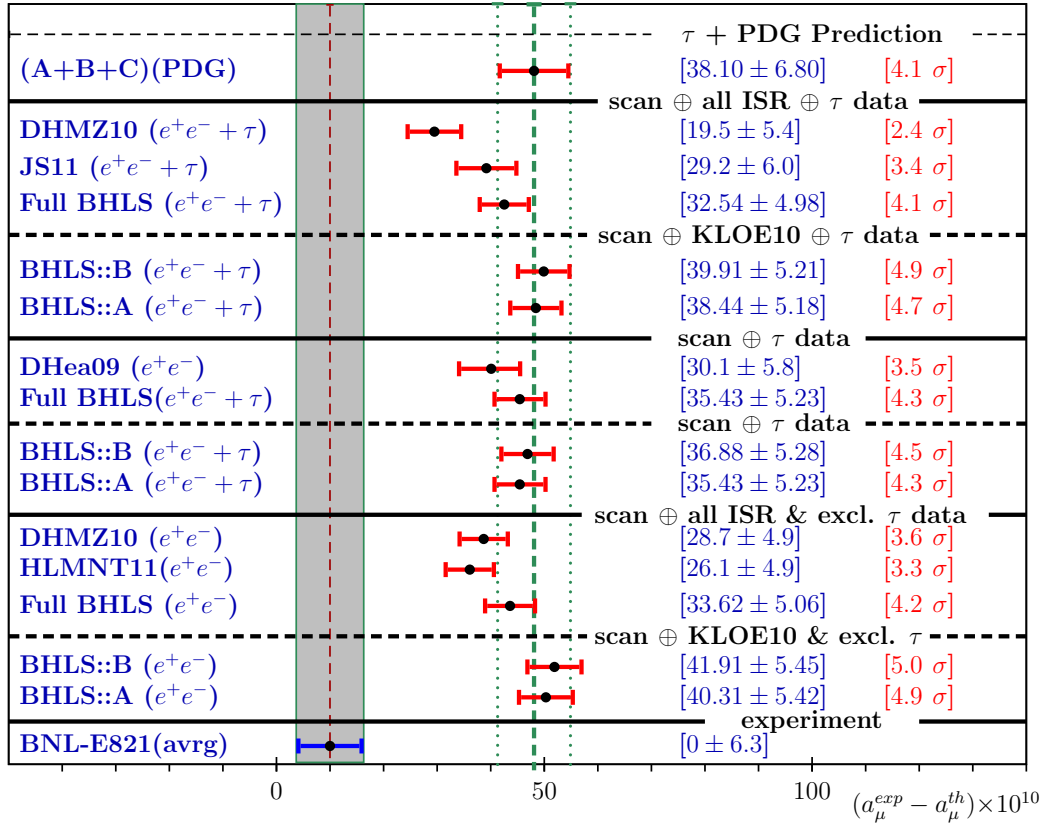}}
\end{center}
\end{minipage}
\begin{center}
\vspace{-0.3cm}
\caption{\label{Fig:gM2_comp}
A set of recent estimates of the muon anomalous magnetic moment
deviation from the BNL average value \cite{BNL,BNL2}. 
Our own (updated) estimates are figured by BHLS::A and BHLS::B for respectively 
 configurations A and B. The results
 obtained using $all$ the ISR data samples inside the global fit procedure
 are displayed under the tag Full BHLS. The statistical significance
of  each $ \Delta a_\mu$ is displayed on the right hand side of the Figure. 
 See the text for more information.
}
\end{center}
\end{figure}

\begin{figure}[!ht]
\begin{minipage}{\textwidth}
\begin{center}
\resizebox{\textwidth}{!}
{\includegraphics*{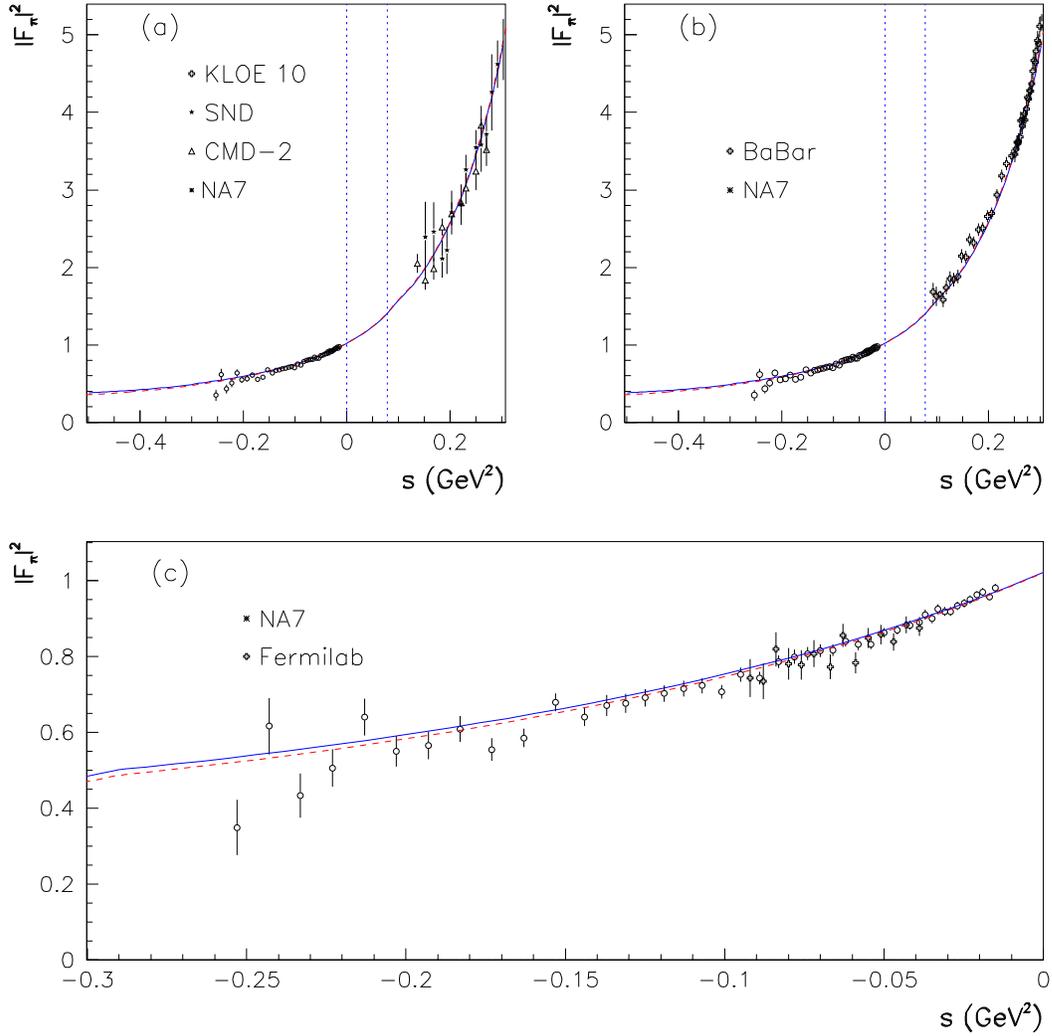}}
\end{center}
\end{minipage}
\begin{center}
\vspace{-0.3cm}
\caption{\label{Fig:na7}
The  Pion Form Factor  around $s=0$. The $\pi^+\pi^- $ annihilation data
submitted to the global fit are $only$ the NSK and KLOE10  samples. The full curve
is the fit function derived when including also the spacelike data from NA7 \cite{NA7} 
and Fermilab \cite{fermilab2}; the dashed curve ris associated with the fit   excluding the spacelike data.
 The BaBar data -- shown in (b) for illustration -- are not submitted to the fit. KLOE08 has no 
 data point located within the plotted window. For clarity,  only 
 the data from NA7 \cite{NA7} are shown in (a) and (b). In (c), one magnifies  
 the spacelike region and plot the 
 data \cite{NA7,fermilab2}, used or not within the fit procedure. The vertical lines in
 (a) and (b) show the locations $s=0$ and  $s=4 m_\pi^2$.
}
\end{center}
\end{figure}


                    \bibliographystyle{h-physrev}
                     \bibliography{vmd4}
\end{document}